\documentclass[twocolumn,apjl]{emulateapj} 
\usepackage{times}
\usepackage{natbib}
\usepackage{rotating}
\newcommand{\oiii}{{[O\,{\sc iii}]}}
\newcommand{\oii}{{[O\,{\sc ii}]}}
\newcommand{\nii}{{[N\,{\sc ii}]}}

\newcommand{\cii}{{[C\,{\sc ii}]}}
\newcommand{\si}{{[Si\,{\sc ii}]}}
\newcommand{\hii}{H\,{\sc ii}\rm}
\newcommand{\hb}{{H$\beta$}}
\newcommand{\ha}{{H$\alpha$}}

\newcommand{\lam}{$\,\lambda$}
\newcommand{\llam}{$\,\lambda\lambda$}
\newcommand{\avg}[1]{\left< #1 \right>} 

\begin{document}
\title{The Metallicity Evolution of Star-Forming Galaxies from redshift 0 to 3: Combining Magnitude Limited Survey with Gravitational Lensing}
\author{T.-T. Yuan\altaffilmark{1,2},  L. J. Kewley\altaffilmark{1,2,3}, J. Richard\altaffilmark{4}}
\altaffiltext{1}{Institute for Astronomy, University of Hawaii, 2680 Woodlawn Drive, Honolulu, HI 96822}
\altaffiltext{2}{Research School of Astronomy and Astrophysics, The Australian National University, Cotter Road, Weston Creek, ACT 2611}
\altaffiltext{3}{ARC Future Fellow}
\altaffiltext{4}{CRAL, Observatoire de Lyon, Universit\'e Lyon 1, 9 avenue Charles Andr\'e, 69561 Saint Genis Laval Cedex, France}

\begin{abstract}
We present a comprehensive observational study of the gas phase metallicity of  star-forming galaxies from $z\sim0\rightarrow3$. 
We combine our new sample of gravitationally lensed galaxies with existing lensed and non-lensed samples to conduct  a large
investigation into the mass-metallicity (MZ) relation at $z>1$. We apply a self-consistent metallicity calibration scheme 
to investigate the metallicity evolution of star-forming galaxies as a function of redshift.
The lensing magnification ensures that our sample spans an unprecedented range of stellar mass (3 $\times$ 10$^{7}$$-$6$\times$10$^{10}$  M$_{\odot}$).
We find that at the median redshift of $z=2.07$,  the median metallicity of the lensed sample is 0.35 dex lower than the local SDSS star-forming galaxies and  0.18 dex lower than the $z\sim0.8$ DEEP2 galaxies.  We also present the $z\sim2$ MZ relation using 19 lensed galaxies.  A more rapid evolution is seen between $z\sim1 \rightarrow 3$  than $z\sim0 \rightarrow 1$ for the high-mass galaxies (10$^{9.5}$ M$_{\odot}<$M$_{\star}<$10$^{11}$ M$_{\odot}$),  with almost twice as much enrichment between $z\sim1 \rightarrow 3$ 
than between $z\sim1 \rightarrow 0$.  We compare this evolution with the most recent cosmological hydrodynamic simulations with momentum driven winds. 
We find that the model metallicity is consistent with the observed metallicity within the observational error for the low mass bins. 
However, for higher masses, the model over-predicts the metallicity at all redshifts. The over-prediction is most significant in the highest mass bin of  10$^{10-11}$ M$_{\odot}$.
\end{abstract}


\section{Introduction}
Soon after the pristine clouds of primordial gas collapsed to assemble a protogalaxy, 
star formation ensued, leading to the production of heavy elements (metals). 
Metals were synthesized exclusively in stars, and were ejected into the interstellar medium (ISM) through
stellar winds or supernovae explosions.  Tracing the heavy element abundance (metallicity)  
in star-forming galaxies provides a ``fossil record" of galaxy formation and evolution.

When considered as a closed system,  the metal content of a galaxy 
 is directly related to the yield  and gas fraction \citep{Searle72,Pagel75,Pagel81,Edmunds90}.  In reality, a galaxy interacts with its surrounding intergalactic medium (IGM), hence both the overall and
  local metallicity distribution of a galaxy is modified by feedback processes such as galactic winds, inflows, and gas accretions \citep[e.g.,][]{Lacey85,Edmunds95, Koppen99,Dalcanton07}.  Therefore, observations of  the chemical abundances in galaxies offer crucial constraints on the star formation history and various mechanisms responsible for galactic inflows and outflows. 

The well-known correlation between galaxy mass (luminosity) and metallicity  was first proposed by 
\citet{Lequeux79}.  Subsequent studies confirmed the existence of the luminosity-metallicity (LZ) relation \citep[e.g.,][]{Rubin84,Skillman89,Zaritsky94,Garnett02}. 
 Luminosity was used as a proxy for stellar mass in these studies as luminosity is a direct observable.   Aided by new sophisticated stellar population models, stellar
 mass can be robustly calculated and a tighter correlation is found in the mass-metallicity (MZ) relation.   
 \citet{Tremonti04} have established the MZ relation for local star-forming galaxies based on $\sim$ 5$\times$10${^5}$ Sloan Digital Sky Survey (SDSS) galaxies.
 At intermediate redshifts ($0.4<z<1$), the MZ relation has also been observed for a large number of galaxies ($>$100) \citep[e.g.,][]{Savaglio05,Cowie08,Lamareille09}.
  \citet{Zahid11} derived the MZ relation for $\sim$ 10$^{3}$ galaxies  from the Deep Extragalactic Evolutionary Probe 2 (DEEP2) survey, validating the MZ relation on a statistically significant level at $z\sim0.8$. 
 
Current cosmological hydrodynamic simulations and semi-analytical models can predict the metallicity history of galaxies on a cosmic timescale \citep{Nagamine01,DeLucia04,Bertone07,Brooks07,Dave07,Dave11b,Dave11a}.  These models show that the shape of the MZ relation is particularly sensitive to the adopted feedback mechanisms.  The cosmological hydrodynamic simulations with momentum-driven winds models provide better match with observations than energy-driven wind models \citep{Oppenheimer08,Finlator08,Dave11b}.  However, these models have not been tested thoroughly in observations, especially at high redshifts ($z>1$), where the MZ relation is still largely uncertain. 

As we move to higher redshifts, selection effects and small number statistics haunt observational metallicity history studies.  The difficulty becomes more severe in the so-called ``redshift desert" ($1\lesssim z \lesssim 3$), where the metallicity sensitive optical emission lines have shifted to the sky-background dominated near infrared (NIR).  Ironically, this redshift range harbors the richest information about galaxy evolution. It is during this redshift period ($\sim$ 2$-$6 Gyrs after the Big Bang)  that the first massive structures condensed;  the star formation rate (SFR),  major merger activity, and black hole accretion rate peaked;  much of today's stellar mass was assembled, and heavy elements were produced \citep{Fan01,Dickinson03, Chapman05,HopkinsAM06,Grazian07,Conselice07,Reddy08}.  It is therefore of crucial importance to explore NIR spectra for galaxies in this redshift range. 

Many  spectroscopic redshift surveys have been carried out to study star-forming galaxies at $z>$1 in recent years \citep[e.g.,][]{Steidel04,Law09}. 
However, due to the low efficiency in the NIR, those spectroscopic surveys almost inevitably have to rely on color-selection criteria and the biases in UV-selected galaxies tend to  select the most massive and less dusty systems \citep[e.g.,][]{Capak04, Steidel04, Reddy06}.  Space telescopes can  observe much deeper in the NIR and  are able to probe a wider mass range.  For example, the narrow-band \ha\ surveys based on the new WFC3 camera aboard the Hubble Space Telescope ($\emph{HST}$) have located hundreds of \ha\ emitters up to z = 2.23, finding much fainter systems than observed from the ground \citep{Sobral09b}. However, the
low-resolution spectra from the narrow band filters forbid  derivations of  physical properties such as metallicities that can only currently be acquired from  ground-based spectral analysis.

Thanks to the advent of long-slit/multi-slit NIR spectrographs on 8$-$10 meter class telescopes, enormous progress has been made in the last decade to capture galaxies in the redshift desert.  For  chemical abundance studies, a full coverage of rest-frame optical spectra  (4000$-$9000$\AA$) is usually mandatory for the most robust diagnostic analysis.  For  1.5 $\lesssim z \lesssim 3$, the
rest-frame optical spectra have shifted into the J, H, and K bands.  It remains challenging and observationally expensive  to obtain high  signal-to-noise ($S/N$) NIR spectra  from the ground, especially for  ``typical" targets at high-$z$ that are less massive than conventional color-selected galaxies. Therefore, previous investigations into the metallicity properties
between $1\lesssim z \lesssim 3$ focused on stacked spectra,  samples of massive luminous individual galaxies, or very small numbers of lower-mass galaxies \citep[e.g.,][]{Erb06,FS06,Law09,Erb10b,Yabe12}.

The first mass-metallicity (MZ) relation 
for galaxies at z $\sim$ 2 was found by \citet{Erb06} using the stacked spectra of 87 UV selected galaxies divided into 6 mass bins.  Subsequently, mass and metallicity measurements  have been reported for 
numerous individual galaxies  at 1.5 $<z<$ 3  \citep{FS06,Genzel08,Hayashi09,Law09,Erb10b}.
These galaxies are selected using broadband colors in the UV \citep[Lyman Break technique;][]{Steidel96,Steidel03} or using B, z, and K-band colors \citep[BzK selection;][]{Daddi04}.  The Lyman break and BzK selection techniques favor galaxies that are luminous in the UV or blue and may therefore be biased against low luminosity (low-metallicity) galaxies, and dusty (potentially metal-rich) galaxies. Because of these biases, galaxies selected in this way may not sample the full range in metallicity at redshift z $>$1.

A powerful alternative method to avoid these selection effects is to use strong gravitationally lensed galaxies.  In the case of  galaxy cluster lensing, 
 the total luminosity and area of the background sources can easily be boosted by $\sim 10-50$ times, providing  
 invaluable opportunities to obtain high S/N spectra and probe intrinsically fainter systems within a reasonable amount of telescope time.  In some cases, sufficient S/N  can even be obtained for spatially resolved pixels  to study the resolved metallicity of high-$z$ galaxies \citep{Swinbank09a,Jones10b,Yuan11,Jones12}.  
Before 2011, metallicities have been reported for a handful of individually lensed galaxies using optical emission lines at 1.5 $<z<$ 3  \citep{Pettini01,LB03, Stark08, Quider09,Yuan09,Jones10b}. 
Fortunately, lensed galaxy samples with metallicity measurements have increased significantly thanks to reliable lensing mass modeling and  larger dedicated spectroscopic surveys of lensed galaxies
 on 8-10 meter telescopes \citep{Richard11a, Wuyts12b,Christensen12}. 

In 2008, we began a spectroscopic observational survey designed specifically to capture metallicity sensitive lines for lensed galaxies. 
Taking advantage of the multi-object cryogenic NIR spectrograph (MOIRCS) on Subaru,  
we targeted  well-known strong lensing galaxy clusters to obtain metallicities  for galaxies between  0.8 $<z<$ 3. 
 In this paper, we present the first metallicity measurement results from our survey.  

Combining our new data with existing data from the literature, we present a coherent observational picture of the 
metallicity history and mass-metallicity evolution of star-forming galaxies from $z\sim0$ to $z\sim3$. 
 \citet{Kewley08} have shown that the metallicity offsets in the diagnostic methods can easily exceed the intrinsic trends. It is of paramount importance to 
make sure that relative metallicities are compared on the same metallicity calibration scale. In MZ relation studies, the methods used to derive the stellar mass can also cause systematic offsets \citep{Zahid11}.  Different SED fitting codes can yield a non-negligible mass offset, hence mimicking or hiding evolution in the MZ relation. 
In this paper,  we  derive the mass and metallicity of all samples using the same methods, ensuring that the 
observational data are compared in a  self-consistent way. We compare our observed metallicity history with the latest prediction from cosmological hydrodynamical simulations.

Throughout this paper we use a standard $\Lambda$CDM cosmology with $H_0$= 70 km s$^{-1}$
Mpc$^{-1}$, $\Omega_M$=0.30, and $\Omega_\Lambda$=0.70.
We use solar oxygen abundance 12 + log(O/H)$_{\odot}$=8.69 \citep{Asplund09}.  

The paper is organized as follows: 
Section~\ref{sec:sample} describes our lensed sample survey and observations.   
Data reduction and analysis are summarized in Section~\ref{sec:data}. 
Section~\ref{sec:sampleoverview} presents an overview of all the samples we use in this study.
Section~\ref{sec:metallicity} describes the methodology of derived quantities. 
The metallicity evolution  of star-forming galaxies with redshift is presented in Section~\ref{sec:metalhistory}. 
Section~\ref{sec:mzr} presents the mass-metallicity relation for our lensed galaxies. 
Section~\ref{sec:discuss} compares our results with previous work in literature.
Section~\ref{sec:sum} summarizes our results. 
In the Appendix, we show the morphology, slit layout,  and reduced 1D spectra for the lensed galaxies reported in our survey.

\section{The LEGMS survey and Observations}\label{sec:sample}

\begin{deluxetable*}{lcccccc}
\tabletypesize{\scriptsize}
\tablewidth{0pt}
\tablecolumns{7}
\tablecaption{MOIRCS Observation Summary \label{tabobs}}
\tablehead{
\colhead{Target} & 
\colhead{Dates} & 
\colhead{Exposure Time} & 
\colhead{PA} & 
\colhead{Seeing($K_s$)} & 
\colhead{Slit width} & 
\colhead{Filter/Grism} \\
\colhead{} & 
\colhead{} & 
\colhead{(ks)} & 
\colhead{(deg)} & 
\colhead{$\!\!^{\prime\prime}$} & 
\colhead{ $\!\!^{\prime\prime}$} & 
\colhead{} 
}
\startdata
Abell 1689 &Apr 28,2011&50.0& 60&0.5-0.8&\nodata&K$_{s}$ Imaging\\
Abell 1689 &Apr 28,2010&15.6& -60&0.5-0.8&0.8&HK500\\
Abell 1689 &Apr 29,2010&19.2& 45&0.5-0.6&0.8&HK500\\
Abell 1689 &Mar 24,2010&16.8& 20&0.5-0.6&0.8&HK500\\
Abell 1689 &Mar 25,2010&12.0& -20&0.6-0.7&0.8&HK500\\
Abell 1689 &Apr 23, Mar 24, 2008&15.6& 60&0.5-0.8&0.8&zJ500\\
Abell  68   &Sep 29-30,2009&12.0& 60&0.6-1.0&1.0&HK500, zJ500\\
\enddata
\tablecomments{Log of the observations. We use a dithering length of 2.$\!\!^{\prime\prime}$5 for all the spectroscopic observations.}
\end{deluxetable*}

\subsection{The Lensed Emission-Line Galaxy Metallicity Survey (LEGMS)}\label{sec:survey}
Our survey  (LEGMS) aims to obtain oxygen abundance of lensed galaxies at 0.8$<$z$<$3.
LEGMS has taken enormous advantage of the state-of-the-art instruments on Mauna Kea. Four instruments have been utilized so far: (1) the Multi-Object InfraRed Camera and Spectrograph (MOIRCS; \citealt{Ichikawa06}) on Subaru;
(2) the OH-Suppressing Infra-Red  Imaging Spectrograph (OSIRIS; \citealt{Larkin06}) on Keck II;
(3) the Near Infrared Spectrograph (NIRSPEC; \citealt{McLean98})  on Keck II;
(4) the Low Dispersion Imaging Spectrograph (LRIS;  \citealt{Oke95}) on Keck I. The scientific objective of each instrument is as follows:
MOIRCS is used to obtain the NIR images and spectra for multiple targets behind lensing clusters;
NIRSPEC is used to capture occasional single field lensed targets (especially galaxy-scale lenses);
LRIS  is used to obtain the  \oii\lam3727 to \oiii\lam5007  spectral range  for targets with z $<$ 1.5. From the slit spectra, we select targets that are have sufficient fluxes and angular sizes to be spatially resolved with OSIRIS.  In this paper we focus on the MOIRCS observations of the lensing cluster Abell 1689 for targets between redshifts 1.5 $\lesssim$ z $\lesssim$ 3.  Observations for other clusters are ongoing and will  be presented in future papers. 
 
The first step to construct a lensed sample for slit spectroscopy is to find the lensed candidates (arcs) that have spectroscopic redshifts from
optical surveys. The number of known spectroscopically identified lensed galaxies at z $>$ 1 is still on the order of a few tens. The limited number
of lensed candidates makes it impractical to build a sample that is complete and well defined in mass. A mass complete sample is the future goal 
of this project.   Our strategy for now is to observe as many arcs with known redshifts as possible. If we assume  the AGN fraction  is similar to local star-forming galaxies, then we expect $\sim$ 10\%  of our targets to be AGN dominated \citep{Kewley04}. Naturally, lensed sample is biased towards highly magnified sources.  However, because the largest magnifications are not biased towards intrinsically bright targets, lensed samples are  less biased towards the intrinsically most luminous galaxies. 

Abell 1689 is chosen as the primary target for MOIRCS observations because it has the largest number ($\sim$ 100 arcs, or $\sim$ 30 source galaxies) of spectroscopically identified lensed arcs \citep{Broadhurst05,Frye07,Limousin07}.  

Multi-slit spectroscopy of NIR lensing surveys greatly enhances the efficiency of spectroscopy of lensed galaxies in clusters.  Theoretically,  $\sim$ 40 slits can be observed simultaneously on the two chips of  MOIRCS with a total field of view (FOV) of 4$^{\prime} \times 7^{\prime}$.   In practice, the number of lensed targets on the slits is restricted by the strong lensing area, slit orientations, and spectral coverage.  For A1689, the lensed candidates cover an area of $\sim$2$^{\prime}$ $\times$ 2$^{\prime}$, well within the FOV of one chip. We design slit masks for chip 2, which has better sensitivity and less bad pixels than chip 1. There are $\sim$ 40 lensed images ($\sim$ 25 individual galaxies) that fall in the range of 1.5 $\lesssim$ z $\lesssim$ 3 in our slit masks.  We use the  MOIRCS low-resolution (R$\sim$ 500) grisms which have a spectral coverage of 0.9 -1.78 $\mu$m in ZJ and 1.3-2.5 $\mu$m in HK.  To maximize the detection efficiency, we give priority to targets with the specific redshift range such that  all the strong emission lines from \oii\lam3727 to \nii\lam6584  can be captured in one grism configuration.  For instance,  the redshift range of 2.5 $\lesssim$ z $\lesssim$ 3 is optimized for the HK500 grism, and 1.5 $\lesssim$ z $\lesssim$ 1.7 is optimized for the ZJ500 grism. 
  
From UT March 2008 to UT April 2010, we used 8 MOIRCS nights (6 usable nights) with 4 position angles (PAs) and 6 masks to observe 25 galaxies. 
Metallicity quality spectra were obtained for 12 of the 25 targets.   We also include one $z>1.5$ galaxy from our observations of Abell 68\footnote{Most of the candidates in A68 are at $z<1$.  Due to the low spectral resolution in this observation, \ha\ and \nii\ are not resolved  at $z<1$. We do not have sufficient data to obtain reliable metallicities for the $z<1$ targets in A68 and therefore exclude them from this study.}. The PA is chosen to optimize the slit orientation along the targeted arcs' elongated directions. For arcs that are not oriented to match the PA, the slits are configured to center on  the brightest knots of the arcs.  We use slit widths of  0.8 $\!\!^{\prime\prime}$  and 1.0 $\!\!^{\prime\prime}$, with a variety of slit lengths for each lensed arc.   For each mask, a bright galaxy/star is placed on one of the slits to trace the slit curvature and determine the offsets among individual exposures.  Typical integrations for individual frames are 400 s, 600 s, and 900 s, depending on levels of skyline saturation.  We use an ABBA dithering sequence along the slit direction, with a dithering length of 2.$\!\!^{\prime\prime}$5. The observational logs are summarized in Table~\ref{tabobs}.

\section{Data Reduction and Analysis}\label{sec:data}
\subsection{Reduce 1D spectrum}\label{sec:1D}
The data reduction procedures from the raw mask data to the final wavelength and flux calibrated 1D spectra were realized by a set of IDL codes called
MOIRCSMOSRED. The codes were scripted originally by Youichi Ohyama. T.-T Yuan  extended  the code  
to  incorporate new skyline subtraction \citep[e.g.,][for a description of utilizing MOIRCSMOSRED]{HenryP10}.

We use the newest version (Apr, 2011) of MOIRCSMOSRED to reduce the data in this work. The sky subtraction is optimized as follows.
For each A$_{i}$ frame, we subtract a sky frame denoted as $\alpha$((B$_{i-1}$+B$_{i+1}$)/2), where B$_{i-1}$ and B$_{i+1}$ are the science frames before and after the 
A$_{i}$ exposure. The scale parameter $\alpha$ is obtained by searching through a parameter range  of 0.5-2.0, with an increment of 0.0001. The 
best  $\alpha$ is obtained where the root mean square (RMS) of the residual  R$=$ A$_{i}$- $\alpha$((B$_{i-1}$+B$_{i+1}$)/2) is minimal for a user defined wavelength region $\lambda_{1}$ and $\lambda_{2}$.   We find that this sky subtraction method yields smaller sky OH line residuals ($\sim$ 20\%) than conventional A-B methods.
  We also  compare with other  skyline subtraction methods in literature \citep{Kelson03,Davies07}. We find the sky residuals from our method are comparable to those from the 
   \citet{Kelson03} and \citet{Davies07} methods within 5\% in general cases. However, in cases where the emission line falls on top of a strong skyline, our method is more stable 
   and improves the skyline residual by   $\sim$ 10\% than the other two methods.

Wavelength calibration is carried out by identifying skylines for the ZJ grism. For the HK grism,  we use argon lines 
to calibrate the wavelength since only a few skylines are available in the HK band. 
 The argon-line calibrated wavelength is then  re-calibrated with the available skylines in HK  to determine the instrumentation shifts between lamp and science exposures. 
Note that the RMS of the wavelength calibration using a 3rd order polynomial fitting is $\sim$  10-20 \AA, corresponding to a systematic redshift uncertainty 
of  0.006.

A sample of A0 stars selected from the UKIRT photometric standards were observed at similar airmass as the targets. 
These stars were used for both telluric absorption corrections and  flux calibrations.   We use the prescriptions of \citet{Erb03}
 for flux calibration. As noted in   \citet{Erb03}, the absolute flux calibration in the NIR is difficult with  typical
  uncertainties of $\sim$20\%. We note that this uncertainty is even larger for  lensed samples observed in multi-slits because of the complicated 
 aperture effects.  The uncertainties in the flux calibration are not a concern for our metallicity analysis where only line ratios are involved. However, 
these errors are a major concern for calculating SFRs.  The uncertainties from the multi-slit aperture effects can cause the SFRs to change by a factor of 2-3.
 For this reason, we refrain from any quantitative analysis of SFRs in this work.

\subsection{Line Fitting}\label{sec:fit}
The emission lines are fitted with Gaussian profiles.  
For the spatially unresolved spectra, the aperture used to extract the spectrum is determined by measuring 
the Gaussian profile of the wavelength collapsed spectrum.  Some of the lensed targets ($\sim$ 10\%) are elongated and spatially resolved in the slit spectra, however, because of the low surface brightness and thus very low S/N per pixel, we are unable to obtain usable spatially resolved spectra. For those targets, we make an initial guess for the width of the spatial profile and force a Gaussian fit, then we extract the integrated spectrum using the aperture determined from the FWHM of the Gaussian profile. 

For widely separated lines such as \oii\lam3727, \hb\lam4861, single Gaussian functions are fitted with 4 free parameters: the centroid (or the redshift), the line width,  the line flux, and the continuum. The doublet \oiii\lam\lam4959,5007 are initially fitted as a double Gaussian function with 6 free parameters: the centroids 1 and 2 , line widths 1 and 2, fluxes 1 and 2, and the continuum. In cases where the \oiii\lam4959 line is too weak, its centroid and line velocity width are fixed to be the same as \oiii\lam5007 and the flux is fixed to be 1/3 of the  \oiii\lam5007 line \citep{Osterbrock89}.  A triple-Gaussian function is fitted simultaneously to the three adjacent emission lines:  \nii\lam6548, 6583 and \ha. The centroid and velocity width of \nii\lam6548, 6583 lines are constrained by the velocity width of \ha\lam6563, and  the ratio of \nii\lam6548 and \nii\lam6583 is constrained  to be the theoretical value of 1/3 given in  \citet{Osterbrock89}.  The line profile fitting is conducted using a $\chi^2$ minimization procedure which uses the inverse of the sky  OH emission as the weighting function.
The  S/N per pixel is calculated from the $\chi^2$ of the fitting. The measured emission line fluxes and line ratios are listed in Table~\ref{tabmetal}.
The final reduced 1D spectra are shown in the Appendix.

\subsection{Lensing Magnification}\label{sec:magnification}
Because the lensing magnification ($\mu$) is not a direct function of wavelength,  line ratio measurements do not require pre-knowledge of the lensing magnification. 
However, $\mu$ is needed for inferring other physical properties such as the intrinsic fluxes, masses and source morphologies. 
  Parametric models of the mass distribution in the clusters Abell 68 and Abell 1689 were constructed using the Lenstool software {\tt Lenstool}\footnote{\tt
  http://www.oamp.fr/cosmology/lenstool} \citep{Kneib93, Jullo07}. The best-fit models have been previously published in \citet{Richard07} and \citet{Limousin07}. As detailed in \citet{Limousin07},   {\tt Lenstool}  uses Bayesian optimization with a Monte-Carlo Markov Chain (MCMC) sampler which provides a family of best models sampling the posterior probability distribution of each parameter. In particular, we use this family of best models to derive the magnification and relative error on magnification $\mu$ associated to each lensed source. Typical errors on $\mu$ are  $\sim$10\% for Abell 1689 and Abell 68.

\subsection{Photometry}\label{sec:sed}
We  determine the photometry for the lensed galaxies in A1689 using  4-band $\emph{HST}$ imaging data, 1-band MOIRCS imaging data, and 2-channel 
$\emph{Spitzer}$  IRAC data at 3.6 and 4.5 $\mu$m. 

We obtained a 5,000 s image exposure for A1689 on the MOIRCS K$_{s}$ filter, at a depth of 24 mag, using a scale of 0.117 $\!\!^{\prime\prime}$ per pixel.
The image was reduced using MCSRED in IRAF written by the MOIRCS supporting astronomer Ichi Tanaka\footnote{\url{http://www.naoj.org/staff/ichi/MCSRED/mcsred.html}}.
The photometry is calibrated using the 2MASS stars located in the field. 

The ACS F475W, F625W, F775W, F850LP data are obtained from the $\emph{HST}$ archive.  The $\emph{HST}$ photometry are determined using SExtractor \citep{Bertin96} with parameters adjusted to detect the faint background sources. The  F775W filter is used as the detection image using a 1.$\!\!^{\prime\prime}$0 aperture.  

The IRAC data are obtained from the $\emph{Spitzer}$ archive and
 are reduced and drizzled to a pixel scale of 0.$\!\!^{\prime\prime}$6 pixel$^{-1}$.
  In order to include the IRAC photometry, 
  we convolved the $\emph{HST}$ and MOIRCS images with the IRAC point spread functions (PSFs) derived from unsaturated stars.  
   All photometric data are measured using a 3.$\!\!^{\prime\prime}$0 radius aperture.  
  Note that we only consider sources that are not contaminated by nearby bright galaxies: $\sim70\%$ of our sources have IRAC photometry (Table 5).
  Typical errors for the IRAC band photometry are 0.3 mag, with uncertainties mainly from the aperture correction and contamination of neighboring galaxies. 
    Typical errors  for the ACS and MOIRCS bands are 0.15 mag, with uncertainties mainly from the Poisson noise and absolute zero-point uncertainties \citep{Wuyts12b}. 
 We refer to Richard et al. (2012, in prep) for the full catalog of the lensing magnification and photometry of the lensed sources in Abell 1689.

\section{Supplementary Samples}\label{sec:sampleoverview}
In addition to our lensed targets observed in LEGMS, we also include literature data  for complementary lensed and non-lensed samples at
both local and high-$z$. The observational data for individually measured metallicities at z $>$ 1.5 are still scarce and caution needs to be taken when using them for comparison.  The different metallicity and mass derivation methods used in different samples can give large systematic discrepancies and provide misleading results. 
For this reason, we only include the literature data that have robust measurements and sufficient data for consistently recalculating the stellar mass and metallicities using our  own methods. Thus, in general,  stacked data, objects with lower/upper limits  in either line ratios or masses are {\it not} chosen.  The one exception is the stacked data of \citet{Erb06}, as it is the most widely used comparison sample at $z\sim2$.

The samples used in this work are:

{\it (1) The Sloan Digital Sky Survey (SDSS) sample ($z\sim$ 0.07).}
We use the SDSS sample \citep[http://www.mpa-garching.mpg.de/SDSS/DR7/]{Abazajian09} 
defined by \citet{Zahid11}.   The mass derivation method used in \citet{Zahid11} is the same as we use in this work. 
All SDSS metallicities are recalculated using the PP04N2 method, which uses an empirical fit to 
the \nii\ and \ha\ line ratios of \hii\ regions \citep{Pettini04}.   \\

{\it (2) The The Deep Extragalactic Evolutionary Probe 2 (DEEP2) sample  ($z\sim$ 0.8).}
 The DEEP2 sample \citep[http://www.deep.berkeley.edu/DR3/]{Davis03} is defined 
in \citet{Zahid11}. At $z\sim$ 0.8, the \nii\ and \ha\  lines are not available in the optical.  We convert the KK04 R$_{23}$ metallicity to the PP04N2 metallicity using the prescriptions of \citet{Kewley08}. \\

 {\it (3) The UV-selected sample ($z \sim$ 2).}
We use the stacked data of \citet{Erb06}.  The metallicity diagnostic used by \citet{Erb06} is the PP04N2 method and no recalculation is needed. We offset the stellar mass scale of \citet{Erb06} by -0.3 dex to match the mass derivation method used in this work \citep{Zahid12b}.  This offset accounts for the different initial mass function (IMF) and stellar 
evolution model parameters applied by \citet{Erb06}.
\\

{\it (4) The lensed sample (1 $<$ z $<$ 3).}  
Besides  the 11 lensed galaxies from our LEGMS survey in Abell 1689, we include 1 lensed source ($z=$1.762) from our MOIRCS data on Abell 68 and 1 lensed spiral  ($z=$1.49)
from \citet{Yuan11}.    We also include 10 lensed  galaxies from  \citet{Wuyts12b} and 3 lensed galaxies from \citet{Richard11a}, since these 13 galaxies have \nii\ and \ha\ measurements, 
as well as photometric data for recalculating stellar masses.   We require all emission lines from literature to have S/N $>$ 3 for quantifying the metallicity of 1 $<$ z $<$ 3 galaxies.
Upper-limit metallicities are found for 6 of the lensed targets from our LEGMS survey. Altogether, the lensed sample is composed of 25 sources, 12 (6/12 upper limits) of which are new observations from this work.  Upper-limit metallicities are not used in our quantitative analysis.
\\

 The methods used to derive stellar mass and metallicity are discussed in detail in Section~\ref{sec:metallicity}.

\begin{figure}[!ht]
\begin{center}
\includegraphics[trim = 5mm 1mm 5mm 5mm, clip, width=8.6cm,angle=0]{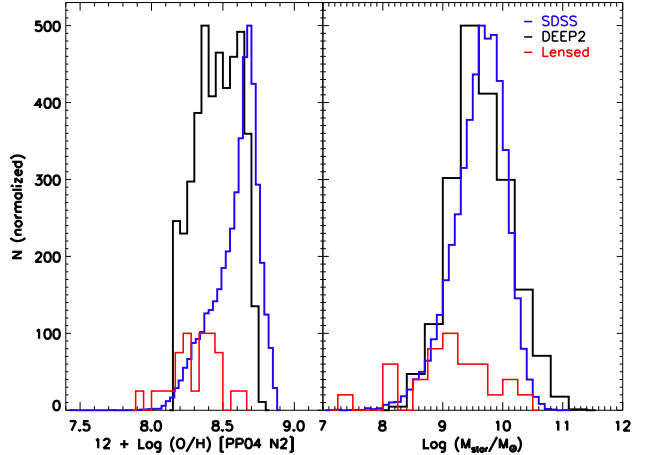}
\caption{Left panel:  the metallicity distribution of the local SDSS (blue), intermediate-$z$ DEEP2 (black), and high-$z$ lensed galaxy samples (red). 
Right panel:  the stellar mass distribution of the same samples. To  present all three samples on the same figure, the SDSS (20577 points) and DEEP2 (1635 points) samples are normalized to 500, and the lensed sample (25 points) is normalized to 100. }
\label{fig:hist}
\end{center}
\end{figure}

\begin{figure*}[!ht]
\begin{center}
\includegraphics[trim = 5mm 1mm 5mm 0mm, clip, width=17cm,angle=0]{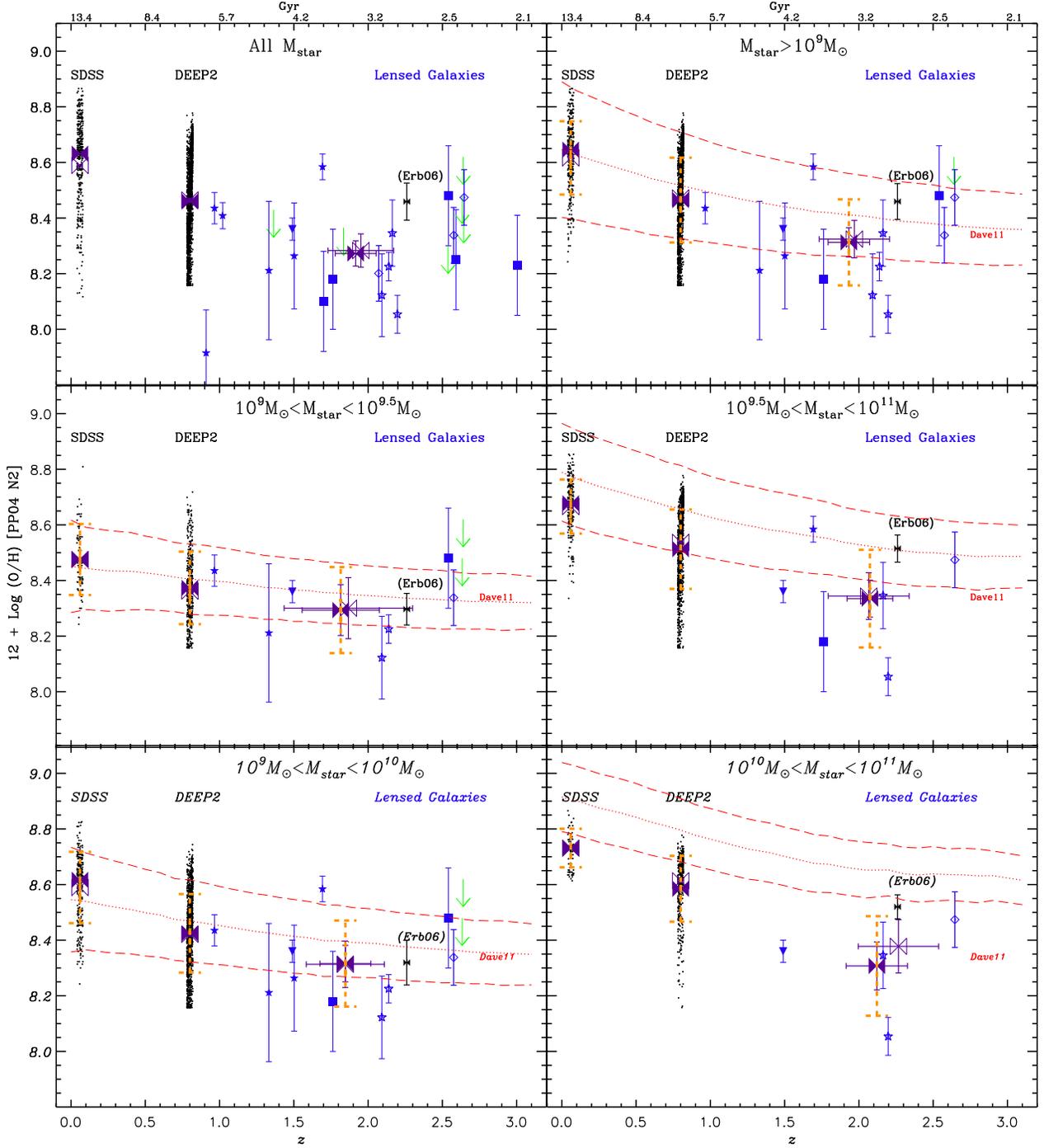}
\caption{The $\rm{Z}\emph{z}$ plot: metallicity history of star-forming galaxies from  redshift 0 to 3. 
The  SDSS and DEEP2 samples (black dots)  are taken from   \citet{Zahid11}.  
The SDSS data are plotted in bins to reduce visual crowdedness.   The lensed galaxies are plotted in blue (upper-limit objects in green arrows), with different lensed samples showing in different symbols (see Figure~\ref{fig:mzobs} for the legends of the different lensed samples). 
The purple ``bowties"  show the bootstrapping mean (filled symbol) and median (empty symbol) metallicities and the 1$\sigma$ standard deviation of the mean and median, 
 whereas the orange dashed error bars show the 1$\sigma$ scatter of the data.
For the SDSS and DEEP2 samples, the $1\sigma$ errors of the median metallicities are 0.001 and 0.006 (indiscernible from the figure), whereas for the lensed sample the $1\sigma$ scatter of the median metallicity is  0.067.  Upper limits are excluded from the median and error calculations. 
 For comparison, we also show the mean metallicity of the UV-selected galaxies from \citet{Erb06}  (symbol: the black bowtie). 
The 6 panels  show samples in different mass ranges. The red dotted and dashed lines are the model predicted median and $1\sigma$ scatter (defined as including 68\% of the data) of the  SFR-weighted gas metallicity in simulated galaxies \citep{Dave11a}. 
}
\label{fig:zredz}
\end{center}
\end{figure*}

\section{Derived Quantities}\label{sec:metallicity}
\subsection{Optical Classification}\label{sec:BPT2}
We use the standard optical diagnostic diagram (BPT) to exclude targets that are dominated by AGN  \citep{Baldwin81, Veilleux87, Kewley06}.
For all 26 lensed targets in our LEGMS sample, we find 1 target that could be contaminated by AGN (B8.2). The fraction of AGN 
in our sample is therefore $\sim$8\%, which is similar to the fraction ($\sim$7\%) of the local SDSS sample \citep{Kewley06}. We also find that the line ratios of the high-z lensed sample has a systematic offset on the BPT diagram, as found in \citet{Shapley05,Erb06,Kriek07,Brinchmann08,Liu08,Richard11a}. 
The redshift evolution of the BPT diagram  will be reported in Kewley et al (2013, in preparation).  

\subsection{Stellar Masses}\label{sec:AGN}
We use  the software \verb+LE PHARE+\footnote{$\url{www.cfht.hawaii.edu/{}_{\textrm{\symbol{126}}}arnouts/LEPHARE/lephare.html}$} \citep{Ilbert09} to determine the stellar mass. 
\verb+LE PHARE+
 is a photometric redshift and simulation package based on the
population synthesis models of \citet{BC03}.  If the redshift is known and held fixed, \verb+LE PHARE+ finds the best fitted SED on a $\chi^2$ minimization 
process and returns physical parameters such as stellar mass, SFR and extinction. 
 We choose the initial mass function (IMF) by
\citet{Chabrier03} and the \citet{Calzetti00} attenuation law, with E$\rm{(B-V)}$ ranging from 0 to 2 and an exponentially decreasing SFR 
(SFR $\varpropto$ e$^{-t/\tau}$) with $\tau$ varying between 0 and 13 Gyrs.    The errors caused by emission line contamination are taken into account by manually increasing the uncertainties in the photometric bands where emission lines are located.  The uncertainties are scaled according to the emission line fluxes measured by MOIRCS. 
The stellar masses derived from the emission line corrected photometry are consistent with those without emission line correction, albeit with larger errors in a few cases ($\sim$ 0.1 dex in log space). We use the emission-line corrected photometric stellar masses in the following analysis.

\subsection{Metallicity Diagnostics}\label{sec:diagnostics}
The abundance of oxygen (12\,+\,log(O/H)) is used as a proxy for the overall metallicity of \hii\ regions in galaxies.   The oxygen abundance can be inferred from the strong recombination lines of hydrogen atoms and collisionally excited metal lines \citep[e.g.,][]{Kewley02}. Before doing any metallicity comparisons across different samples and redshifts, it is essential to convert all metallicities to the same base calibration.
The discrepancy among different diagnostics can be as large as 0.7 dex for a given mass, large enough to mimic or hide any intrinsic observational trends. \citet{Kewley08} (KE08) have shown that both the shape and the amplitude of the MZ relation change substantially with different diagnostics.  
 For this work, we convert  all metallicities to the PP04N2 method using  the prescriptions from KE08.

For our lensed targets with only \nii\ and \ha, we use the N2\,=\,log(\nii\lam6583/\ha) index, as calibrated by \citet{Pettini04} (the PP04N2 method). All lines are required to have S/N$>$3 for reliable metallicity estimations. Lines that have S/N$<$3 are presented as 3-$\sigma$  upper limits.   For  targets with only \oii\ to \oiii\ lines, we use the indicator R$_{23}$\,=\,(\oii\lam3727 + \oiii\llam4959, 5007)/\hb\/ to calculate metallicity.  The formalization is given in  \citet{Kobulnicky04} (KK04 method).   The upper and lower branch degeneracy of R$_{23}$ can be broken by the value/upper limit of \nii/\ha.   If the upper limit of \nii/\ha\ is not sufficient or available to break the degeneracy, we calculate both the upper and lower 
branch metallicities and assign the statistical errors of the metallicities as the range of the upper and lower branches.  The KK04 R$_{23}$ metallicity is then converted to the
PP04N2 method using the  KE08 prescriptions.  The line fluxes and metallicity are listed in Table~\ref{tabmetal}. For the literature data, we have recalculated the metallicities in the PP04N2 scheme. 

The statistical metallicity uncertainties are calculated by propagating the flux errors of the \nii\ and \ha\ lines.  
The metallicity calibration of the PP04N2 method itself has  a 1$\sigma$ dispersion of 0.18 dex \citep{Pettini04,Erb06}. 
Therefore, for individual galaxies that have statistical  metallicity uncertainties of less than 0.18 dex,  we assign errors of  0.18 dex.

Note that we are not comparing absolute metallicities between galaxies as they depend on the accuracy of the calibration methods.
However, by re-calculating all metallicities to the same calibration diagnostic, relative metallicities can be compared reliably. 
The systematic error of relative metallicities is $<$ 0.07 dex for strong-line methods \citep{Kewley08}.


\begin{deluxetable*}{cccccccc}
\tabletypesize{\footnotesize}
\tablewidth{0pt}
\tablecolumns{6}
\tablecaption{Median/Mean Redshift and Metallicity of the Samples \label{tabsample}}
\tablehead{
\colhead{Sample} & 
\colhead{Redshift} & 
\multicolumn{5}{c}{Metallicity (12\,+\,log(O/H))} & \\
\cline{3-8}
\colhead{} & 
\colhead{} &
\multicolumn{1}{c}{$>10^{7}M_{\odot}$(all)} &
\multicolumn{1}{c}{$>10^{9}M_{\odot}$} &
\multicolumn{1}{c}{$10^{9-9.5}M_{\odot}$} &
\multicolumn{1}{c}{$10^{9.5-11}M_{\odot}$} & 
\multicolumn{1}{c}{$10^{9-10}M_{\odot}$} &
\multicolumn{1}{c}{$10^{10-11}M_{\odot}$}
}
\startdata
&&&& Mean&&&\\
\hline\hline
SDSS & 0.071$\pm$0.016    & 8.589$\pm$0.001 & 8.616 $\pm$0.001&8.475$\pm$0.002&8.666$\pm$0.001&8.589$\pm0.001$&8.731$\pm$0.001\\
DEEP2 &0.782$\pm$0.018   & 8.459$\pm$0.004 & 8.464$\pm$0.004&8.373$\pm$0.006 &8.512$\pm$0.005&8.425$\pm$0.004&8.585$\pm$0.006\\
Erb06& 2.26$\pm$0.17         & 8.418$\pm$0.051 &8.418$\pm$0.050&8.265$\pm$0.046 &8.495$\pm$0.030&8.316$\pm$0.052&8.520$\pm$0.028\\
Lensed &1.91$\pm$ 0.63      & 8.274$\pm$0.045 &8.309$\pm$0.049 &8.296$\pm$0.090&8.336$\pm$0.066&8.313$\pm$0.083&8.309$\pm$0.086\\
\hline\hline
&&&& Median&&&\\
\hline\hline
SDSS & 0.072  &  8.631 $\pm$0.001  &8.646 $\pm$0.001&8.475$\pm$0.003&8.677$\pm$0.001&8.617$\pm$0.001&8.730$\pm$0.001\\
DEEP2 &0.783  & 8.465$\pm$0.005   &8.472 $\pm$0.006&8.362$\pm$0.009 &8.537$\pm$0.008&8.421$\pm$0.008&8.614$\pm$0.006\\
Erb06 & \nodata &8.459$\pm$0.065   &8.459 $\pm$0.065&8.297$\pm$0.056 &8.515$\pm$0.048&8.319$\pm$0.008&8.521$\pm$0.043\\
Lensed &2.07    &8.286$\pm$0.059   &8.335 $\pm$0.063&8.303$\pm$0.106&8.346$\pm$0.085&8.313$\pm$0.083&8.379$\pm$0.094\\
\enddata
\tablecomments{The errors for the redshift are the  $1\sigma$ standard deviation of the sample redshift distribution (not the $\sigma$ of the mean/median). 
  The errors for the metallicity are the $1\sigma$ standard deviation of the mean/median from bootstrapping.                 
}
\end{deluxetable*}

\begin{figure*}[!ht]
\begin{center}
\includegraphics[trim = 2mm 1mm 2mm 0mm, clip, width=14cm,angle=0]{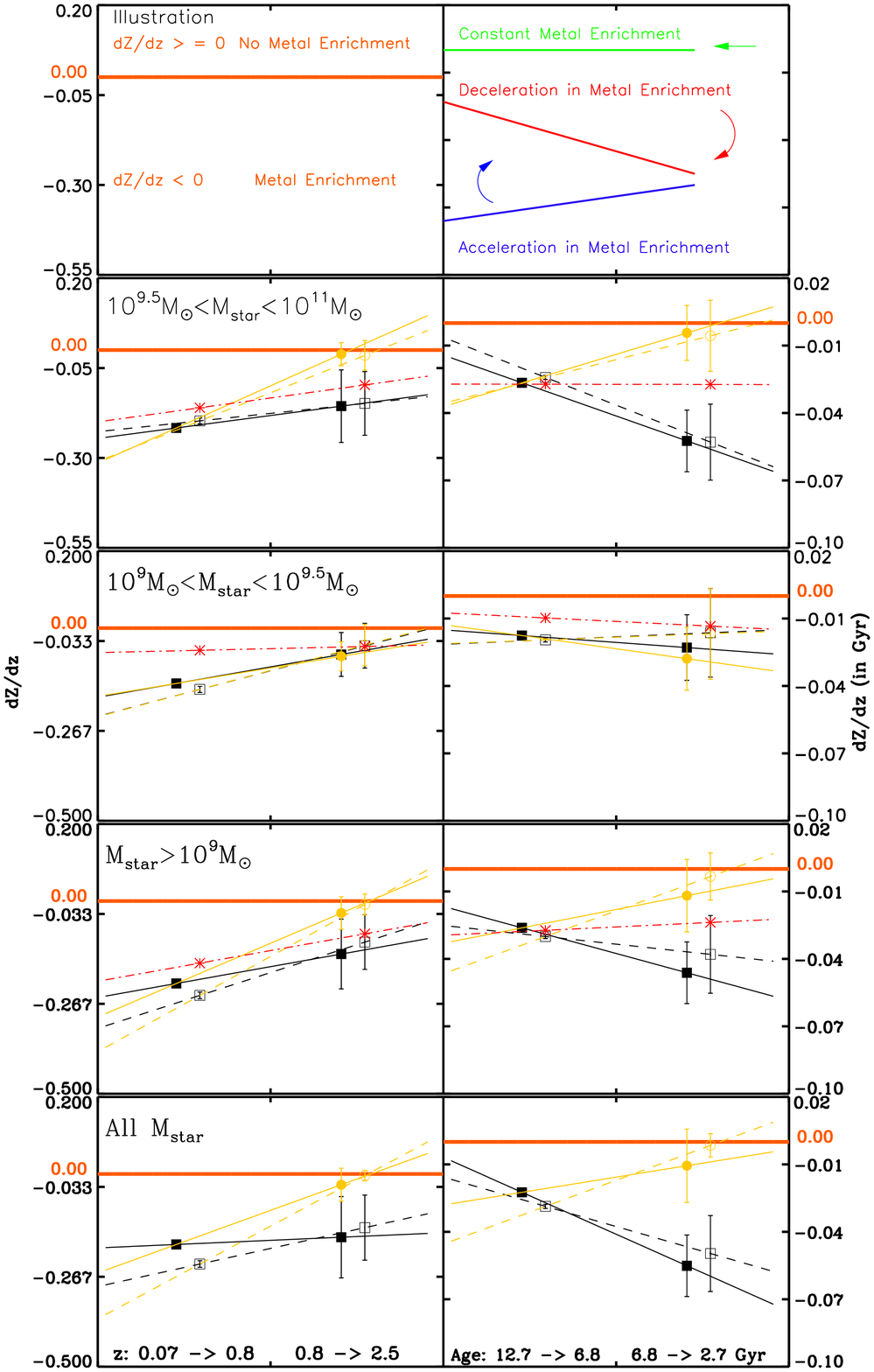}
\caption{Cosmic metal enrichment rate ($\frac{d\rm{Z}}{d\emph{z}}$) in two redshift (cosmic time) epochs. 
$\frac{d\rm{Z}}{d\emph{z}}$ is defined as the slope of the $\rm{Z}\emph{z}$ relation.
Left coloumn shows $\frac{d\rm{Z}}{d\emph{z}}$ in unit of $\Delta$dex per redshift whereas the right coloumn is  in unit of $\Delta$dex per Gyr.
We derive $\frac{d\rm{Z}}{d\emph{z}}$  for  the SDSS to the DEEP2 ($z\sim 0$ to 0.8), and the DEEP2 to the Lensed ($z\sim 0.8$ to 2.0) samples respectively (black squares/lines). 
As a comparison, we also derive $\frac{d\rm{Z}}{d\emph{z}}$ from $z\sim 0.8$ to 2.0 using the DEEP2 and the Erb06 samples  (yellow circles/lines).
Filled and empty squares are results from the mean and median quantities. The model prediction (using median) from the cosmological hydrodynamical simulation of \citet{Dave11b}
is shown in red stars. The second to fifth rows show $\frac{d\rm{Z}}{d\emph{z}}$ in different mass ranges.  The first row illustrates the interpretation of the 
$\frac{d\rm{Z}}{d\emph{z}}$ in redshift and cosmic time frames.  A negative value of $\frac{d\rm{Z}}{d\emph{z}}$ means a positive metal enrichment from high-redshift to local universe.
The negative slope of $\frac{d\rm{Z}}{d\emph{z}}$ versus cosmic time (right column) indicates a deceleration in metal enrichment from from high-$z$ to low-$z$.
}
\label{fig:evolv}
\end{center}
\end{figure*}

\section{The Cosmic Evolution of Metallicity for Star-Forming Galaxies}\label{sec:metalhistory}
\subsection{The $\rm{Z}\emph{z}$ Relation}\label{sec:zz}
In this section, we present the observational investigation into the cosmic evolution of metallicity for star-forming galaxies from redshift 0 to 3.
The metallicity in the local universe is represented by the SDSS sample (20577 objects, $\avg{z}=0.072\pm0.016$).  
The metallicity in the intermediate-redshift universe is represented by the DEEP2 sample (1635 objects, $\avg{z}=0.78\pm0.02$).  
For redshift $1\lesssim z \lesssim 3$, we use 19 lensed galaxies  (plus 6 upper limit measurements)   ($\avg{z}=1.91\pm0.61$) to infer the metallicity
range.  

The redshift distributions for the SDSS and DEEP2 samples are very narrow ($\Delta z \sim 0.02$),  and  the mean and median redshifts are identical within 0.001 dex.
Whereas for the lensed sample, the median redshift is 2.07, and is 0.16 dex higher than the mean redshift. 
There are two  z $\sim$ 0.9 objects in the lensed sample, and if these two objects are excluded, the mean and median redshifts for the lensed sample are 
$\avg{z}=2.03\pm0.54$, $z_{median}=2.09$ (see Table~\ref{tabsample}). 

The overall metallicity distributions of the SDSS, DEEP2, and lensed samples are shown in Figure~\ref{fig:hist}.  Since the $z>1$ sample size is 2-3 orders of magnitude 
smaller than the $z<1$ samples, we use a bootstrapping process to derive the mean and median metallicities of each sample. 
Assuming the measured metallicity distribution of each sample is representative of their parent population, we draw from the initial sample a random subset and  repeat the process 
for 50000 times. We use the 50000 replicated  samples to measure the mean, median and standard deviations of the initial sample.  This method prevents artifacts from small-number statistics and provides robust estimation of the median, mean and errors, especially for the high-$z$ lensed sample. 

The fraction of low-mass (M$_{\star}<$10$^{9}$ M$_{\odot}$) galaxies is largest (31\%) in the lensed sample, compared to  
9\% and 5\% in the SDSS and DEEP2 samples respectively.  Excluding the low-mass galaxies does not notably change the median metallicity of the SDSS and DEEP2 samples ($\sim$ 0.01 dex), while it increases the median metallicity of the lensed sample by $\sim$ 0.05 dex. To investigate whether the metallicity evolution is different for 
various stellar mass ranges, we separate the samples in different mass ranges and derive the mean and median metallicities  (Table~\ref{tabsample}).
 The mass bins of 10$^{9}$ M$_{\odot}<$M$_{\star}<$10$^{9.5 }$ M$_{\odot}$ and 10$^{9.5}$ M$_{\odot}<$M$_{\star}<$10$^{11}$ M$_{\odot}$
are chosen such that there are similar number of lensed galaxies in each bin.  Alternatively, the mass bins of  
 10$^{9}$ M$_{\odot}<$M$_{\star}<$10$^{10 }$ M$_{\odot}$ and 10$^{10}$ M$_{\odot}<$M$_{\star}<$10$^{11}$ M$_{\odot}$ are chosen to span equal mass scales.

We plot the metallicity ($\rm{Z}$) of all samples as a function of redshift $\emph{z}$ in Figure~\ref{fig:zredz} (dubbed the $\rm{Z}\emph{z}$ plot hereafter). 
The first panel shows the complete observational data used in this study. The following three panels show the data and model predictions in different mass ranges.
The samples at local and intermediate redshifts are large enough such that the 1$\sigma$ errors of the mean and median metallicity are smaller than the symbol sizes on the $\rm{Z}\emph{z}$ plot (0.001-0.006 dex). Although the $z>1$ samples are still composed of a relatively small number of objects, 
we suggest that the lensed galaxies and their bootstrapped mean and median values more closely 
  represent the average metallicities of star-forming galaxies at $z>1$ than Lyman break, or B-band magnitude limited samples because the lensed galaxies are selected based on magnification rather than 
  colors.  Although we do note that there is still a magnitude limit and flux limit for each lensed galaxy. 
   
We derive the metallicity evolution in units of ``dex per redshift" and ``dex per Gyr" using both the mean and median values.  The metallicity evolution can be characterized by  
the slope  ($\frac{d\rm{Z}}{d\emph{z}}$) of  the  $\rm{Z}\emph{z}$ plot. We compute $\frac{d\rm{Z}}{d\emph{z}}$ for two redshift ranges:  $z\sim 0\rightarrow$ 0.8  (SDSS  to DEEP2) and $z\sim 0.8\rightarrow $ $\sim$2.5 (DEEP2 to Lensed galaxies).     As a comparison, we also derive $\frac{d\rm{Z}}{d\emph{z}}$ from $z\sim 0.8$ to 2.5 using the DEEP2 and the Erb06 samples  (yellow circles/lines in Figure~\ref{fig:evolv}).
We derive separate evolutions for different mass bins. We show our result in Figure~\ref{fig:evolv}.

 A positive metallicity evolution, i.e.,  metals enrich galaxies from high-$z$ to the local universe, is robustly found 
in all mass bins from  $z\sim 0.8 \rightarrow$ 0.  This positive evolution is indicated by 
the negative values of $\frac{d\rm{Z}}{d\emph{z}}$ (or $\frac{d\rm{Z}}{d\emph{z}(Gyr)}$)   in Figure~\ref{fig:evolv}. 
The negative signs (both mean and median) of $\frac{d\rm{Z}}{d\emph{z}}$ are significant at $>$5 $\sigma$ of the measurement errors from $z\sim 0.8 \rightarrow$ 0. 
From  $z\sim 2.5 $ to 0.8, however,  $\frac{d\rm{Z}}{d\emph{z}}$ is marginally smaller than zero at the $\sim$1 $\sigma$ level from the Lensed $\rightarrow$ DEEP2 samples.  
If using the Erb06 $\rightarrow$ DEEP2 samples,  the metallicity evolution ($\frac{d\rm{Z}}{d\emph{z}}$ ) from $z\sim 2.5 $ to 0.8 is consistent with zero within $\sim$1 $\sigma$ of the measurement errors. 
The reason that there is no metallicity evolution from the $z\sim2$ Erb06 $\rightarrow$ $z\sim 0.8$ DEEP2 samples may be due to the UV-selected sample of Erb06 being biased towards more metal-rich
galaxies.

The right column of Figure~\ref{fig:evolv} is used to interpret the deceleration/acceleration in metal enrichment. 
Deceleration means the metal enrichment rate ($\frac{d\rm{Z}}{d\emph{z}(Gyr)}$=$\Delta$ dex Gyr$^{-1}$) is dropping from  high-$z$ to low-$z$.
Using our lensed galaxies, the mean rise in metallicity is $0.055\pm0.014$ dex Gyr$^{-1}$ for $z\sim 2.5\rightarrow0.8$, and $0.022\pm0.001$ dex Gyr$^{-1}$ for  $z\sim0.8\rightarrow0$.
The Mann-Whitney test shows that the mean rises in metallicity are larger for $z\sim 2.5\rightarrow0.8$ than for  $z\sim0.8\rightarrow0$ at a significance level of 95\% 
for the high mass bins (10$^{9.5}$ M$_{\odot}<$M$_{\star}<$10$^{11}$ M$_{\odot}$).  For lower mass bins,  the hypothesis that
the metal enrichment rates are the same  for $z\sim 2.5\rightarrow0.8$ and  $z\sim0.8\rightarrow0$ 
 can not be rejected at the 95\% confidence level, i.e, there is no difference in the metal enrichment rates for the lower mass bin. 
 Interestingly, if  the Erb06 sample is used instead of the lensed sample, the hypothesis that
the metal enrichment rates are the same  for $z\sim 2.5\rightarrow0.8$ and  $z\sim0.8\rightarrow0$ 
 can not be rejected at the 95\% confidence level  for all mass bins.  This means that statistically,
  the metal enrichment rates are the same  for $z\sim 2.5\rightarrow0.8$ and  $z\sim0.8\rightarrow0$ for all mass bins from the Erb06 $\rightarrow$ DEEP2 $\rightarrow$ SDSS  samples.

The clear trend of the average/median metallicity in galaxies rising from high-redshift to the local universe is not surprising.
Observations based on  absorption lines  have shown a continuing  fall 
in metallicity using the damped Ly$\alpha$ absorption (DLA) galaxies at higher redshifts ($z \sim 2 - 5$) \citep[e.g.,][]{Songaila02,Rafelski12}. 
There are several physical reasons to expect that high-$z$ galaxies are less metal-enriched:
(1) high-$z$ galaxies are younger, have higher gas fractions, and have gone through less generations of star formation than local galaxies; 
(2) high-$z$ galaxies may be still accreting a large amount of metal-poor pristine gas from the environment, hence have lower
average metallicities; 
(3) high-$z$ galaxies may have more powerful outflows that 
drive the metals out of the galaxy. 
It is likely that all of these mechanisms have played a role in diluting the metal content at high redshifts.

\subsection{Comparison between the  $\rm{Z}\emph{z}$ Relation and  Theory}\label{sec:zzmd}
We compare our observations with model predictions from the 
cosmological hydrodynamic simulations of \citet{Dave11b,Dave11a}.  These models are built within a canonical hierarchical structure  formation context. The models take into account the important feedback of outflows by implementing an observation-motivated momentum-driven wind model  \citep{Oppenheimer08}. 
The effect of inflows and mergers 
are  included in the  hierarchical structure formation of the simulations. Galactic outflows are dealt specifically in the momentum-driven wind models. 
\citet{Dave07} found that the outflows are key to regulating metallicity, while inflows play a second-order regulation role.

 The model of \citet{Dave11b} focuses on the metal content of  star-forming galaxies. Compared with the previous work of \citet{Dave07}, 
 the new simulations employ the most up-to-date treatment 
 for supernova and AGB star enrichment, and include an improved version of the 
 momentum-driven wind models (the $vzw$ model) where the wind properties are derived based on host galaxy masses \citep{Oppenheimer08}.  
  The model metallicity in \citet{Dave11b} is defined as the SFR-weighted metallicity of all gas particles in the identified simulated galaxies.
This model metallicity can be compared directly with the metallicity we observe in star-forming galaxies after a constant offset normalization to account for 
the uncertainty in the absolute metallicity scale \citep{Kewley08}.
The offset is obtained by matching the model metallicity with  the SDSS metallicity.    
Note that the model has a galaxy mass resolution limit of M$_{\star}\sim$10$^{9}$ ~M$_{\odot}$.  
 For the  $\rm{Z}\emph{z}$ plot, we normalize the model metallicity with the median SDSS metallicity 
computed from all SDSS galaxies $>$10$^{9}$ M$_{\odot}$. For the MZ relation in Section~\ref{sec:mzr}, 
we normalize the model metallicity with the SDSS metallicity at the  stellar mass of  10$^{10}$ M$_{\odot}$.

We compute the median metallicities of the \citet{Dave11b} model outputs in redshift bins from $z=0$ to $z=3$ with an increment of 0.1.    
The median metallicities with 1$\sigma$ spread (defined as including 68\% of the data)  of the model at each redshift are overlaid on the observational data in the $\rm{Z}\emph{z}$ plot. 

 We compare our observations with the model prediction in 3 ways: 

 (1) We compare the observed median metallicity  with the model median metallicity. 
We see that for the lower mass bins (10$^{9-9.5}$, 10$^{9-10}$  M$_{\odot}$),  the median of the model metallicity is consistent with the median of the observed metallicity within the observational errors.  However, 
for higher mass bins, the model over-predicts the metallicity at all redshifts. The over-prediction is most significant in the highest mass bin of  10$^{10-11}$ M$_{\odot}$, 
where the Student's t-statistic shows that  the model distributions have significantly different means than the observational data at all redshifts, with 
a probability of being a chance difference  of $<10^{-8}$, $<10^{-8}$, 1.7\%, 5.7\%  for SDSS, DEEP2, the Lensed, and the Erb06 samples respectively.
For the alternative high-mass bin of 10$^{9.5-11}$ M$_{\odot}$, the model also over-predicts  the observed metallicity except for the Erb06 sample, with a 
chance difference between the model and observations of $<10^{-8}$, $<10^{-8}$, 1.7\%, 8.9\%,  93\%  for SDSS, DEEP2, the Lensed, and the Erb06 samples respectively.

(2) We compare the scatter of the observed metallicity (orange error bars on $\rm{Z}\emph{z}$ plot)  with the scatter of the models  (red dashed lines). 
For all the samples, the 1$\sigma$ scatter of the data from the SDSS ($z\sim0$), DEEP2($z\sim0.8$), and the Lensed sample ($z\sim2$) are: 0.13, 0.15, and 0.15 dex;
whereas the  1$\sigma$ model scatter is 0.23, 0.19, and 0.14 dex.   
We find  that the observed metallicity scatter is increasing systematically as a function of redshift for the high mass bins whereas the model does not predict such a trend: 
   0.10, 0.14, 0.17 dex c.f. model 0.17,  0.15, 0.12 dex; 10$^{9.5-11}$ M$_{\odot}$ and 0.07, 0.12, 0.18 dex  c.f. model 0.12, 0.11, 0.10 dex ; 10$^{10-11}$ M$_{\odot}$ from SDSS $\rightarrow$ DEEP2
   $\rightarrow$  the Lensed sample.   Our observed scatter is  in tune with the work of \citet{Nagamine01} in which the predicted stellar metallicity scatter  increases with redshift. 
  Note that our lensed samples are still small and have large measurement errors in metallicity ($\sim 0.2$ dex). The discrepancy between the 
observed scatter and models needs to be further confirmed with a larger sample.

(3) We compare  the  observed slope  ($\frac{d\rm{Z}}{d\emph{z}}$) of  the  $\rm{Z}\emph{z}$ plot with the model predictions (Figure~\ref{fig:evolv}). 
We find the observed $\frac{d\rm{Z}}{d\emph{z}}$  is consistent with the  model prediction within the observational errors  for
 the undivided sample of all masses $>$10$^{9.0}$ M$_{\odot}$.   However, when divided into mass bins, 
  the model predicts a slower enrichment than observations from $z\sim0\rightarrow0.8$ for  
the  lower mass bin of 10$^{9.0-9.5}$ M$_{\odot}$,  and from $z\sim0.8\rightarrow2.5$ for  
the  higher mass bin of 10$^{9.5-11}$ M$_{\odot}$ at a 95\% significance level.  

Dave et al. (2011) showed that their models over-predict the metallicities for the highest mass galaxies in the SDSS. 
They suggested that either (1) an additional feedback mechanism might be needed to suppress star formation in the most massive galaxies; or (2) wind recycling may be bringing in highly enriched material that elevates the galaxy metallicities.  It is unclear from our data which (if any) of these interpretations is correct.  Additional theoretical investigations specifically focusing on metallicities in the most massive active galaxies are needed to determine the true nature of this discrepancy.

\section{Evolution of the Mass-Metallicity Relation}\label{sec:mzr}
\subsection{The Observational Limit of the Mass-Metallicity Relation}\label{sec:obslimit}
For the N2 based metallicity,  there is a limiting metallicity below which the \nii\ line is too weak to be detected. 
Since \nii\ is the weakest of the \ha\,+\nii\ lines, it is therefore the 
flux of \nii\ that drives the metallicity detection limit.  
Thus, for a given instrument sensitivity, there is a region on the mass-metallicity relation that is observationally unobtainable. 
Based on a few simple assumptions, we can derive the boundary of this region as follows.

Observations have shown that there is a positive correlation between the stellar mass $M_{\star}$ and $SFR$ \citep{Noeske07a,Elbaz11,Wuyts_S11}.
One explanation for the $M_{\star}$ vs. SFR relation is that  more massive galaxies have earlier onset of initial star formation with shorter timescales of exponential decay \citep{Noeske07b,Zahid12b}.
The shape and amplitude of the SFR vs. $M_{\star}$  relation at different redshift $z$ 
can be characterized by two parameters  $\delta(z)$ and $\gamma(z)$,  
where $\delta(z)$ is the logarithm of the SFR at $10^{10} M_{\star}$ and $\gamma(z)$ is the power law index \citep{Zahid12b}.
 
\begin{figure}[!ht]
\begin{center}
\includegraphics[trim = 5mm 5mm 10mm 13mm, clip, width=6.25cm,angle=90]{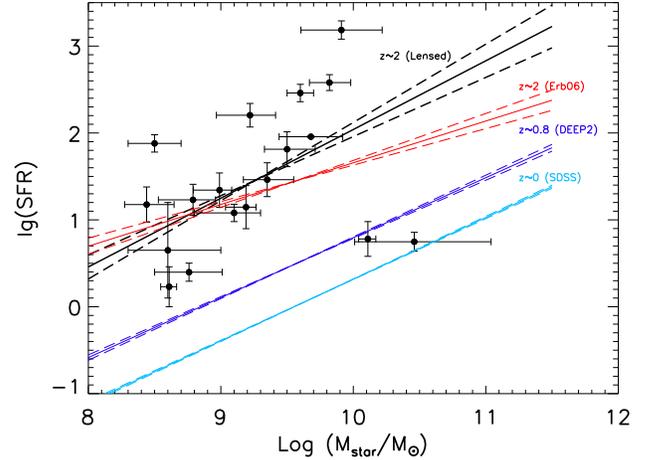}
\caption{ SFR vs. stellar mass relation.  
  The light-blue, blue, and red lines 
show the best-fit SFR vs. stellar mass relation from the  SDSS, DEEP2, and Erb06 samples respectively (\citep{Zahid11}, see also Table~\ref{tabms}). 
Back dots are the lensed sample used in this work. The SFR for the lensed sample is derived from the \ha\ flux with dust extinction corrected from the SED fitting.
The errors on the SFR of the lensed sample are statistical errors of the \ha\ fluxes. Systematic errors of the SFR can be large  (a factor of 2-3) for our lensed galaxies  due to complicated 
aperture effects (Section 3.1). 
}
 \label{fig:ms}
 \end{center}
\end{figure}

\begin{figure*}[!ht]
\begin{center}
\includegraphics[trim = 5mm 4mm 8mm 8mm, clip, width=10cm,angle=90]{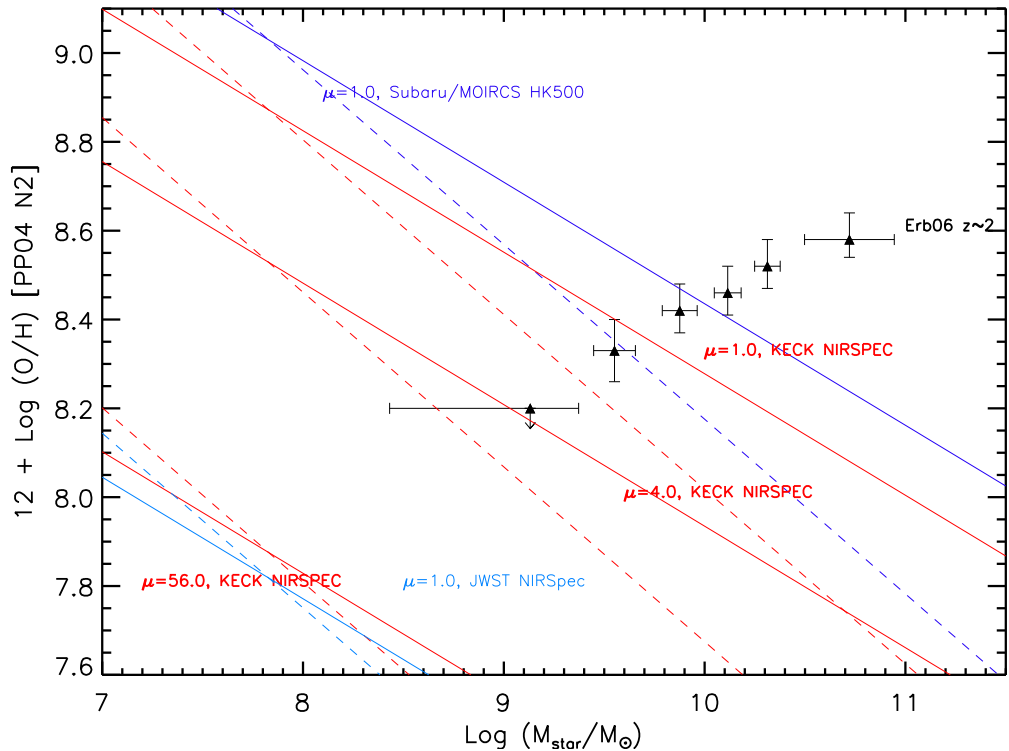}
\caption{The instrument detection limit on the MZ relation.  We give the dependence of this detection limit in Equation 4.  Shown here
are examples of the detection limit based on given parameters specified as follows.  The solid lines use the parameters based on the mass-SFR relation of the Erb06 sample: 
$\delta=1.657$ and $\gamma=0.48$ at $z=2.26$.  The dashed lines  use the parameters   based on the mass-SFR relation of the Lensed sample:
$\delta=2.02$ and $\gamma=0.69$ at $z=2.07$ (see Figure~\ref{fig:ms}; Table~\ref{tabms}).  The parameters adopted for the instrument flux limit are given in Section 7.1.
The lensing magnification ($\mu$) are fixed at 1.0 (i.e., non-lensing cases) for 
Subaru/MOIRCS (blue lines) and JWST/NIRSpec (light blue). 
The red lines show the detection limits for KECK/NIRSPEC with different magnifications. 
Black filled triangles show the \citet{Erb06} sample.   We show that stacking and/or lensing magnification can help to push the observational boundary
of the MZ relation to lower mass and metallicity regions.  For example, \citet{Erb06} used stacked NIRSPEC spectra with $N\sim$ 15 spectra in each mass bin. The effect of stacking  ($N\sim$ 15 per bin) is similar  to observing with a lensing magnification of $\mu \sim 4$.
 }
 \label{fig:mzall}
\end{center}
\end{figure*}

The relationship between the SFR and M$_{\star}$ then becomes:
\begin{equation}\label{ms}
\log_{10} (SFR(z)) = \delta(z) + \gamma(z) [\log_{10} (M_{\star}/M_{\odot})-10]
\end{equation}
 
As an example, we show in Figure~\ref{fig:ms} the SFR vs. M$_{\star}$ relation at three  redshifts ($z\sim 0, 0.8, 2$).
The  best-fit values of $\delta(z)$ and $\gamma(z)$  are listed in Table~\ref{tabms}.
 
Using the \citet{Kennicutt98b} relation between SFR and \ha\ : 
\begin{equation}\label{k08}
SFR = 7.9 \times 10^{-42} L(H\alpha) [ergs~s^{-1}]
\end{equation}

and  the N2 metallicity calibration \citep{Pettini04}:
 \begin{equation}\label{pp04}
12+{\rm log (O/H)}=8.90+0.57\times {\log_{10} [NII]/H\alpha},
\end{equation}
we can then derive a metallicity detection limit.  
We combine Equations (\ref{ms}), (\ref{k08}) and (\ref{pp04}), and assume 
the  \nii\  flux is greater than the instrument flux detection limit. We provide the detection limit for the PP04N2 diagnosed  MZ relation:

\begin{equation}
\begin{array}{l}
\displaystyle Z_{met} \ge [\log_{10} (f_{inst}/\mu)+2\log_{10} D_{L}(z)-\gamma(z)\\
\displaystyle ~~~~~~~~~~\\
\displaystyle ~~~~~~~~~~~~~~M_{\star}-\beta(z)+\log_{10}(4 \pi)] 0.57 + 8.9
\end{array} 
\label{eq:inst}
\end{equation}

where:
\begin{equation}\label{beta}
\beta(z) \equiv \delta(z)-\gamma(z) 10+42-\log_{10}7.9;
\end{equation}
$\delta(z)$, $\gamma(z)$ are defined in Equation (\ref{ms});
$f_{inst}$ is the instrument flux detection limit in $ergs~s^{-1}~cm^{-2}$; $\mu$ is the lensing magnification in flux;
$D_{L}(z)$ is the luminosity distance in $cm$.

\begin{deluxetable}{lccccc}
\tabletypesize{\scriptsize}
\tablewidth{0pt}
\tablecolumns{4}
\tablecaption{Fit to the SFR-Stellar Mass Relation \label{tabms}}
\tablehead{
\colhead{Sample} & 
\colhead{Redshift (Mean)} & 
\colhead{$\delta$} & 
\colhead{$\gamma$}
}
\startdata
SDSS & 0.072& 0.317$\pm$0.003 & 0.71 $\pm$0.01\\
DEEP2 &0.78& 0.795$\pm$0.009 & 0.69$\pm$0.02\\
Erb06& 2.26 & 1.657$\pm$0.027&0.48$\pm$0.06\\
Lensed (Wuyts12)&1.69 & 2.93$\pm$1.28 & 1.47$\pm$0.14\\
Lensed (all) &2.07 & 2.02$\pm$0.83 & 0.69$\pm$0.09\\
\enddata
\tablecomments{The SFR vs. stellar mass relations at different redshifts can be characterize by
two parameters $\delta(z)$ and $\gamma(z)$, where $\delta(z)$ is the logarithm of the SFR at $10^{10} M_{\star}$, and $\gamma(z)$ is the power law index.
The best fits for the non-lensed samples are adopted from \citet{Zahid12b}.   The best fits for the lensed sample are calculated for the \citet{Wuyts12b} sample and the 
whole lensed sample separately.                                         
}
\end{deluxetable}

The slope of the mass-metallicity detection 
limit is related to the slope of the SFR-mass relation, whereas the y-intercept of the slope depends on the instrument flux limit (and flux 
magnification for gravitational lensing), redshift, and the y-intercept  of the SFR-mass relation.

Note that the exact location of the boundary depends on the input parameters of Equation 8. As 
an example, we use the $\delta(z)$ and $\gamma(z)$ values  of the Erb06 and Lensed samples respectively (Table~\ref{tabms}).
We show the detection boundary for  three current and future NIR instruments: Subaru/MOIRCS, KECK/NIRSPEC and JWST/NIRSpec. 
The  instrument flux detection limit  is based on background limited estimation in 10$^{5}$ seconds (flux in units of 10$^{-18}$~ergs~s$^{-1}$~cm$^{-2}$ below).   
For Subaru/MOIRCS (low resolution mode, HK500), we adopt $f_{inst}$  = 23.0 based on the 1$\sigma$ uncertainty of our MOIRCS spectrum (flux=4.6 in 10 hours), scaled to 3$\sigma$ in 10$^{5}$ seconds.
For KECK/NIRSPEC,  we use $f_{inst}$  = 12.0, based on the 1$\sigma$ uncertainty of Erb et al. (2006) (flux=3.0 in 15 hours), scaled to 3$\sigma$ in 10$^{5}$ seconds.
For JWST/NIRSpec,  we use $f_{inst}$  = 0.17,  scaled to 3$\sigma$ in 10$^{5}$ seconds\footnote{\url{http://www.stsci.edu/jwst/instruments/nirspec/sensitivity/}}.

Since lensing flux magnification  is equivalent to lowering the instrument flux detection limit, we see that with a lensing magnification of $\sim$55, 
we reach the sensitivity of JWST using KECK/NIRSPEC.  Stacking can also push the observations below the instrument flux limit. For instance, the $z\sim 2$ Erb et al. (2006) sample
was obtained from stacking the NIRSPEC spectra of 87 galaxies, with $\sim$ 15 spectra in each mass bin, thus the Erb06 sample has been able to probe $\sim$ 4 times deeper
than the nominal detection boundary of NIRSPEC. 

The observational detection limit on the MZ relation is important for understanding the incompleteness and biases of samples due to observational constraints.
However, we caution that  the relation between $Z_{met}$ and M$_{\star}$ in Equation 4 will have significant intrinsic dispersion due to variations in the 
observed properties of individual galaxies. This includes scatter in the M$_{\star}$-SFR relation, the N2 metallicity calibration, the amount of dust extinction, 
and variable slit losses in spectroscopic observations. For example, a scatter of 0.8 dex in $\delta$ for the lensed sample (Table 3) 
implies a scatter of approximately 0.5 dex in $Z_{met}$.  In addition, 
Equations 2 and 4 include implicit assumptions of zero dust extinction and no slit loss, such that the derived line flux is overestimated (and $Z_{met}$ is underestimated). 
Because of  the above uncertainties and biases in the assumptions we made, Equation 4 should be used with due caution.

\subsection{The Evolution of the  MZ Relation }

\begin{figure*}[!ht]
\begin{center}
\includegraphics[trim = 10mm 2mm 4mm 8mm, clip, width=18.cm,angle=0]{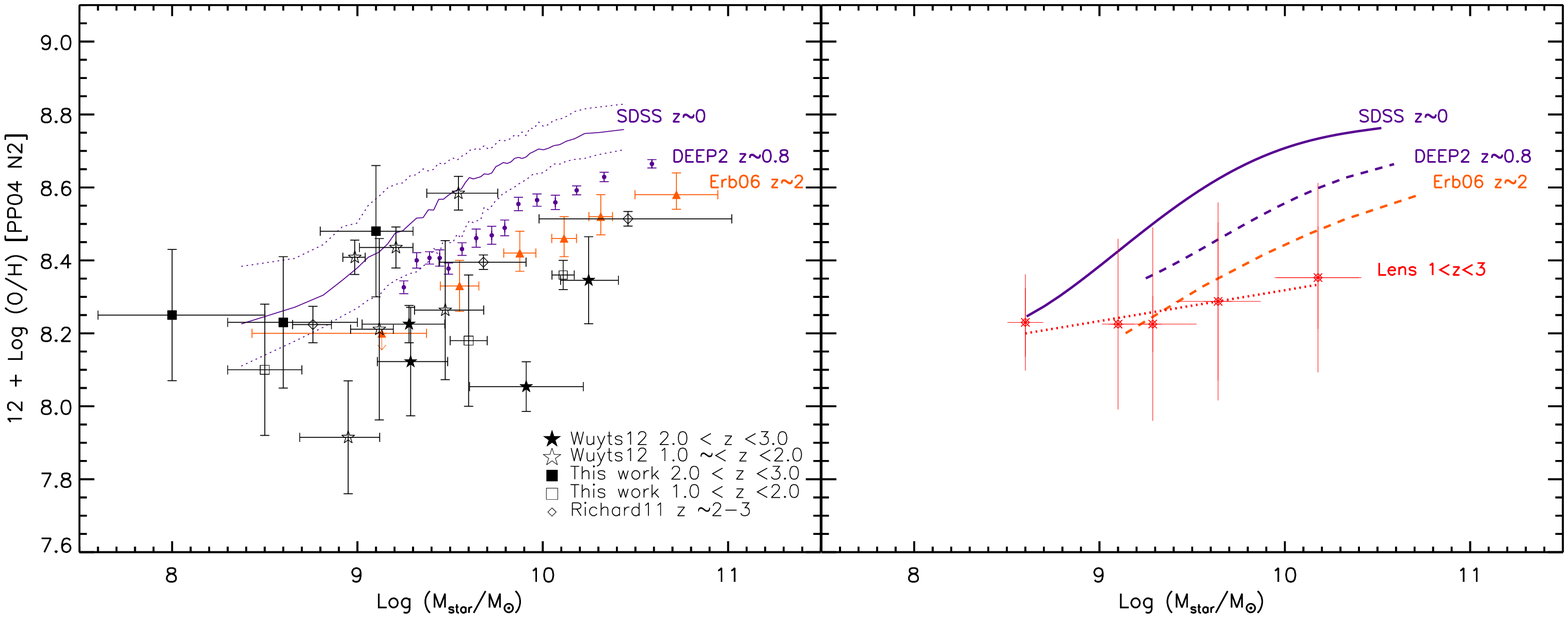}
\caption{Left: the observed MZ relation.   Black symbols are the lensed galaxy sample at $z>1$. Specifically,  the squares are from this work;
the stars are  from \citet{Wuyts12b}, and the diamonds are from \citet{Richard11a}. 
The orange triangles show the \citet{Erb06} sample.  The local SDSS relation and its 1-sigma range are drawn in purple lines. 
    The $z\sim$ 0.8 DEEP2 relations from \citet{Zahid11} are drawn in purple dots. 
 Right:  the best fit to the MZ relation. A second degree polynomial function is fit to the SDSSS, DEEEP2, and Erb06 samples. A simple linear function is fit to the lensed sample. The $z>1$ lensed data are binned in 5 mass bins (symbol: red star) and  the   median and 1$\sigma$ standard deviation of each bin  are plotted on top of the linear fit.
 }
 \label{fig:mzobs}
 \end{center}
\end{figure*}

\begin{figure*}[!ht]
\begin{center}
\includegraphics[trim = 10mm 2mm 4mm 8mm, clip, width=18.cm,angle=0]{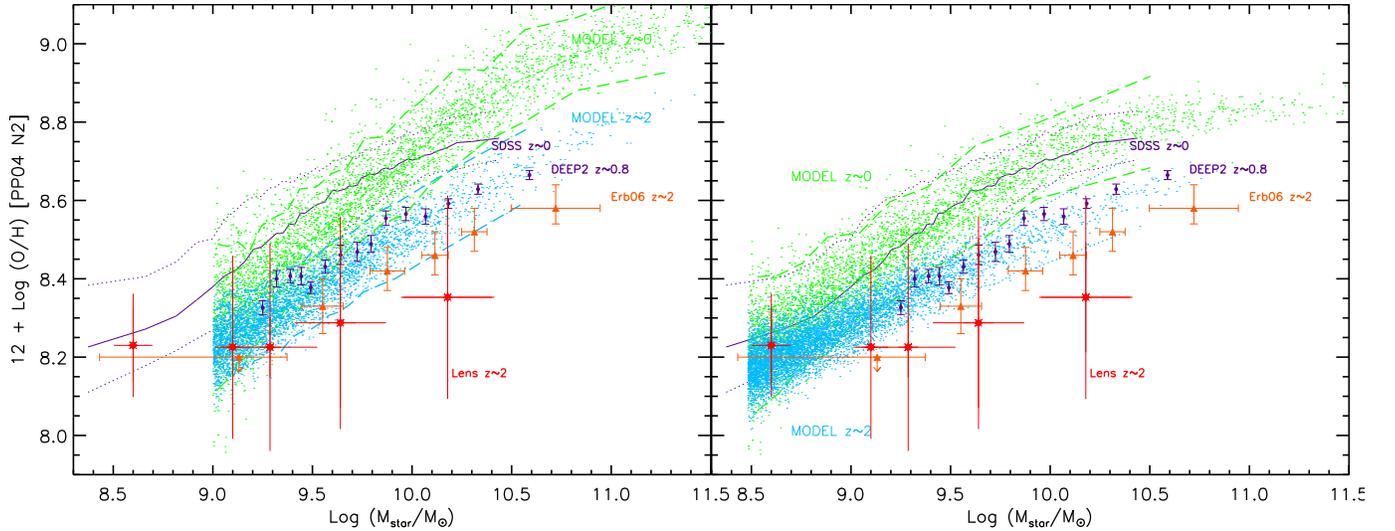}
\caption{Model predictions of the MZ relation. 
The data symbols are the same as  those used  in Figure~\ref{fig:mzobs}.
The small green and light blue dots are the  cosmological hydrodynamic simulations with momentum-conserving wind models
 from  \citet{Dave11b}.  The difference between the left and right panels are the different normalization methods used. The left panel normalizes the model metallicity  to the observed SDSS  values by applying a constant offset at $M_{star}\sim 10^{10}M_\odot$,  whereas the right panel normalizes the model metallicity  to the observed SDSS metallicity by allowing a constant shift in the slope, amplitude and stellar mass.  Note that the model has a mass cut off at $1.1\times 10^{9}M_\odot$. 
    }
 \label{fig:mz}
\end{center}
\end{figure*}

 Figure~\ref{fig:mzobs} shows the mass and metallicity measured from 
 the SDSS, DEEP2, and our lensed samples. The Erb et al. (2006) (Erb06) stacked data are also included for comparison.
 We highlight a few interesting features in Figure~\ref{fig:mz}: \\

(1) To first order,  the  MZ relation still exists at $z\sim2$, i.e., more massive systems are more metal rich. 
The Pearson correlation coefficient is $r = 0.33349$, with a probability of being a chance correlation of $P =$ 17\%. 
A simple linear fit to the lensed sample yields a slope of 0.164$\pm$0.033, with a y-intercept of 6.8$\pm$0.3.
\\

(2) All $z>1$ samples show evidence of evolution to lower metallicities at fixed stellar masses.  
At high stellar mass (M$_{\star}>$10$^{10}$ M$_{\odot}$),  the lensed sample has  a mean metallicity and a standard deviation of the mean 
of 8.41$\pm$0.05, whereas the mean and standard deviation of the mean for the Erb06 sample is 8.52$\pm$0.03.  The lensed sample is offset to lower metallicity by 0.11$\pm$0.06 dex compared to the 
Erb06 sample.  This slight offset may indicate the selection difference between the UV-selected (potentially more dusty and metal rich) sample and the lensed sample (less biased towards UV bright systems).\\

(3) At lower mass  (M$_{\star}<$10$^{9.4}$ M$_{\odot}$), our lensed sample provides  12 individual metallicity measurements 
at $z>1$.  The mean metallicity of the galaxies with M$_{\star}<$10$^{9.4}$ M$_{\odot}$ is 8.25$\pm$0.05,  roughly consistent with the $<$8.20 upper limit of the 
stacked metallicity  of the lowest mass bin  (M$_{\star}\sim$10$^{9.1}$ M$_{\odot}$) of  the   Erb06 galaxies.\\

(4) Compared with the Erb06  galaxies,
there is a lack of the highest mass galaxies in our lensed sample. 
We note that there is only 1 object with M$_{\star}>$10$^{10.4}$ M$_{\odot}$ among all three lensed samples combined.
The lensed sample is less affected by the color selection and may be more representative of the mass distribution of  high-$z$ galaxies.
In the hierarchical galaxy formation paradigm, galaxies grow their masses with time.  The number density of massive galaxies at high redshift 
is smaller than at $z\sim0$, thus the number of massive lensed galaxies is small.
Selection criteria such as the UV-color selection of the Erb06 and SINs \citep{Genzel11} galaxies can be applied to 
target the high-mass galaxies on the MZ relation at high redshift.\\

\begin{figure}[!ht]
\begin{center}
\includegraphics[scale=0.354,angle=90]{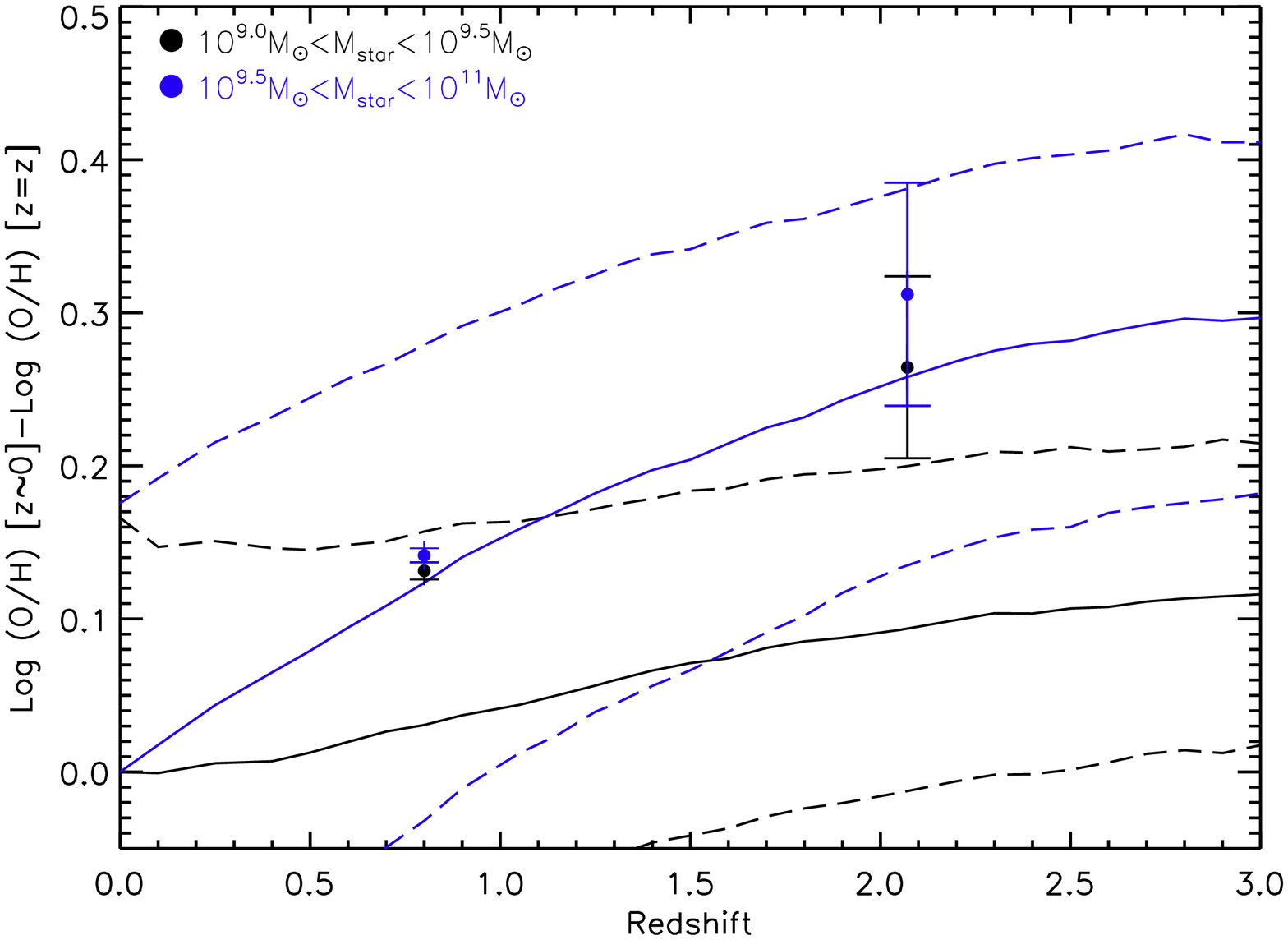}
\caption{  The ``mean evolved metallicity"  as a function of redshift for two mass bins (indicated by four colors). Dashed lines show the 
median and 1$\sigma$ scatter of the model prediction from  \citet{Dave11b}.   The observed data from DEEP2  and our lensed sample are plotted as filled circles.
}
 \label{fig:mz1}
 \end{center}
\end{figure}

\subsection{Comparison with Theoretical MZ Relations}
Understanding the origins of the MZ relation has been the driver of copious theoretical work. 
Based on the idea that metallicities are mainly driven by an equilibrium among stellar enrichment, infall and outflow, 
\citet{Finlator08} developed smoothed particle  hydrodynamic simulations. They found that the inclusion of a momentum-driven wind model  ($vzw$)
fits best to the $z\sim2$ MZ relations compared to other outflow/wind models. The updated version of their $vzw$ model is described in detail in \citet{Dave11b}. 
We overlay the \citet{Dave11b} $vzw$ model outputs on the MZ relation in Figure~\ref{fig:mz}.  
We  find that the model does not  reproduce the MZ redshift evolution seen in our observations.  We provide possible explanations as follows.

 \citet{Kewley08} found that both the  shape and scatter of the MZ relation  vary significantly among different metallicity diagnostics. 
 This poses a  tricky normalization problem when comparing models to observations. For example,  a model output may fit the MZ relation 
 slope from one strong-line diagnostic, but fail to fit the MZ relation from another  diagnostic, which may have a very different slope.
 This is exactly what we are seeing on the left panel of  Figure~\ref{fig:mz}.  \citet{Dave11b}  applied a constant offset of 
 the model metallicities by matching the amplitude of the model  MZ relation at $z\sim0$ with the observed local MZ relation  of \citet[][T04]{Tremonti04} at the stellar mass of 10$^{10}$ M$_{\odot}$.   \citet{Dave11b} found that the characteristic shape and scatter of the MZ relation from the  $vzw$ model matches the T04 MZ relation between
 10$^{9.0}$ M$_{\odot}<$M$_{\star}<$10$^{11.0}$  within the 1$\sigma$ model and observational scatter. However, since both the slope and amplitude of the T04 SDSS MZ relation are significantly larger than the SDSS MZ relation derived using  the PP04N2 method \citep{Kewley08},
 the PP04N2-normalized MZ relation from the model does not recover  the local MZ relation within $1\sigma$. 
 
 In addition, the stellar mass measurements
 from different methods may cause a systematic offsets in the x-direction of the MZ relation \citep{Zahid11}. As a result, 
 even though the  shape, scatter, and evolution with redshifts are independent predictions from the model,  systematic uncertainties in metallicity diagnostics and stellar mass estimates do not
allow the shape to be constrained separately. 

In  the right panel of Figure~\ref{fig:mz}, we allow the model slope ($\alpha$), metallicity amplitude ($Z$), and stellar mass (M$_{\ast}$) to change slightly so that it fits the local SDSS MZ relation. Assuming that this change in slope ($\Delta \alpha$), and x, y amplitudes ($\Delta Z, \Delta M_{\ast}$) are caused by the systematic offsets in observations,  then the same $\Delta \alpha$, $\Delta Z$, and  $\Delta M_{\ast}$ can be applied to model MZ relations at other redshifts.   Although normalizing the model MZ relation in this way 
 will make the model lose prediction power 
for the shape of the MZ relation, it at least leaves the redshift evolution of the MZ relation as a testable model output.  
 
Despite the normalization correction, we see from Figure~\ref{fig:mz} that the models predict less evolution from z $\sim$ 2 to   z $\sim$ 0
than the observed MZ relation.  To quantify, we divide the model data
into two mass bins and derive the mean and 1$\sigma$ scatter in each mass bin as a function of redshift. 
We define the  ``mean evolved metallicity" on the MZ relation as 
the difference between the mean metallicity at redshift $z$ and the mean metallicity at $z \sim 0$ at a fixed stellar mass (log (O/H) [z$ \sim $0] $-$ log (O/H) [z$ \sim $2]).   
The   ``mean evolved metallicity" errors are calculated based on the standard errors of the mean.

In Figure~\ref{fig:mz1} we plot the ``mean evolved metallicity"  as a function of redshift for two mass bins: 
10$^{9.0}$ M$_{\odot}<$M$_{\star}<$10$^{9.5}$ M$_{\odot}$, 10$^{9.5}$ M$_{\odot}<$M$_{\star}<$10$^{11}$ M$_{\odot}$. 
We calculate the  observed ``mean evolved metallicity" for DEEP2 and our lensed sample in the same mass bins. 
 We see that the observed mean evolution of the lensed sample are largely uncertain and no conclusion between the model and observational data can be drawn.
 However, the DEEP2 data are well-constrained and can be compared with the model. 

We find that at $z\sim0.8$,  the  mean evolved metallicity of the high-mass galaxies are consistent with the  mean evolved metallicity of the models. 
The observed mean evolved metallicity of the low-mass bin galaxies is  $\sim$ 0.12 dex larger than the mean evolved metallicity of the models in the same
mass bins. 

\section{Compare with Previous Work in Literature}\label{sec:discuss}

In this Section, we compare our findings with previous work on the evolution of the  MZ relation.

For low masses (10$^{9}$ M$_{\odot}$), we find a larger enrichment (i.e., smaller decrease in metallicity) between 
 $z\sim2 \rightarrow 0$ than either the non-lensed sample of \citet{Maiolino08}  (0.15 dex c.f. 0.6 dex) or the lensed sample of 
\citet{Wuyts12b, Richard11a} (0.4 dex).  These discrepancies may reflect differences in metallicity calibrations applied.  
It is clear that a larger sample is required to characterize the true mean and spread in metallicities at intermediate redshift.  
Note that the lensed samples are still small and have large measurement errors in both stellar masses (0.1 to 0.5 dex) and metallicity ($\sim 0.2$ dex).

For high masses (10$^{10}$ M$_{\odot}$), we find similar  enrichment  (0.4 dex) between 
 $z\sim2 \rightarrow 0$ compare to the non-lensed sample of \citet{Maiolino08}  and the lensed sample of 
\citet{Wuyts12b, Richard11a}.

We find in Section 6.1 that the deceleration  in metal enrichment is significant in the highest mass bin (10$^{9.5}$ M$_{\odot}<$M$_{\star}<$10$^{11}$~M$_{\odot}$) of our samples.
The deceleration in metal enrichment  from $z\sim 2\rightarrow0.8$ to $z\sim0.8\rightarrow0$ is consistent with the picture that the star formation and mass assembly peak
between redshift 1 and 3 \citep{HopkinsAM06}.   The deceleration is larger by 0.019$\pm$0.013 dex Gyr$^{-2}$  in the high mass bin, suggesting a possible mass-dependence
 in chemical enrichment, similar to the  ``downsizing" mass-dependent growth of stellar mass \citep{Cowie96, Bundy06}.  
In the  downsizing picture,  more massive galaxies formed their stars earlier and on shorter timescales compared with less massive galaxies \citep{Noeske07b}.
Our observation of the chemical  downsizing is consistent with previous metallicity evolution work \citep{Panter08,Maiolino08, Richard11a,Wuyts12b}.

 We find that  for higher mass bins, the model of \citet{Dave11b} over-predicts the metallicity at all redshifts. 
The over-prediction is most significant in the highest mass bin of  10$^{10-11}$ M$_{\odot}$. This conclusion similar to the findings in \citet{Dave11b,Dave11a}.
In addition, we point out that when comparing the model metallicity with the observed metallicity, there is a normalization problem stemming from the discrepancy among different metallicity calibrations (Section 7.3).

We note the evolution of the MZ relation is  based on an ensemble of the averaged
SFR weighted metallicity of the star-forming galaxies at each epoch.  
The MZ relation  does not reflect an evolutionary track of individual galaxies. We are probably seeing a
different population of galaxies at each redshift \citep{Brooks07,Conroy08}.  For example, a $\sim$10$^{10.5}$ M$_{\odot}$ massive galaxy at $z\sim$2 will most likely evolve into an elliptical galaxy in the local universe 
and will not appear on the local MZ relation. On the other hand, to trace the progenitor of a $\sim$10$^{11}$ M$_{\odot}$ massive galaxy today, we need to observe 
a $\sim$10$^{9.5}$ M$_{\odot}$ galaxy at $z\sim$2 \citep{Zahid12b}.  

It is clear that gravitational lensing has the power to probe lower stellar masses than current color selection techniques.
Larger  lensed samples with high-quality observations are required to reduce the measurement errors.

\section{Summary}\label{sec:sum}
To study the evolution of the overall metallicity and MZ  relation as a function of redshift, it is critical to remove the systematics among different redshift samples. 
The major caveats in current MZ relation studies at $z>$1 are: (1) metallicity is not based on the same diagnostic method; 
(2) stellar mass is not derived  using the same method;  (3) the samples are selected differently and selection effects on mass and metallicity are poorly understood. 
In this paper, we attempt to minimize these issues  by re-calculating the stellar mass and metallicity consistently, 
 and by expanding the lens-selected sample at $z>$1. 
We aim to present a reliable observational picture of the metallicity evolution of star forming galaxies as a function  of stellar mass between $0 < z <3$. 
We find that: 
\begin{itemize}
    
\item There is a clear evolution in the mean and median metallicities of star-forming galaxies as a function of redshift. The mean metallicity 
 falls by $\sim0.18$ dex from redshift 0 to 1 and falls further by $\sim0.16$ dex from redshift 1 to 2. 
  
 \item A more rapid evolution is seen between $z\sim1 \rightarrow 3$  than $z\sim0 \rightarrow 1$ for the high-mass galaxies (10$^{9.5}$ M$_{\odot}<$M$_{\star}<$10$^{11}$ M$_{\odot}$),
 with almost twice as much enrichment between $z\sim1 \rightarrow 3$ 
than between $z\sim1 \rightarrow 0$.

\item  The deceleration in metal enrichment  from $z\sim 2\rightarrow0.8$ to $z\sim0.8\rightarrow0$ is significant in the high-mass galaxies (10$^{9.5}$ M$_{\odot}<$M$_{\star}<$10$^{11}$~M$_{\odot}$),
consistent with  a mass-dependent  chemical enrichment.

\item We compare the metallicity evolution of star-forming galaxies from $z=0\rightarrow3$ with the most recent cosmological hydrodynamic simulations. 
 We see that the model metallicity is consistent with the observed metallicity within the observational error for the low mass bins. 
However, for higher mass bins, the model over-predicts the metallicity at all redshifts. The over-prediction is most significant in the highest mass bin of  10$^{10-11}$ M$_{\odot}$.
Further theoretical investigation into the metallicity of the highest mass galaxies is required to determine the cause of this discrepancy.

\item The median metallicity of the lensed sample is 0.35$\pm$0.06 dex lower than local SDSS galaxies and  0.28$\pm$0.06 dex lower than the $z\sim0.8$ DEEP2 galaxies. 

\item Cosmological hydrodynamic simulation \citep{Dave11b} does not agree with
the evolutions of the observed MZ relation based on the PP04N2 diagnostic.
Whether the model fits the slope of the MZ relation depends on the normalization methods used.

\end{itemize}
This study is based on 6 clear nights of observations  on a 8-meter telescope, highlighting the efficiency in using lens-selected targets.  
  However, the lensed sample at $z > 1$ is still small.  We aim to significantly increase the sample size over the  years.

\acknowledgments 
We would like to thank the referee for an excellent report that has significantly improved this paper.
 T.-Y. wants to thank the MOIRCS supporting astronomer Ichi Tanaka and Kentaro Aoki for their enormous support  on the
MOIRCS observations.  We thank Youichi Ohyama for scripting the original MOIRCS data reduction pipeline. 
We are grateful to Dave Rommel for providing and explaining to us his most recent models.  T.-Y. wants to thank Jabran Zahid for the SDSS and DEEP2 data and many insightful discussions.  
 T.-Y. acknowledges a Soroptimist Founder Region Fellowship for Women.  L.K. acknowledges a NSF Early CAREER Award AST 0748559 and an 
 ARC Future Fellowship award FT110101052.  JR is supported by the Marie Curie Career Integration Grant 294074.  We wish to recognize and acknowledge the very significant cultural role and reverence that the summit of Mauna Kea has always had within the indigenous Hawaiian community.  
 
{\it Facilities:} \facility{Subaru (MOIRCS)}

\begin{appendix}
\section{Slit layout,  spectra for the lensed sample}\label{spec}
This section presents the slit layouts, reduced and fitted spectra for the newly observed lensed objects in this work. The line fitting procedure is described in Section~\ref{sec:fit}.
For each target, the top panel shows the HST  ACS 475W broad-band image of the lensed target. 
The slit layouts with different positional angles are drawn in white boxes. 
The bottom panel(s) show(s) the final reduced 1D spectrum(a) zoomed in for emission line vicinities. 
The black line is the observed spectrum for the target. The cyan line is the noise spectrum extracted from object-free pixels of the final 2D spectrum. 
Tilted grey mesh lines indicate spectral ranges where the sky absorption is severe.  Emission lines falling in these spectral windows suffer
from large uncertainties in telluric absorption correction.  The blue horizontal line is the continuum fit using first order polynomial
function after blanking out the severe sky absorption region.   The red lines overplotted on the emission lines are the overall Gaussian fit, with the blue
lines show individual components of the multiple Gaussian functions.  Vertical dashed lines show the center of the Gaussian profile for each emission line. 
The S/N of each line are marked under the emission line labels. Note that for lines with S/N $<$3, the fit is rejected and a 3-$\sigma$ upper limit is derived.   

\begin{figure*}[!ht]
\includegraphics[scale=0.55]{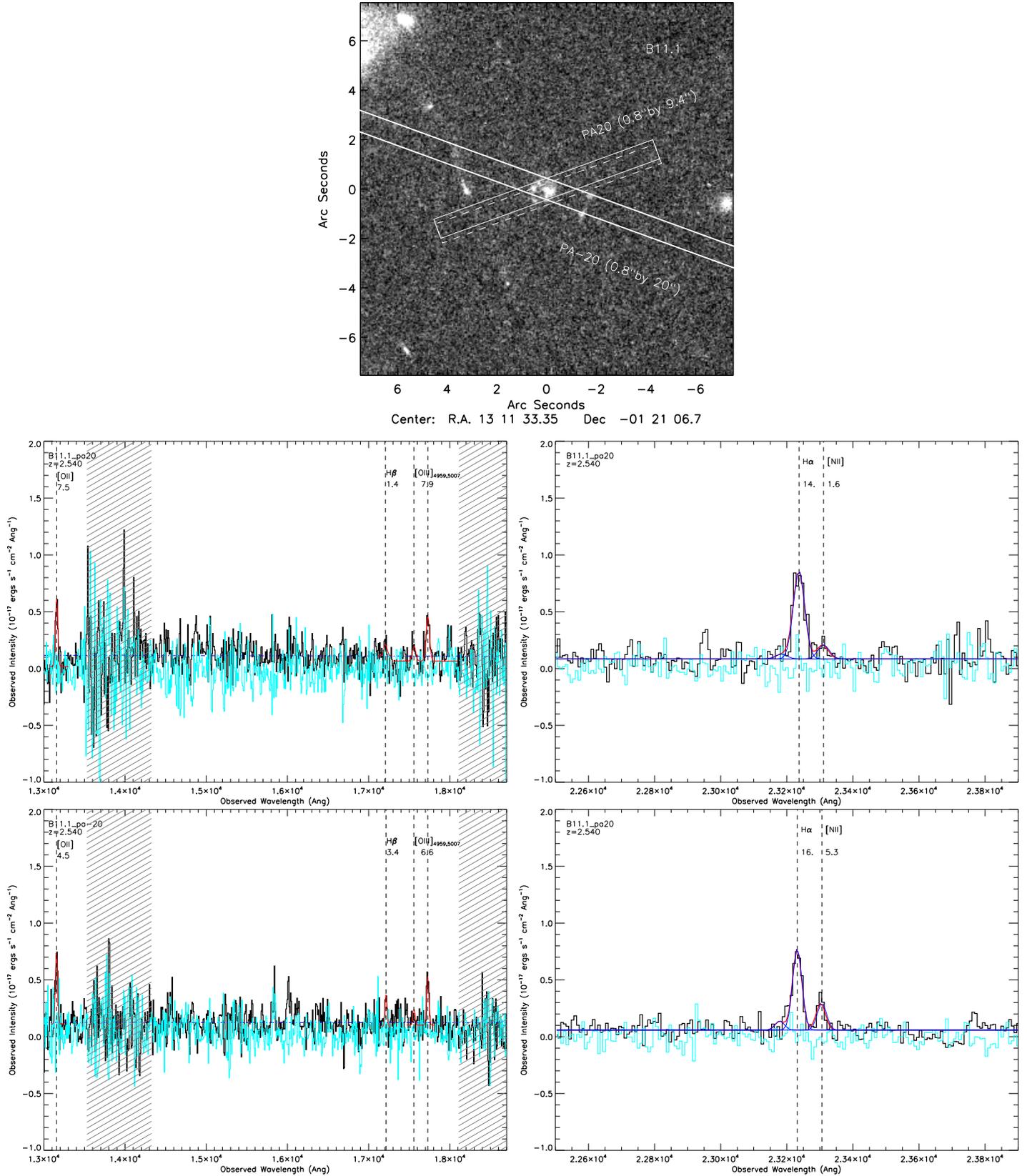}
\caption{z=2.540, B11.1 MOIRCS J, H band spectra. Detail descriptions are given in the Appendix text.
Note that the dashed box indicates the $\sim$0.1 arcsec alignment error of MOIRCS. 
}
 \label{fig:B11.1}
\end{figure*}

\begin{figure*}[!ht]
\includegraphics[scale=0.56]{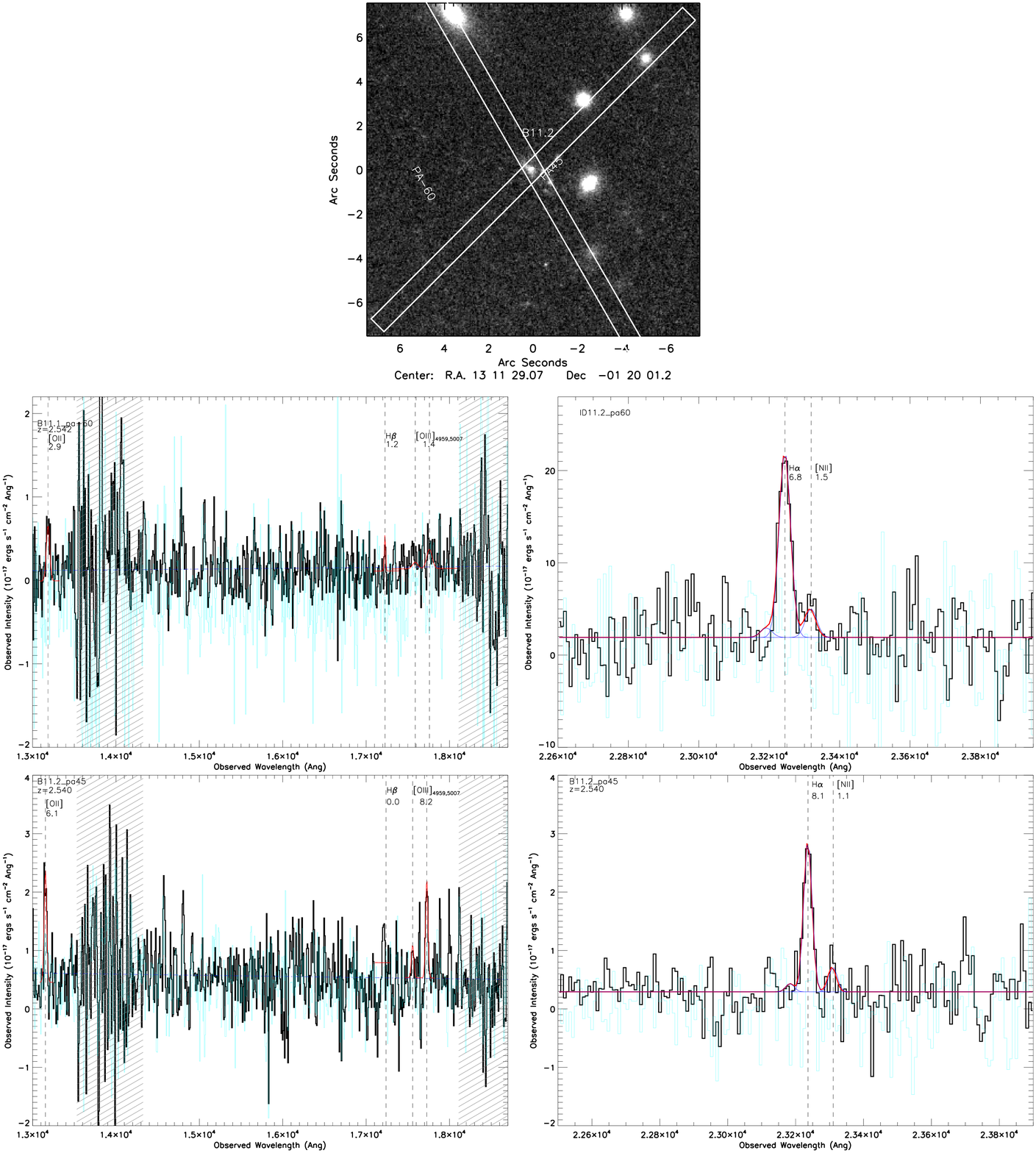}
\caption{z=2.540, B11.2 MOIRCS J, H band spectra. Detail descriptions are given in the Appendix text. 
}
\label{fig:B11.2}
\end{figure*}

\begin{figure*}[!ht]
\begin{center}
\includegraphics[scale=0.5]{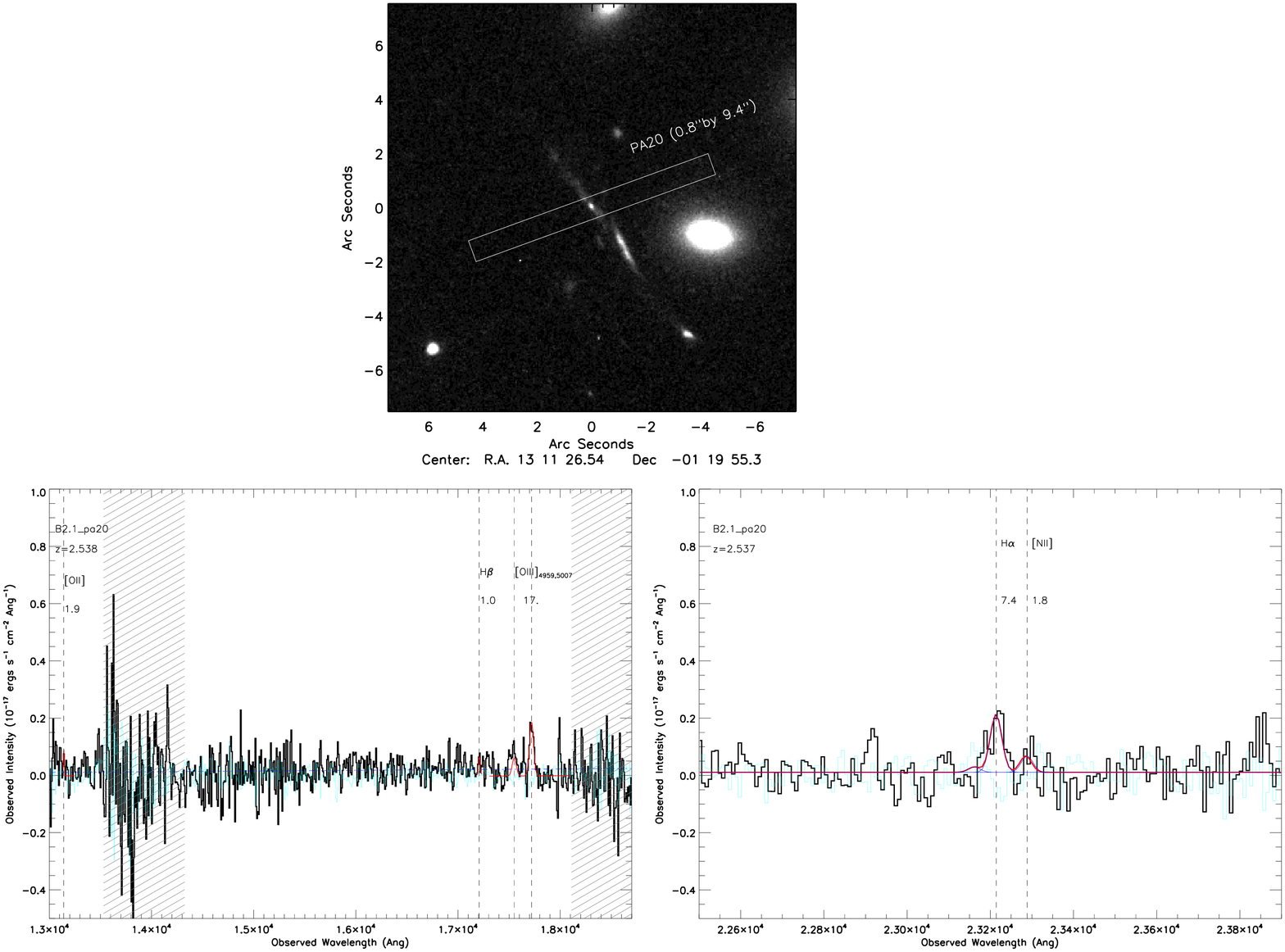}
\caption{z=2.537, B2.1, MOIRCS J, H band spectra. Detail descriptions are given in the Appendix text. 
}
\label{fig:B2.1}
\end{center}
\end{figure*}

\begin{figure*}[!ht]
\begin{center}
\includegraphics[scale=0.54]{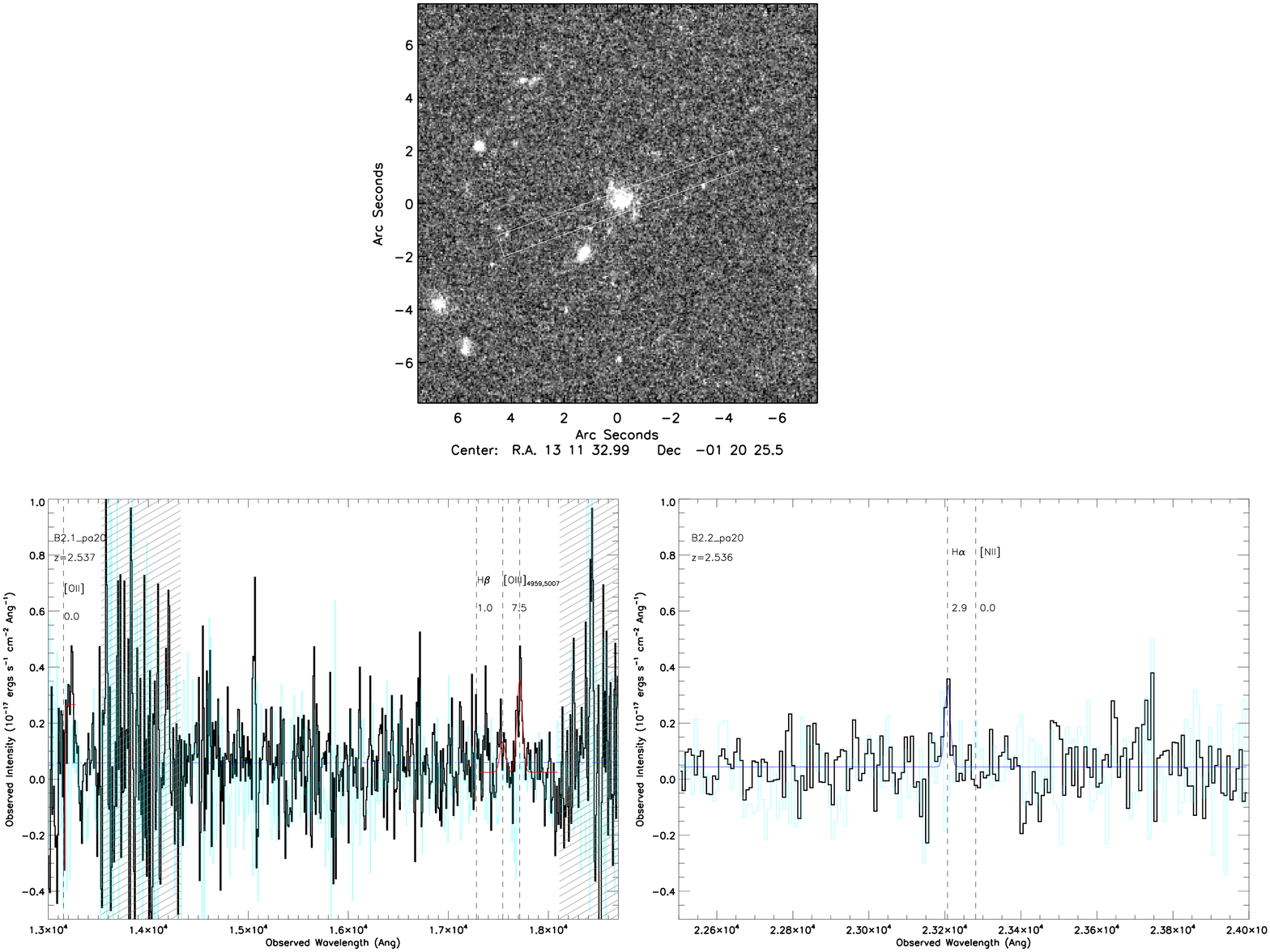}
\caption{z=2.537, B2.2, MOIRCS J, H band spectra. Detail descriptions are given in the Appendix text. 
}
 \label{fig:B2.2}
\end{center}
\end{figure*}

\begin{figure*}[!ht]
\begin{center}
\includegraphics[scale=0.6]{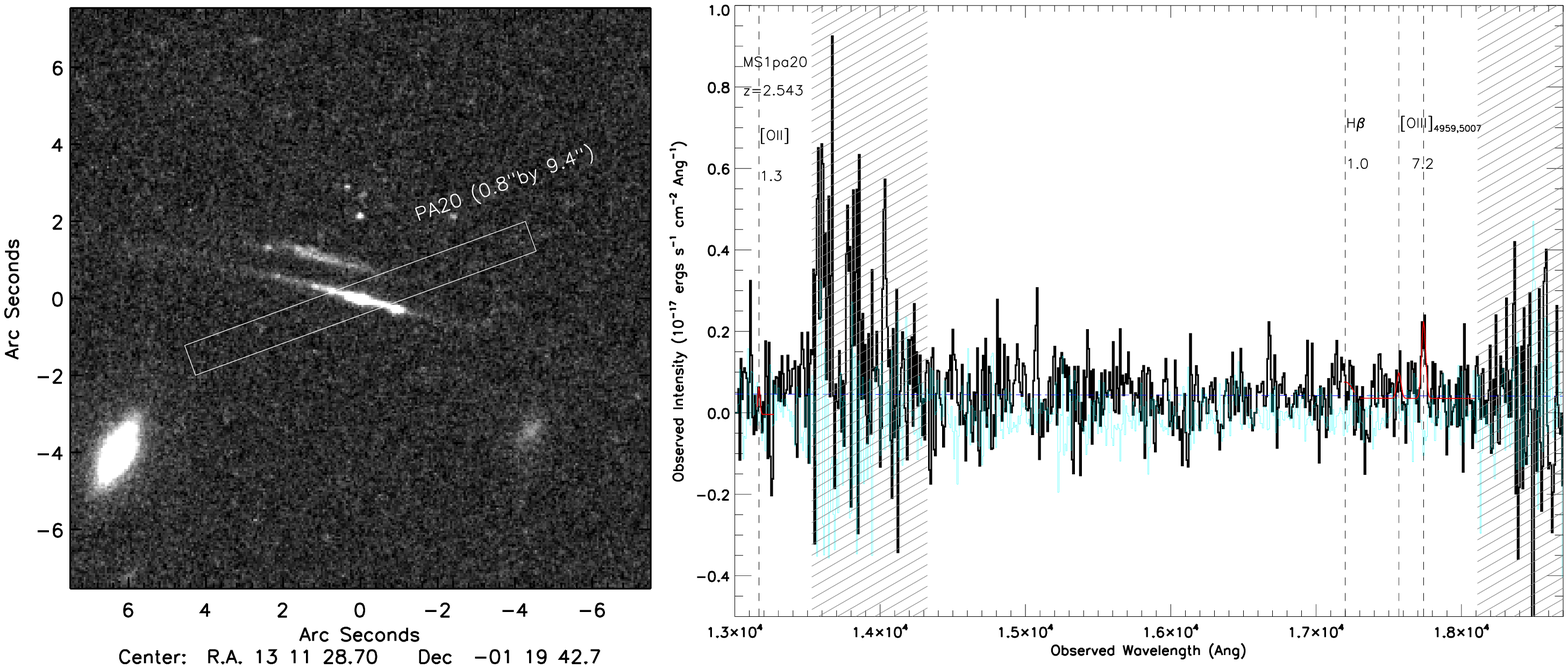}
\caption{z=2.54, MS1, MOIRCS J, H band spectra. Detail descriptions are given in the Appendix text. 
}
 \label{fig:MS1}
\end{center}
\end{figure*}

\begin{figure*}[!ht]
\begin{center}
\includegraphics[scale=0.6]{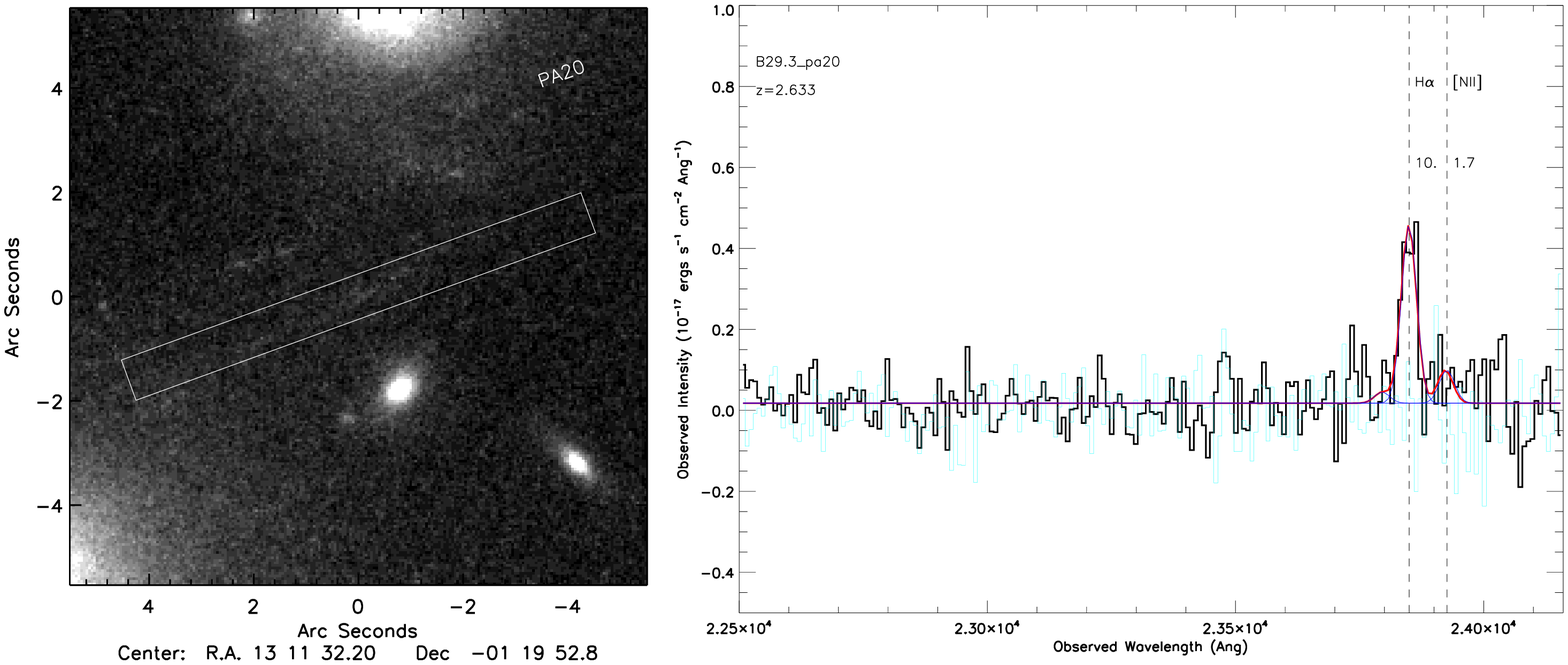}
\caption{z=2.633, B29.3, MOIRCS J, H band spectra. Detail descriptions are given in the Appendix text. 
}
 \label{fig:B29}
\end{center}
\end{figure*}

\begin{figure*}[!ht]
\begin{center}
\includegraphics[scale=0.6]{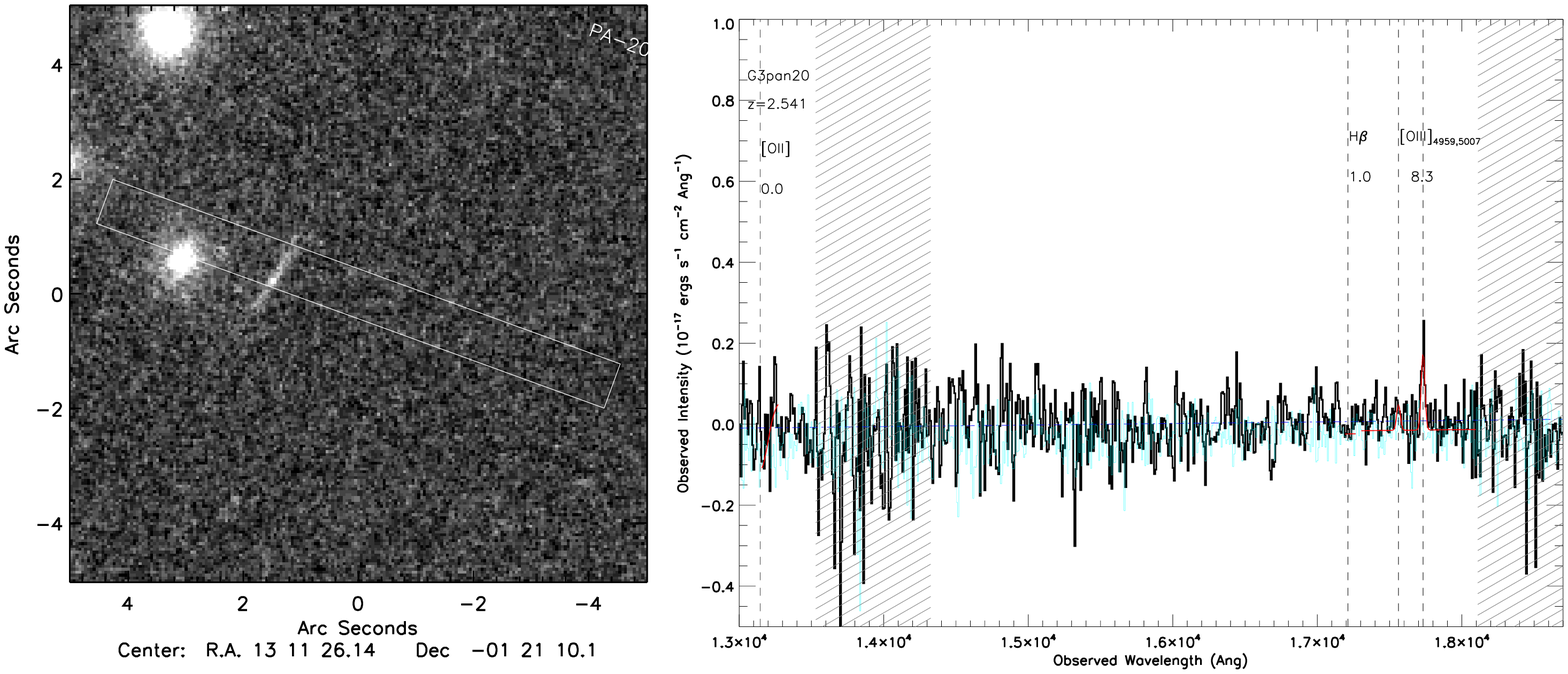}
\caption{z=2.540, G3, MOIRCS J, H band spectra. Detail descriptions are given in the Appendix text. 
}
 \label{fig:G3}
\end{center}
\end{figure*}

\begin{figure*}[!ht]
\begin{center}
\includegraphics[scale=0.5]{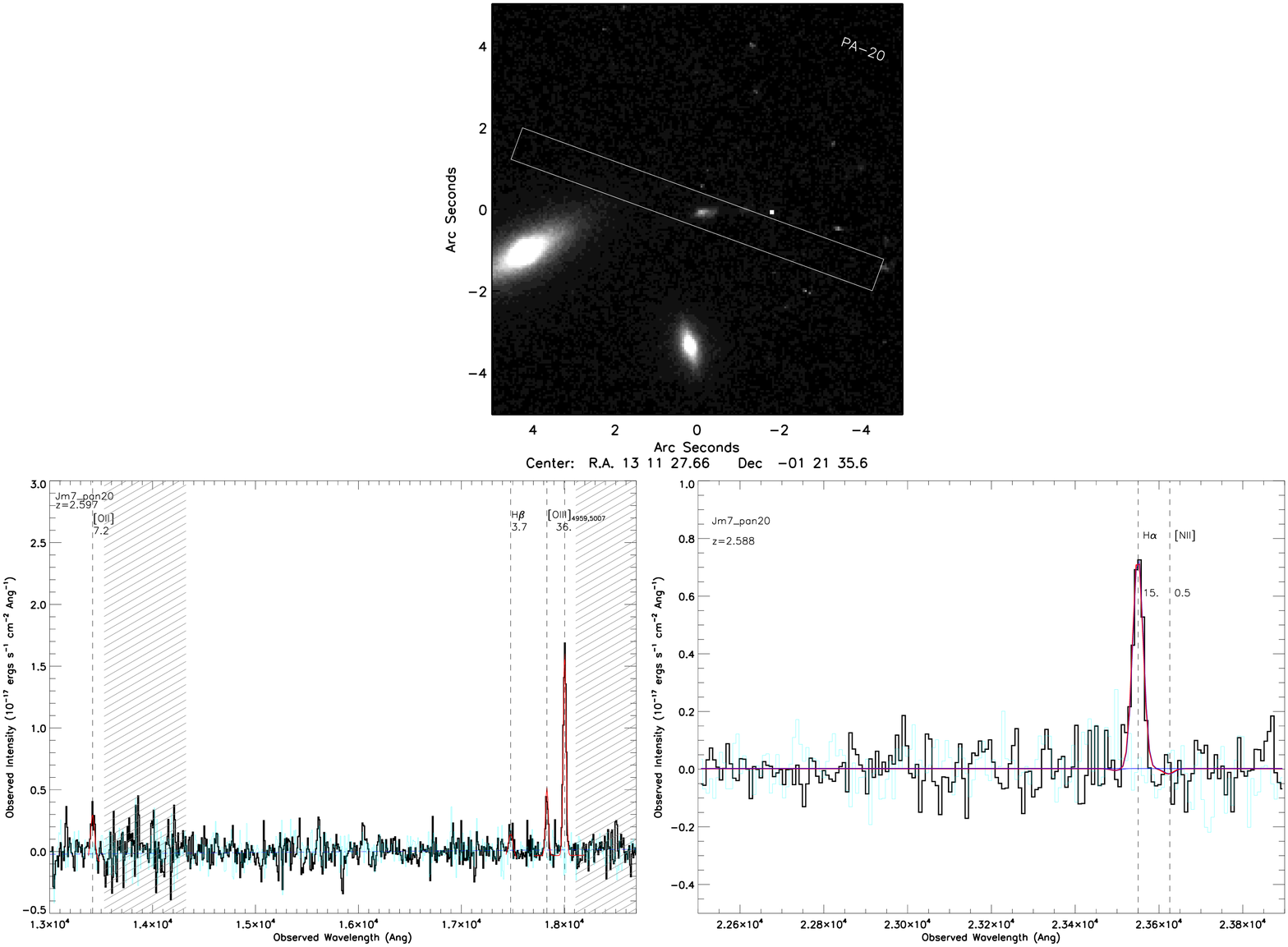}
\caption{z=2.588, Jm7, MOIRCS J, H band spectra. Detail descriptions are given in the Appendix text. 
}
 \label{fig:Jm7}
\end{center}
\end{figure*}

\begin{figure*}[!ht]
\begin{center}
\includegraphics[scale=0.56]{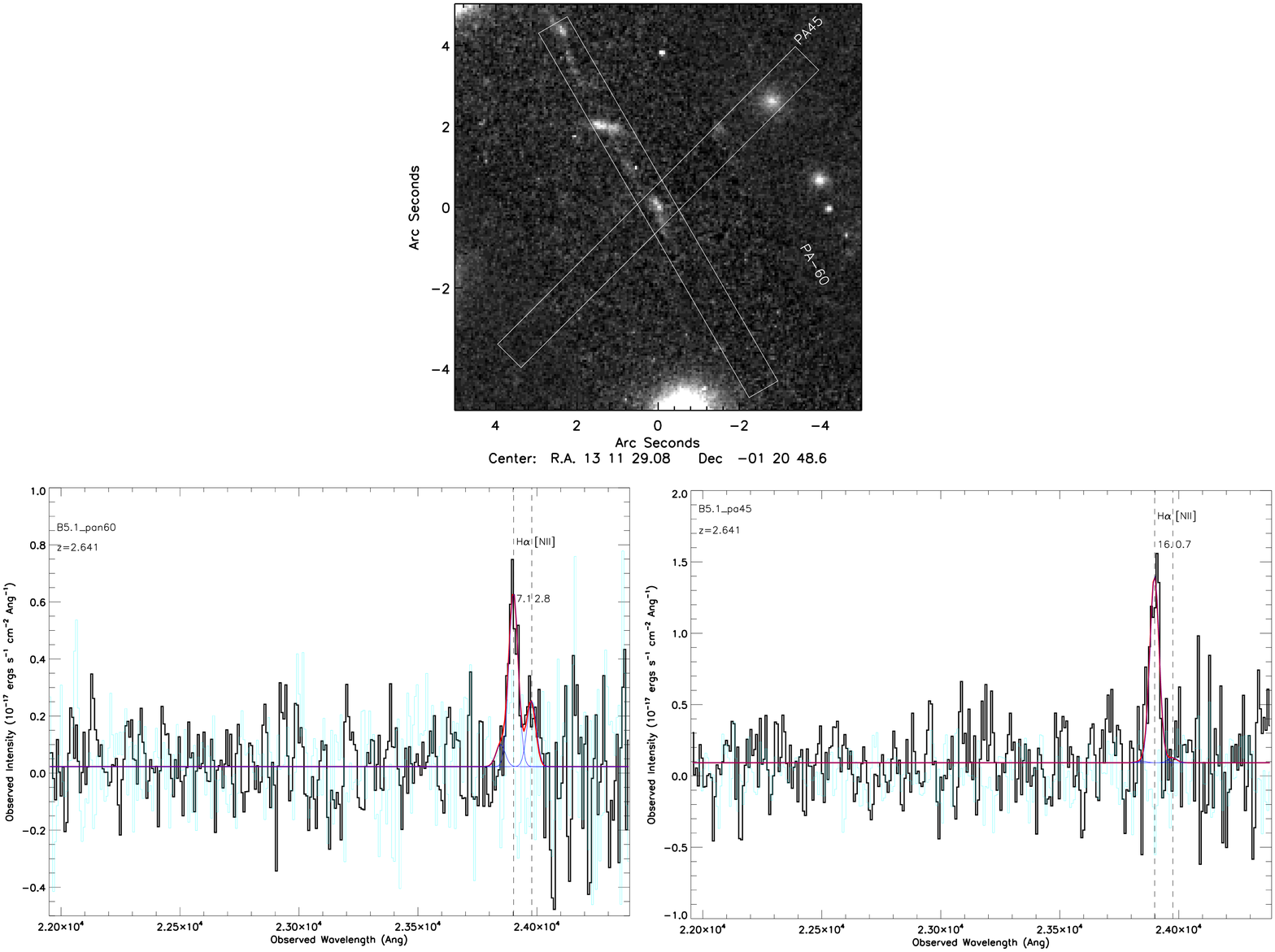}
\caption{z=2.641, B5.1, MOIRCS J, H band spectra. Note that the reason that the flux of 
 B5.1 in slit position PAn60 is less than PA45 (B5.1+B5.2) is that the dithering length of PAn60 was smaller than the separation of 5.1 and 5.2, thus part of the flux 
of PA45 (B5.1+B5.2) was cancelled out during the dithering process.
Detail descriptions are given in the Appendix text. 
}
 \label{fig:B5.1}
 \end{center}
\end{figure*}

\begin{figure*}[!ht]
\begin{center}
\includegraphics[scale=0.6]{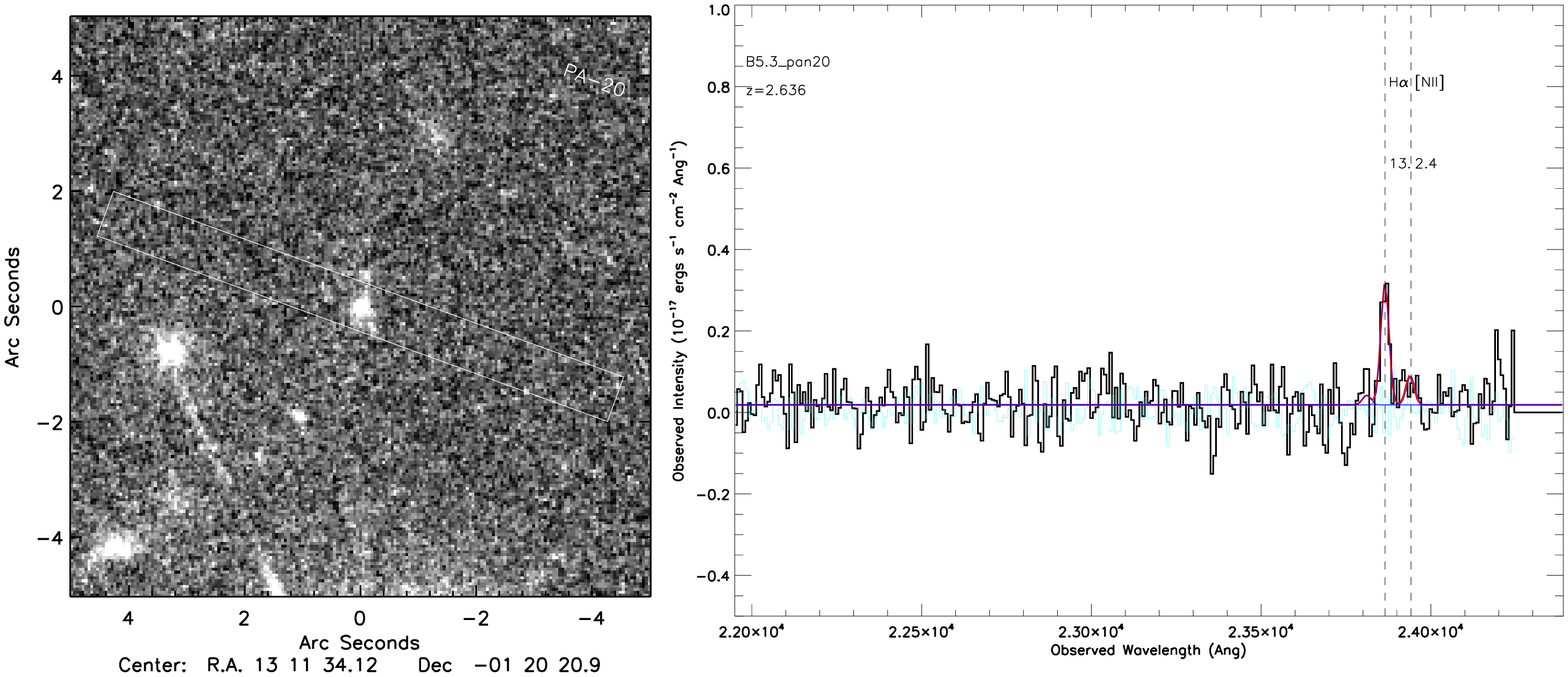}
\caption{z=2.636, B5.3, MOIRCS J, H band spectra. Detail descriptions are given in the Appendix text. 
}
 \label{fig:B5.3}
\end{center}
\end{figure*}

\begin{figure*}[!ht]
\begin{center}
\includegraphics[scale=0.56]{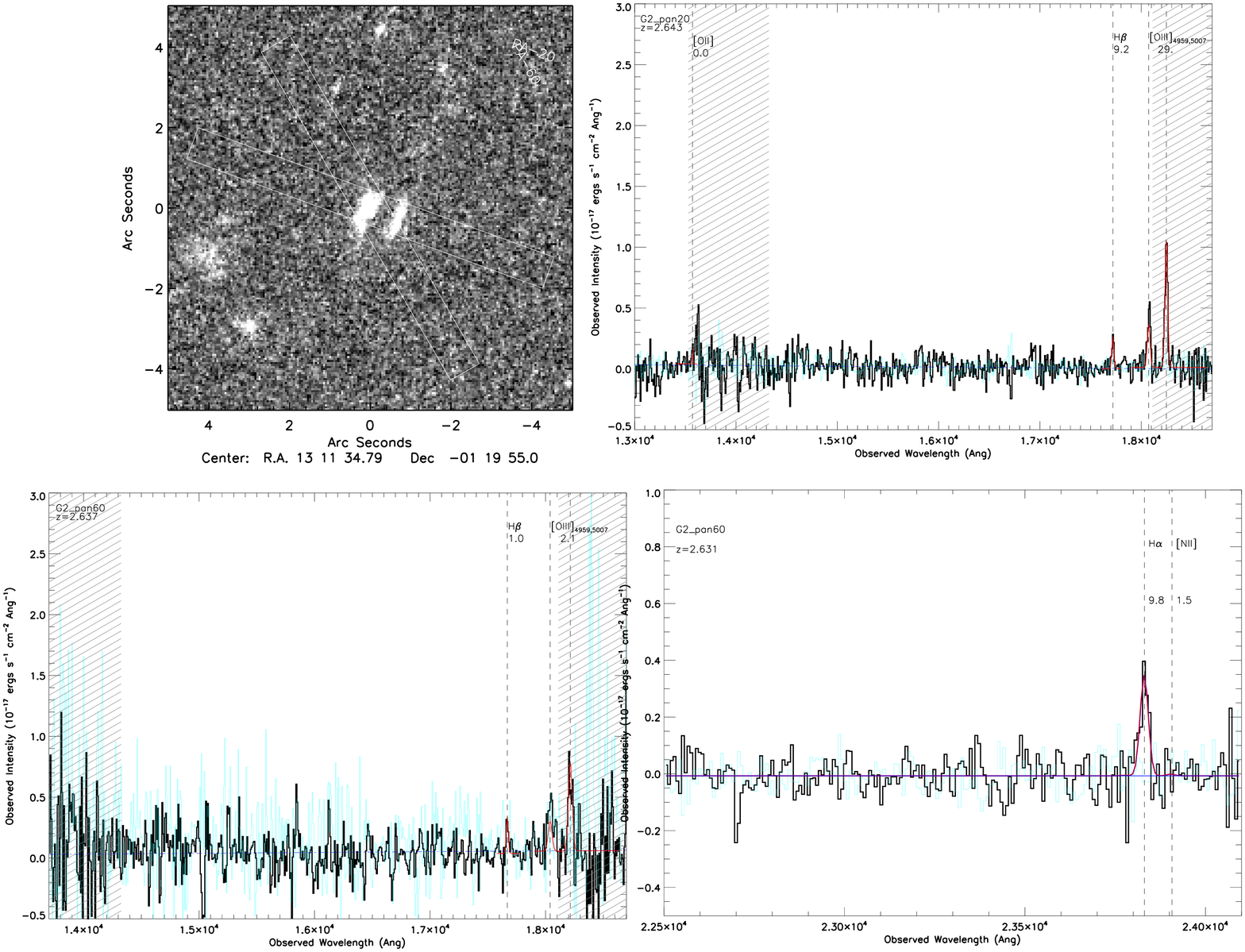}
\caption{z=2.643, G2, MOIRCS J, H band spectra. Detail descriptions are given in the Appendix text. 
}
 \label{fig:G2}
\end{center}
\end{figure*}

\begin{figure*}[!ht]
\begin{center}
\includegraphics[scale=0.5]{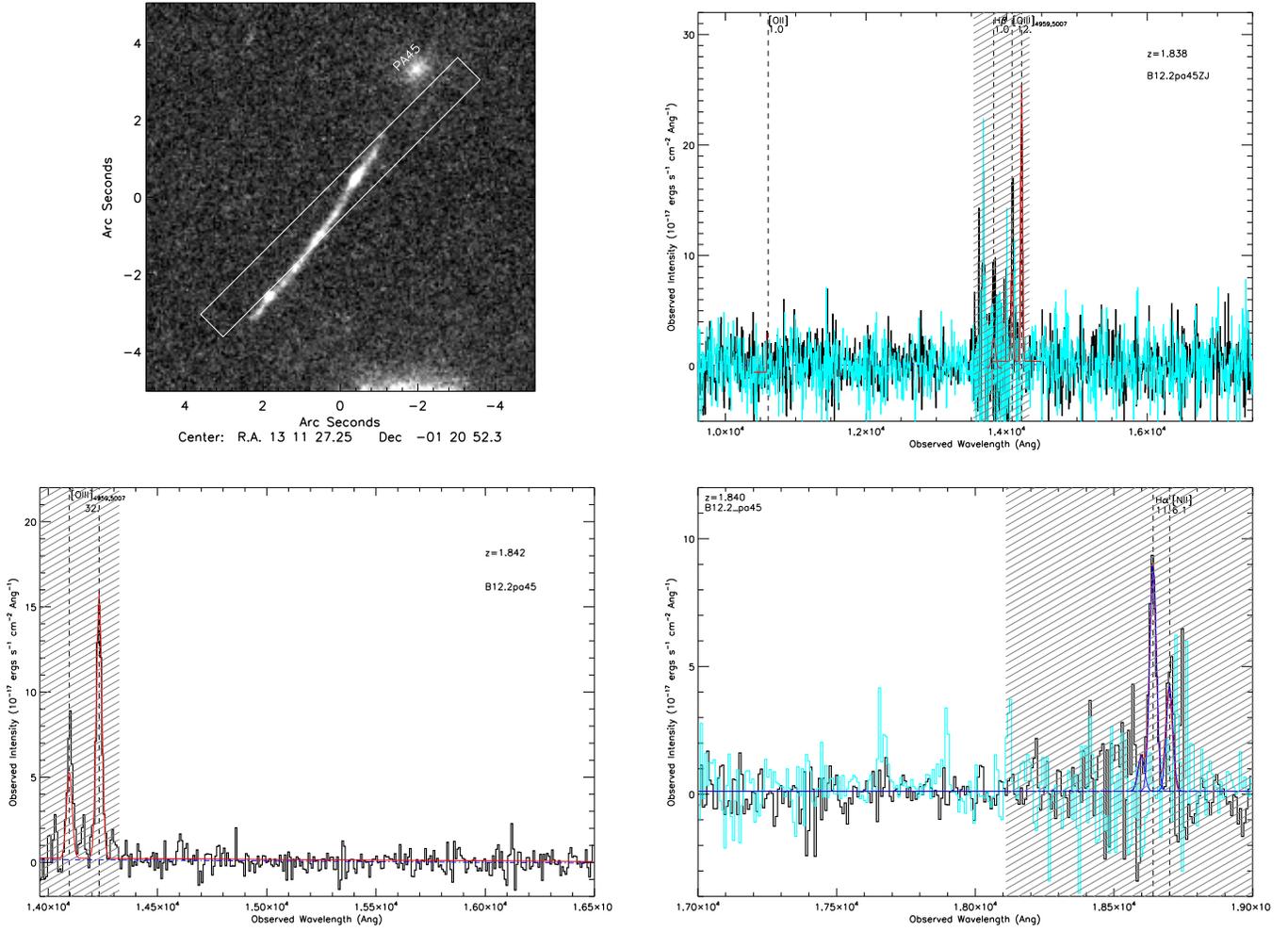}
\caption{z=1.834$\pm$0.002, B12.2, MOIRCS J, H band spectra. Detail descriptions are given in the Appendix text. 
}
 \label{fig:B12.2}
\end{center}
\end{figure*}

\begin{figure*}[!ht]
\begin{center}
\includegraphics[scale=0.6]{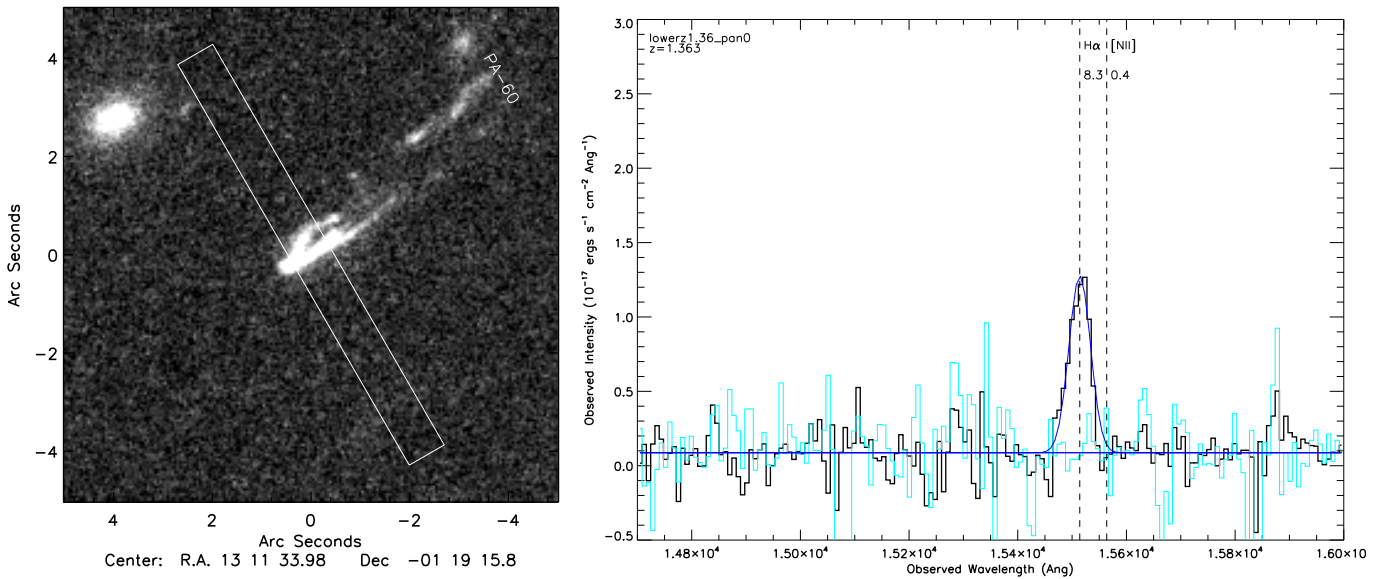}
\caption{z=1.363, Low-z, MOIRCS J, H band spectra. Detail descriptions are given in the Appendix text. 
}
 \label{fig:lowz1.36}
\end{center}
\end{figure*}

\begin{figure*}[!ht]
\begin{center}
\includegraphics[scale=0.6]{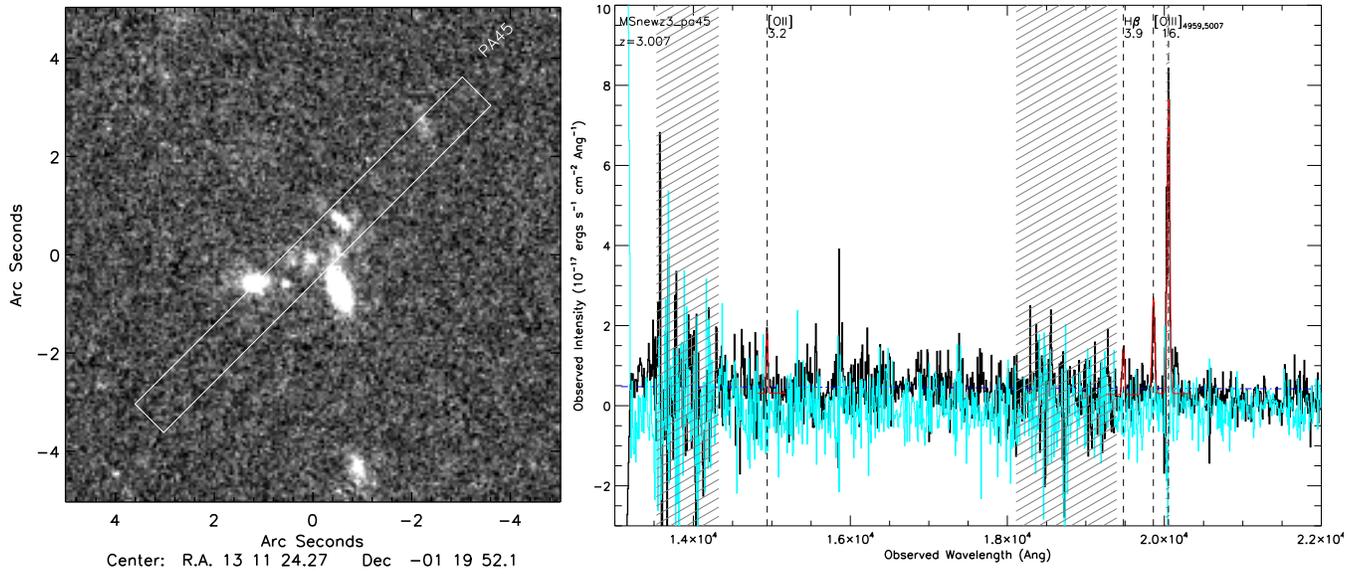}
\caption{z=3.003, new target, MOIRCS J, H band spectra. Detail descriptions are given in the Appendix text. 
}
 \label{fig:newz3}
\end{center}
\end{figure*}

\begin{figure*}[!ht]
\begin{center}
\includegraphics[scale=0.5]{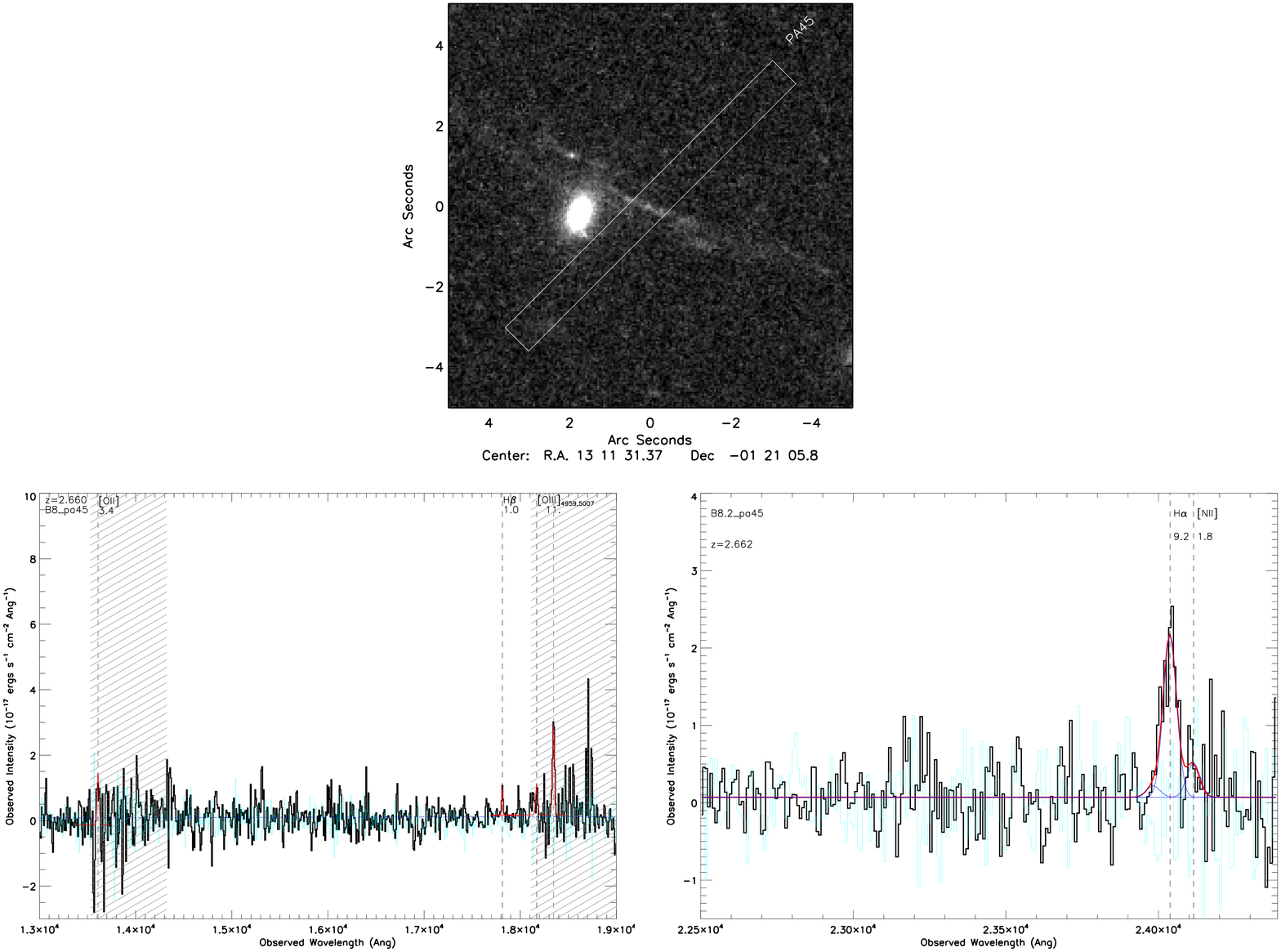}
\caption{z=2.663, B8.2, MOIRCS J, H band spectra. Detail descriptions are given in the Appendix text. 
}
 \label{fig:B8.2}
\end{center}
\end{figure*}

\begin{figure*}[!ht]
\begin{center}
\includegraphics[scale=0.5]{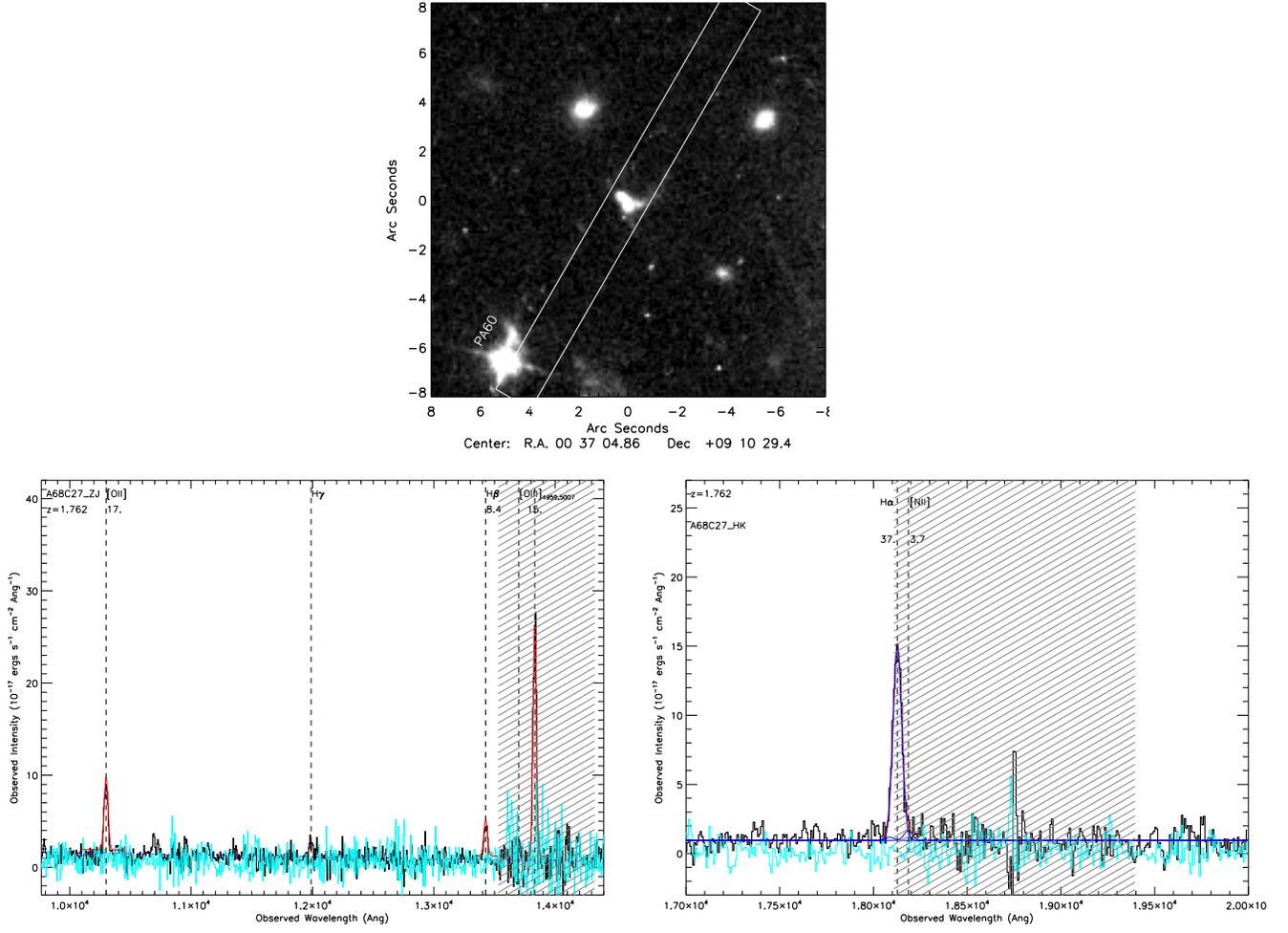}
\caption{z=1.763, A68C27, MOIRCS J, H band spectra. Detail descriptions are given in the Appendix text. 
}
 \label{fig:A68}
\end{center}
\end{figure*}

Brief remarks on individual objects (see also Table 2 and 3 for more information):
\begin{itemize}
\item Figure~\ref{fig:B11.1} and~\ref{fig:B11.2}, 
B11 (888\_351) : this is a resolved galaxy with spiral-like structure at $z=2.540\pm0.006$. As reported in \citet{Broadhurst05}, It is likely to be the most distant known spiral galaxy so far. 
B11 has 3 multiple images. We have observed B11.1, and B11.2, with two slit orientations on each image respectively.  Different slit orientation yields very different line ratios, implying possible gradients. Our IFU follow-up observations are in progress to reveal the details of this 2.6-Gyr-old spiral.

\item Figure~\ref{fig:B2.1} and~\ref{fig:B2.2}, B2 (860\_331): this is one of the interesting systems reported in \citet{Frye07}.   It has 5 multiple images, and is 
only 2~$\!\!^{\prime\prime}$ away from another five-image lensed system, ``The Sextet Arcs" at z$=$3.038. We have observed B2.1 and B2.2 and detected strong \ha\ and \oiii\ lines in both of them, yielding a redshift of $2.537\pm0.006$, consistent with the redshift $z=2.534$ measured from the absorption lines (\cii\lam1334,  \si\lam1527) in \citet{Frye07}. 
 
\item Figure~\ref{fig:MS1}, MS1 (869\_328): 
We have detected a 7-$\sigma$ \oiii\ line and determined its redshift to be $z=2.534\pm0.010$.  

\item Figure~\ref{fig:B29}, B29 (884\_331): this is a lensed system with 5 multiple images. 
We observed B29.3, the brightest of the five images.  The overall surface brightness of the B29.3 arc is very low, 
We have observed a 10-$\sigma$ \ha\ and an upper limit for \nii, placing it  at $z=2.633\pm0.010$.

\item Figure~\ref{fig:G3}, G3:  this lensed arc with a bright knot has no recorded redshift before this study. It was put on one of the extra slits during mask designing.
We have detected a 8-$\sigma$ \oiii\ line and determined its redshift to be $z=2.540\pm0.010$.  

\item Figure~\ref{fig:Jm7}, Ms-Jm7 (865\_359):   We detected \oii\, \hb\, \oiii\, \ha\, and an upper limit for \nii\, placing it at redshift
$z=2.588\pm0.006$.

\item Figure~\ref{fig:B5.1} and~\ref{fig:B5.3}, B5 (892\_339, 870\_346): it has three multiple images, of which we observed B5.1 and B5.3. Two slit orientations were observed for B5.1,  the final spectrum
for B5.1 has combined the two slit orientations weighted by the S/N of \ha\.  Strong \ha\ and upper limit of \nii\ were obtained in both images, yielding a redshift of $z=2.636\pm0.004$.
 
\item  Figure~\ref{fig:G2}, G2 (894\_332): two slit orientations were available for G2, with detections of \hb, \oiii, \ha, and upper limits for \oii\ and \nii. The redshift measured is $z=1.643\pm0.010$.

\item Figure~\ref{fig:B12.2}, B12: this blue giant arc has 5 multiple images, and we observed B12.2.  It shows a series of strong emission lines, with an average redshift of $z=1.834\pm0.006$.

\item Figure~\ref{fig:lowz1.36}, Lensz1.36 (891\_321): it has a very strong \ha\ and \nii\ is at noise level, from \ha\, we derive $z=1.363\pm0.010$. 

\item Figure~\ref{fig:newz3}, MSnewz3: this is a new target observed in Abell 1689, we detect \oii, \hb, and \oiii\ at a significant level, yielding $z=3.007\pm0.003$. 

\item  Figure~\ref{fig:B8.2}, B8: this arc has five multiple images in total, and we observed B8.2, detection of \oii, \oiii, \ha, with \hb\ and \nii\ as upper limit yields 
an average redshift of $z=2.662\pm0.006$. 

\item B22.3: a three-image lensed system at $z=1.703\pm0.004$, this is the first object reported from our LEGMS program,  see \citet{Yuan09}.

\item  Figure~\ref{fig:A68}, A68-C27: this is the only object chosen from our unfinished observations on Abell 68. This target has many strong emission lines. $z=1.762\pm0.006$. The 
morphology of C27 shows signs of merger.  IFU observation on this target is in process.

\end{itemize}

\clearpage
\clearpage

\clearpage


\clearpage
 \end{appendix}
\clearpage

\clearpage
\begin{sidewaystable}
\LongTables 
\begin{deluxetable}{lcccccccccc}
\tabletypesize{\scriptsize}
\tablewidth{0pt}
\tablecolumns{11}
\tablecaption{Measured Emission Line fluxes \label{tabmetal}}
\tablehead{
\colhead{Id} & 
\colhead{\oii\lam3727} & 
\colhead{\hb} & 
\colhead{\oiii\lam5007} &  
\colhead{\ha} & 
\colhead{\nii\lam6584} & 
\colhead{KK04($\to$PP04N2)} & 
\colhead{Branch} & 
\colhead{PP04N2} & 
\colhead{E(B-V)\tablenotemark{a}} &
\colhead{Final Adopted\tablenotemark{b}}
}
\startdata
B11.1:pa20 &21.43$\pm$2.60&5.42$\pm$1.63&21.12$\pm$2.43& 33.54$\pm$2.38&$<$4.59 &8.38(8.16)$\pm$0.14& up&$<$8.41&0.73$\pm$0.29&8.48$\pm$0.18\tablenotemark{e}\\
B11.1:pa-20 &22.02$\pm$4.88&9.40$\pm$2.75 &14.78$\pm$2.21& 28.13$\pm$2.56&8.90$\pm$0.89 &8.74(8.54)$\pm$0.14& up&8.61$\pm$0.05&0.05$\pm$0.29&\\
B11.2:pa-60 &$<$60.5&$<$64.6&$<$60.3& 53.6$\pm$4.09&$<$17.12&\nodata& \nodata&$<$8.62&\nodata&\\
B11.2:pa45 &73.65$\pm$12.01&$<$34.23&61.06$\pm$7.4&80.46$\pm$9.9&$<$13.08&$<$8.74(8.54)& up&$<$8.73&0.&\\
\noalign{\smallskip}
\hline
\noalign{\smallskip}
B2.1:pa20 &$<$2.82&$<$6.3&9.47$\pm$0.56&7.66$\pm$0.69&$<$0.67& \nodata& \nodata&$<$8.30&\nodata&$<$8.30\\
B2.2:pa20 &$<$7.73&$<$20.6&23.2$\pm$3.0&$<$5.45&$<$6.4&\nodata& \nodata& \nodata&\nodata& \nodata\\
\noalign{\smallskip}
\hline
\noalign{\smallskip}
MS1:pa20 &$<3.08$&$<$4.5&6.7$\pm$0.9&\nodata&\nodata&\nodata& \nodata&\nodata&\nodata&\nodata\\
\noalign{\smallskip}
\hline
\noalign{\smallskip}
B29.3:pa20 &\nodata&\nodata&\nodata&17.05$\pm$1.6&$<$3.1&\nodata& \nodata&$<$8.48&\nodata&$<$8.48\\
\noalign{\smallskip}
\hline
\noalign{\smallskip}
G3:pan20 &\nodata&$<$4.0&6.0$\pm$0.7&\nodata&\nodata&\nodata& \nodata&\nodata&\nodata&\nodata\\
\noalign{\smallskip}
\hline
\noalign{\smallskip}
MS-Jm7:pan20 &19.16$\pm$2.74&12.5$\pm$3.6&58.12$\pm$1.64&23.03$\pm$1.6&$<$9.22&8.69(8.33)$\pm$0.12& up&$<$8.67&0.&8.25$\pm$0.18\\
            &&&&&&8.23(8.19)$\pm$0.12& low&&&\\
\noalign{\smallskip}
\hline
\noalign{\smallskip}
B5.3:pan20 &\nodata&\nodata&\nodata&9.07$\pm$0.7&$<$2.94&\nodata& \nodata&$<$8.62&\nodata&$<$8.62\\
B5.1:pan20 &\nodata&\nodata&\nodata&30.38$\pm$4.6&$<$13.34&\nodata& \nodata&$<$8.70&\nodata&\\
B5.1:pa45 &\nodata&\nodata&\nodata&64.39$\pm$3.9&$<$59.5&\nodata& \nodata&$<$8.88&\nodata&\\
\noalign{\smallskip}
\hline
\noalign{\smallskip}
G2:pan20 &$<$4.7&6.84$\pm$0.74&36.49$\pm$1.25&\nodata&\nodata&$<$8.62(8.41)& up&\nodata&\nodata&$<$8.41\\
G2:pan60 &$<$25.8&$<$8.8&$<$98.7&10.09$\pm$1.0&$<$3.1&\nodata&\nodata&$<$8.60&\nodata&\\
\noalign{\smallskip}
\hline
\noalign{\smallskip}
Lowz1.36:pan60 &\nodata&\nodata&\nodata&59.19$\pm$7.1&$<$8.82&\nodata& \nodata&$<$8.43&\nodata&$<$8.43\\
\noalign{\smallskip}
\hline
\noalign{\smallskip}
MSnewz3:pa45 &36.94$\pm$11.5&44.02$\pm$11.06&300.3$\pm$17.8&\nodata&\nodata&8.5(8.29)$\pm$0.11& up&\nodata&\nodata&8.23$\pm$0.18\\
            &                                                    &                                &                              &             &              &8.12(8.16)$\pm$0.11& low&\nodata&\nodata&\\
\noalign{\smallskip}
\hline
\noalign{\smallskip}
B12.2:pa45 &$<$71.58&$<67.79$&141.01$\pm$10.07 &90.45$\pm$6.95 &$<10.6$ &\nodata& \nodata &$<$8.369\tablenotemark{c}&\nodata&$<$8.369\\
\noalign{\smallskip}
\hline
\noalign{\smallskip}
B8.2:pa45                &40.2$\pm$11.8&$<$17.7&75.7$\pm$6.6 &115.26$\pm$12.5 &$<$72.13 &$<$8.51(8.29)& up &$<$8.78\tablenotemark{d}&$>$1.2&$<$8.29\\
            &                                                    &                                &                              &             &              & $<$8.11(8.16)& low&\nodata&\nodata&\\
\noalign{\smallskip}
\hline
B22.3:pa60          &162.1$\pm$20.3  &146.0$\pm$29.2&942.3$\pm$62.8 &734.4$\pm$56.5&$<$3.65 &8.13(8.17)$\pm$0.12& low &$<$8.22&0.54$\pm$0.22&8.10$\pm$0.18\\
\hline
\hline
\noalign{\smallskip}
A68-C27:pa60                &317.01$\pm$17.9&149.2$\pm$17.5&884.4$\pm$58.9 &814.6$\pm$21.8 &40.4$\pm$10.92 &8.26(8.25)$\pm$0.06& low &8.16$\pm$0.07&0.62$\pm$0.11&8.16$\pm$0.18\\
\enddata
\tablecomments{ Observed emission line fluxes for the lensed background galaxies  in A1689.
Fluxes are in units of 10$^{-17}$ ergs~s$^{-1}$~cm$^{-2}$, without lensing magnification correction. Some lines are not detected because of the severe telluric absorption.}
\tablenotetext{a}{E(B-V) calculated from Balmer decrement, if possible.}
\tablenotetext{b}{Final  adopted metallicity, converted to PP04N2 base and extinction corrected using E(B-V) values from Balmer decrement if available, otherwise E(B-V) returned from
SED fitting are assumed.}
\tablenotetext{c}{Based on NIRSPEC spectrum at KECK II (Kewley et al. 2013, in prep)}
\tablenotetext{d}{Possible AGN contamination.}
\tablenotetext{e}{This galaxy shows significant \nii\,/\ha\, ratios in slit position pa-20. The final metallicity is based on the average spectrum over all slit positions. }
\end{deluxetable}
\end{sidewaystable} 

\clearpage
\begin{deluxetable}{lllcccc}
\tabletypesize{\scriptsize}
\tablewidth{0pt}
\tablecolumns{7}
\tablecaption{Physical Properties of the Lensed Sample \label{tabphy}}
\tablehead{
\colhead{ID1} & 
\colhead{ID2\tablenotemark{a}} & 
\colhead{RA, DEC (J2000)} & 
\colhead{Redshift} & 
\colhead{lg(SFR)\tablenotemark{b}} & 
\colhead{Lensing Magnification} & 
\colhead{log(M$_{\ast}$/M$_{\odot}$)} \\
\colhead{ } &
\colhead{ } &
\colhead{ }&
\colhead{ }&
\colhead{(M$_{\odot}$~yr$^{-1}$)}&
\colhead{(flux)} &
\colhead{}
}
\startdata
B11.1:pa20 &888\_351&13:11:33.336, -01:21:06.94 &2.540$\pm$0.006& 1.08$\pm$0.1&11.8$\pm$2.7&9.1$^{+0.2}_{-0.3}$\\
B11.2:pa45 &\nodata&13:11:29.053, -01:20:01.26 &2.540$\pm$0.006& 1.42$\pm$0.11&13.1$\pm$1.8&\\
\noalign{\smallskip}
\hline
\noalign{\smallskip}
B2.1:pa20 &860\_331  &13:11:26.521, -01:19:55.24   &2.537$\pm$0.006 &   0.20$\pm$0.03      &20.6$\pm$1.8&8.2$^{+0.3}_{-0.3}$\\
B2.2:pa20   &\nodata&13:11:32.961, -01:20:25.31   &2.537$\pm$0.006 &   \nodata      &15.0$\pm$2.0&\\
\noalign{\smallskip}
\hline
\noalign{\smallskip}
MS1:pa20  &869\_328 &13:11:28.684,-01:19:42.62  &2.534$\pm$0.01 &  \nodata    &58.3$\pm$2.8&8.5$^{+0.1}_{-0.1}$\\
\noalign{\smallskip}
\hline
\noalign{\smallskip}
B29.3:pa20 &884\_331  &13:11:32.164,-01:19:52.53  &2.633$\pm$0.01 &  0.43$\pm$0.06   &22.5$\pm$6.9&9.0$^{+0.4}_{-0.5}$\tablenotemark{f}\\
\noalign{\smallskip}
\hline
\noalign{\smallskip}
G3:pan20    &\nodata      &13:11:26.219,-01:21:09.64 &2.540$\pm$0.01  & \nodata & 7.7$\pm$0.1 & \nodata\\
\noalign{\smallskip}
\hline
\noalign{\smallskip}
MS-Jm7:pan20&865\_359       &13:11:27.600,-01:21:35.00  &2.588$\pm$0.006  & \nodata & 18.5$\pm$3.2 &8.0$^{+0.5}_{-0.4}$\tablenotemark{f}\\
\noalign{\smallskip}
\hline
\noalign{\smallskip}
B5.3:pan20 &892\_339&13:11:34.109,-01:20:20.90  &2.636$\pm$0.004  & 0.47$\pm$0.05 & 14.2$\pm$1.3 & 9.1$^{+0.4}_{-0.2}$\\
B5.1:pan60 &870\_346&13:11:29.064,-01:20:48.33  &2.641$\pm$0.004  & 1.0$\pm$0.05                   &  14.3$\pm$0.3 & \\
\noalign{\smallskip}
\hline
\noalign{\smallskip}
G2 &894\_332&13:11:34.730,-01:19:55.53  &1.643$\pm$0.01  & 0.45$\pm$0.09 & 16.7$\pm$3.1 & 8.0$^{+0.3}_{-0.4}$\\
\noalign{\smallskip}
\hline
\noalign{\smallskip}
Lowz1.36& 891\_321& 13:11:33.957,-01:19:15.90 &1.363$\pm$0.01  & 0.67$\pm$0.11 & 11.6$\pm$2.7 & 8.9$^{+0.3}_{-0.3}$\\
\noalign{\smallskip}
\hline
\noalign{\smallskip}
MSnewz3:pa45& \nodata&13:11:24.276,-01:19:52.08 &3.007$\pm$0.003  &0.65$\pm$0.55 & 2.9$\pm$1.7 & 8.6$^{+0.3}_{-0.4}$\tablenotemark{f}\\
\noalign{\smallskip}
\hline
\noalign{\smallskip}
B12.2:pa45 &863\_348& 13:11:27.212,-01:20:51.89  &1.834$\pm$0.006  &1.00$\pm$0.05\tablenotemark{c} & 56.0$\pm$4.4& 7.4$^{+0.2}_{-0.0}$\\
\noalign{\smallskip}
\hline
\noalign{\smallskip}
B8.2:pa45 &\nodata    & 13:11:27.212,-01:20:51.89  &2.662$\pm$0.006 &1.36$\pm$0.07\tablenotemark{d}  & 23.7$\pm$3.0&  8.2$^{+0.5}_{-0.6}$\tablenotemark{f}   \\
\noalign{\smallskip}
\hline
B22.3:pa60\tablenotemark{e}   &\nodata    & 13:11:32.4150,-01:21:15.917  &1.703$\pm$0.006 &1.88$\pm$0.04  & 15.5$\pm$0.3&  8.5$^{+0.2}_{-0.2}$   \\
\noalign{\smallskip}
\hline
\hline
\noalign{\smallskip}
A68-C27:pa60       &\nodata& 00:37:04.866,+09:10:29.26 &1.762$\pm$0.006 &2.46$\pm$0.1 & 4.9$\pm$1.1&  9.6$^{+0.1}_{-0.1}$   \\
\enddata
\tablecomments{ The redshift errors in Table~\ref{tabphy} is determined  from RMS of different emission line centroids. If the RMS is 
smaller than 0.006 (for most targets) or if there is only one line fitted, we adopt the systematic error of 0.006 as a conservative estimation for absolute redshift measurements.
}
\tablenotetext{a}{ID used in Richard et al. (2012, in prep). The name tags of the objects are chosen to be consistent with the \citet{Broadhurst05} conventions if overlapping.}
\tablenotetext{b}{Corrected for lensing magnification, but without dust extinction correction. We note that the systematic errors of  SFR in this work are extremely uncertain due to 
complicated aperture correction and flux calibration in the multi-slit of MOIRCS.}
\tablenotetext{c}{Based on NIRSPEC observation}
\tablenotetext{d}{Possible AGN contamination.}
\tablenotetext{e}{See also \citet{Yuan09}}
\tablenotetext{f}{The IRAC photometry for these sources are not included in the stellar mass calculation due to the difficulty in resolving the lensed image from the adjacent foreground galaxies.}

\end{deluxetable}
\clearpage


\begin{thebibliography}{113}
\expandafter\ifx\csname natexlab\endcsname\relax\def\natexlab#1{#1}\fi

\bibitem[{{Abazajian} {et~al.}(2009){Abazajian}, {Adelman-McCarthy},
  {Ag{\"u}eros}, {Allam}, {Allende Prieto}, {An}, {Anderson}, {Anderson},
  {Annis}, {Bahcall}, {Bailer-Jones}, {Barentine}, {Bassett}, {Becker},
  {Beers}, {Bell}, {Belokurov}, {Berlind}, {Berman}, {Bernardi}, {Bickerton},
  {Bizyaev}, {Blakeslee}, {Blanton}, {Bochanski}, {Boroski}, {Brewington},
  {Brinchmann}, {Brinkmann}, {Brunner}, {Budav{\'a}ri}, {Carey}, {Carliles},
  {Carr}, {Castander}, {Cinabro}, {Connolly}, {Csabai}, {Cunha}, {Czarapata},
  {Davenport}, {de Haas}, {Dilday}, {Doi}, {Eisenstein}, {Evans}, {Evans},
  {Fan}, {Friedman}, {Frieman}, {Fukugita}, {G{\"a}nsicke}, {Gates},
  {Gillespie}, {Gilmore}, {Gonzalez}, {Gonzalez}, {Grebel}, {Gunn},
  {Gy{\"o}ry}, {Hall}, {Harding}, {Harris}, {Harvanek}, {Hawley}, {Hayes},
  {Heckman}, {Hendry}, {Hennessy}, {Hindsley}, {Hoblitt}, {Hogan}, {Hogg},
  {Holtzman}, {Hyde}, {Ichikawa}, {Ichikawa}, {Im}, {Ivezi{\'c}}, {Jester},
  {Jiang}, {Johnson}, {Jorgensen}, {Juri{\'c}}, {Kent}, {Kessler}, {Kleinman},
  {Knapp}, {Konishi}, {Kron}, {Krzesinski}, {Kuropatkin}, {Lampeitl},
  {Lebedeva}, {Lee}, {Lee}, {French Leger}, {L{\'e}pine}, {Li}, {Lima}, {Lin},
  {Long}, {Loomis}, {Loveday}, {Lupton}, {Magnier}, {Malanushenko},
  {Malanushenko}, {Mandelbaum}, {Margon}, {Marriner},
  {Mart{\'{\i}}nez-Delgado}, {Matsubara}, {McGehee}, {McKay}, {Meiksin},
  {Morrison}, {Mullally}, {Munn}, {Murphy}, {Nash}, {Nebot}, {Neilsen},
  {Newberg}, {Newman}, {Nichol}, {Nicinski}, {Nieto-Santisteban}, {Nitta},
  {Okamura}, {Oravetz}, {Ostriker}, {Owen}, {Padmanabhan}, {Pan}, {Park},
  {Pauls}, {Peoples}, {Percival}, {Pier}, {Pope}, {Pourbaix}, {Price},
  {Purger}, {Quinn}, {Raddick}, {Re Fiorentin}, {Richards}, {Richmond},
  {Riess}, {Rix}, {Rockosi}, {Sako}, {Schlegel}, {Schneider}, {Scholz},
  {Schreiber}, {Schwope}, {Seljak}, {Sesar}, {Sheldon}, {Shimasaku}, {Sibley},
  {Simmons}, {Sivarani}, {Allyn Smith}, {Smith}, {Smol{\v c}i{\'c}}, {Snedden},
  {Stebbins}, {Steinmetz}, {Stoughton}, {Strauss}, {SubbaRao}, {Suto},
  {Szalay}, {Szapudi}, {Szkody}, {Tanaka}, {Tegmark}, {Teodoro}, {Thakar},
  {Tremonti}, {Tucker}, {Uomoto}, {Vanden Berk}, {Vandenberg}, {Vidrih},
  {Vogeley}, {Voges}, {Vogt}, {Wadadekar}, {Watters}, {Weinberg}, {West},
  {White}, {Wilhite}, {Wonders}, {Yanny}, {Yocum}, {York}, {Zehavi}, {Zibetti},
  \& {Zucker}}]{Abazajian09}
{Abazajian}, K.~N., {et~al.} 2009, \apjs, 182, 543

\bibitem[{{Asplund} {et~al.}(2009){Asplund}, {Grevesse}, {Sauval}, \&
  {Scott}}]{Asplund09}
{Asplund}, M., {Grevesse}, N., {Sauval}, A.~J., \& {Scott}, P. 2009, \araa, 47,
  481

\bibitem[{{Baldwin} {et~al.}(1981){Baldwin}, {Phillips}, \&
  {Terlevich}}]{Baldwin81}
{Baldwin}, J.~A., {Phillips}, M.~M., \& {Terlevich}, R. 1981, \pasp, 93, 5

\bibitem[{{Bertin} \& {Arnouts}(1996)}]{Bertin96}
{Bertin}, E., \& {Arnouts}, S. 1996, \aaps, 117, 393

\bibitem[{{Bertone} {et~al.}(2007){Bertone}, {De Lucia}, \&
  {Thomas}}]{Bertone07}
{Bertone}, S., {De Lucia}, G., \& {Thomas}, P.~A. 2007, \mnras, 379, 1143

\bibitem[{{Brinchmann} {et~al.}(2008){Brinchmann}, {Pettini}, \&
  {Charlot}}]{Brinchmann08}
{Brinchmann}, J., {Pettini}, M., \& {Charlot}, S. 2008, \mnras, 385, 769

\bibitem[{{Broadhurst} {et~al.}(2005){Broadhurst}, {Ben{\'{\i}}tez}, {Coe},
  {Sharon}, {Zekser}, {White}, {Ford}, {Bouwens}, {Blakeslee}, {Clampin},
  {Cross}, {Franx}, {Frye}, {Hartig}, {Illingworth}, {Infante}, {Menanteau},
  {Meurer}, {Postman}, {Ardila}, {Bartko}, {Brown}, {Burrows}, {Cheng},
  {Feldman}, {Golimowski}, {Goto}, {Gronwall}, {Herranz}, {Holden}, {Homeier},
  {Krist}, {Lesser}, {Martel}, {Miley}, {Rosati}, {Sirianni}, {Sparks},
  {Steindling}, {Tran}, {Tsvetanov}, \& {Zheng}}]{Broadhurst05}
{Broadhurst}, T., {et~al.} 2005, \apj, 621, 53

\bibitem[{{Brooks} {et~al.}(2007){Brooks}, {Governato}, {Booth}, {Willman},
  {Gardner}, {Wadsley}, {Stinson}, \& {Quinn}}]{Brooks07}
{Brooks}, A.~M., {Governato}, F., {Booth}, C.~M., {Willman}, B., {Gardner},
  J.~P., {Wadsley}, J., {Stinson}, G., \& {Quinn}, T. 2007, \apjl, 655, L17

\bibitem[{{Bruzual} \& {Charlot}(2003)}]{BC03}
{Bruzual}, G., \& {Charlot}, S. 2003, \mnras, 344, 1000

\bibitem[{{Bundy} {et~al.}(2006){Bundy}, {Ellis}, {Conselice}, {Taylor},
  {Cooper}, {Willmer}, {Weiner}, {Coil}, {Noeske}, \& {Eisenhardt}}]{Bundy06}
{Bundy}, K., {et~al.} 2006, \apj, 651, 120

\bibitem[{{Calzetti} {et~al.}(2000){Calzetti}, {Armus}, {Bohlin}, {Kinney},
  {Koornneef}, \& {Storchi-Bergmann}}]{Calzetti00}
{Calzetti}, D., {Armus}, L., {Bohlin}, R.~C., {Kinney}, A.~L., {Koornneef}, J.,
  \& {Storchi-Bergmann}, T. 2000, \apj, 533, 682

\bibitem[{{Capak} {et~al.}(2004){Capak}, {Cowie}, {Hu}, {Barger}, {Dickinson},
  {Fernandez}, {Giavalisco}, {Komiyama}, {Kretchmer}, {McNally}, {Miyazaki},
  {Okamura}, \& {Stern}}]{Capak04}
{Capak}, P., {et~al.} 2004, \aj, 127, 180

\bibitem[{{Chabrier}(2003)}]{Chabrier03}
{Chabrier}, G. 2003, \pasp, 115, 763

\bibitem[{{Chapman} {et~al.}(2005){Chapman}, {Blain}, {Smail}, \&
  {Ivison}}]{Chapman05}
{Chapman}, S.~C., {Blain}, A.~W., {Smail}, I., \& {Ivison}, R.~J. 2005, \apj,
  622, 772

\bibitem[{{Christensen} {et~al.}(2012){Christensen}, {Laursen}, {Richard},
  {Hjorth}, {Milvang-Jensen}, {Dessauges-Zavadsky}, {Limousin}, {Grillo}, \&
  {Ebeling}}]{Christensen12}
{Christensen}, L., {et~al.} 2012, ArXiv e-prints

\bibitem[{{Conroy} {et~al.}(2008){Conroy}, {Shapley}, {Tinker}, {Santos}, \&
  {Lemson}}]{Conroy08}
{Conroy}, C., {Shapley}, A.~E., {Tinker}, J.~L., {Santos}, M.~R., \& {Lemson},
  G. 2008, \apj, 679, 1192

\bibitem[{{Conselice} {et~al.}(2007){Conselice}, {Bundy}, {Trujillo}, {Coil},
  {Eisenhardt}, {Ellis}, {Georgakakis}, {Huang}, {Lotz}, {Nandra}, {Newman},
  {Papovich}, {Weiner}, \& {Willmer}}]{Conselice07}
{Conselice}, C.~J., {et~al.} 2007, \mnras, 381, 962

\bibitem[{{Cowie} \& {Barger}(2008)}]{Cowie08}
{Cowie}, L.~L., \& {Barger}, A.~J. 2008, \apj, 686, 72

\bibitem[{{Cowie} {et~al.}(1996){Cowie}, {Songaila}, {Hu}, \&
  {Cohen}}]{Cowie96}
{Cowie}, L.~L., {Songaila}, A., {Hu}, E.~M., \& {Cohen}, J.~G. 1996, \aj, 112,
  839

\bibitem[{{Daddi} {et~al.}(2004){Daddi}, {Cimatti}, {Renzini}, {Fontana},
  {Mignoli}, {Pozzetti}, {Tozzi}, \& {Zamorani}}]{Daddi04}
{Daddi}, E., {Cimatti}, A., {Renzini}, A., {Fontana}, A., {Mignoli}, M.,
  {Pozzetti}, L., {Tozzi}, P., \& {Zamorani}, G. 2004, \apj, 617, 746

\bibitem[{{Dalcanton}(2007)}]{Dalcanton07}
{Dalcanton}, J.~J. 2007, \apj, 658, 941

\bibitem[{{Dav{\'e}} {et~al.}(2011{\natexlab{a}}){Dav{\'e}}, {Finlator}, \&
  {Oppenheimer}}]{Dave11b}
{Dav{\'e}}, R., {Finlator}, K., \& {Oppenheimer}, B.~D. 2011{\natexlab{a}},
  \mnras, 416, 1354

\bibitem[{{Dav{\'e}} \& {Oppenheimer}(2007)}]{Dave07}
{Dav{\'e}}, R., \& {Oppenheimer}, B.~D. 2007, \mnras, 374, 427

\bibitem[{{Dav{\'e}} {et~al.}(2011{\natexlab{b}}){Dav{\'e}}, {Oppenheimer}, \&
  {Finlator}}]{Dave11a}
{Dav{\'e}}, R., {Oppenheimer}, B.~D., \& {Finlator}, K. 2011{\natexlab{b}},
  \mnras, 415, 11

\bibitem[{{Davies}(2007)}]{Davies07}
{Davies}, R.~I. 2007, \mnras, 375, 1099

\bibitem[{{Davis} {et~al.}(2003){Davis}, {Faber}, {Newman}, {Phillips},
  {Ellis}, {Steidel}, {Conselice}, {Coil}, {Finkbeiner}, {Koo}, {Guhathakurta},
  {Weiner}, {Schiavon}, {Willmer}, {Kaiser}, {Luppino}, {Wirth}, {Connolly},
  {Eisenhardt}, {Cooper}, \& {Gerke}}]{Davis03}
{Davis}, M., {et~al.} 2003, in Society of Photo-Optical Instrumentation
  Engineers (SPIE) Conference Series, Vol. 4834, Society of Photo-Optical
  Instrumentation Engineers (SPIE) Conference Series, ed. P.~{Guhathakurta},
  161--172

\bibitem[{{De Lucia} {et~al.}(2004){De Lucia}, {Kauffmann}, \&
  {White}}]{DeLucia04}
{De Lucia}, G., {Kauffmann}, G., \& {White}, S.~D.~M. 2004, \mnras, 349, 1101

\bibitem[{{Dickinson} {et~al.}(2003){Dickinson}, {Papovich}, {Ferguson}, \&
  {Budav{\'a}ri}}]{Dickinson03}
{Dickinson}, M., {Papovich}, C., {Ferguson}, H.~C., \& {Budav{\'a}ri}, T. 2003,
  \apj, 587, 25

\bibitem[{{Edmunds}(1990)}]{Edmunds90}
{Edmunds}, M.~G. 1990, \mnras, 246, 678

\bibitem[{{Edmunds} \& {Greenhow}(1995)}]{Edmunds95}
{Edmunds}, M.~G., \& {Greenhow}, R.~M. 1995, \mnras, 272, 241

\bibitem[{{Elbaz} {et~al.}(2011){Elbaz}, {Dickinson}, {Hwang},
  {D{\'{\i}}az-Santos}, {Magdis}, {Magnelli}, {Le Borgne}, {Galliano},
  {Pannella}, {Chanial}, {Armus}, {Charmandaris}, {Daddi}, {Aussel}, {Popesso},
  {Kartaltepe}, {Altieri}, {Valtchanov}, {Coia}, {Dannerbauer}, {Dasyra},
  {Leiton}, {Mazzarella}, {Alexander}, {Buat}, {Burgarella}, {Chary}, {Gilli},
  {Ivison}, {Juneau}, {Le Floc'h}, {Lutz}, {Morrison}, {Mullaney}, {Murphy},
  {Pope}, {Scott}, {Brodwin}, {Calzetti}, {Cesarsky}, {Charlot}, {Dole},
  {Eisenhardt}, {Ferguson}, {F{\"o}rster Schreiber}, {Frayer}, {Giavalisco},
  {Huynh}, {Koekemoer}, {Papovich}, {Reddy}, {Surace}, {Teplitz}, {Yun}, \&
  {Wilson}}]{Elbaz11}
{Elbaz}, D., {et~al.} 2011, \aap, 533, A119

\bibitem[{{Erb} {et~al.}(2010){Erb}, {Pettini}, {Shapley}, {Steidel}, {Law}, \&
  {Reddy}}]{Erb10b}
{Erb}, D.~K., {Pettini}, M., {Shapley}, A.~E., {Steidel}, C.~C., {Law}, D.~R.,
  \& {Reddy}, N.~A. 2010, \apj, 719, 1168

\bibitem[{{Erb} {et~al.}(2006){Erb}, {Shapley}, {Pettini}, {Steidel}, {Reddy},
  \& {Adelberger}}]{Erb06}
{Erb}, D.~K., {Shapley}, A.~E., {Pettini}, M., {Steidel}, C.~C., {Reddy},
  N.~A., \& {Adelberger}, K.~L. 2006, \apj, 644, 813

\bibitem[{{Erb} {et~al.}(2003){Erb}, {Shapley}, {Steidel}, {Pettini},
  {Adelberger}, {Hunt}, {Moorwood}, \& {Cuby}}]{Erb03}
{Erb}, D.~K., {Shapley}, A.~E., {Steidel}, C.~C., {Pettini}, M., {Adelberger},
  K.~L., {Hunt}, M.~P., {Moorwood}, A.~F.~M., \& {Cuby}, J.-G. 2003, \apj, 591,
  101

\bibitem[{{Fan} {et~al.}(2001){Fan}, {Strauss}, {Schneider}, {Gunn}, {Lupton},
  {Becker}, {Davis}, {Newman}, {Richards}, {White}, {Anderson}, {Annis},
  {Bahcall}, {Brunner}, {Csabai}, {Hennessy}, {Hindsley}, {Fukugita}, {Kunszt},
  {Ivezi{\'c}}, {Knapp}, {McKay}, {Munn}, {Pier}, {Szalay}, \& {York}}]{Fan01}
{Fan}, X., {et~al.} 2001, \aj, 121, 54

\bibitem[{{Finlator} \& {Dav{\'e}}(2008)}]{Finlator08}
{Finlator}, K., \& {Dav{\'e}}, R. 2008, \mnras, 385, 2181

\bibitem[{{F{\"o}rster Schreiber} {et~al.}(2006){F{\"o}rster Schreiber},
  {Genzel}, {Lehnert}, {Bouch{\'e}}, {Verma}, {Erb}, {Shapley}, {Steidel},
  {Davies}, {Lutz}, {Nesvadba}, {Tacconi}, {Eisenhauer}, {Abuter}, {Gilbert},
  {Gillessen}, \& {Sternberg}}]{FS06}
{F{\"o}rster Schreiber}, N.~M., {et~al.} 2006, \apj, 645, 1062

\bibitem[{{F{\"o}rster Schreiber} {et~al.}(2011){F{\"o}rster Schreiber},
  {Shapley}, {Erb}, {Genzel}, {Steidel}, {Bouch{\'e}}, {Cresci}, \&
  {Davies}}]{FS11}
{F{\"o}rster Schreiber}, N.~M., {Shapley}, A.~E., {Erb}, D.~K., {Genzel}, R.,
  {Steidel}, C.~C., {Bouch{\'e}}, N., {Cresci}, G., \& {Davies}, R. 2011, \apj,
  731, 65

\bibitem[{{Frye} {et~al.}(2007){Frye}, {Coe}, {Bowen}, {Ben{\'{\i}}tez},
  {Broadhurst}, {Guhathakurta}, {Illingworth}, {Menanteau}, {Sharon}, {Lupton},
  {Meylan}, {Zekser}, {Meurer}, \& {Hurley}}]{Frye07}
{Frye}, B.~L., {et~al.} 2007, \apj, 665, 921

\bibitem[{{Garnett}(2002)}]{Garnett02}
{Garnett}, D.~R. 2002, \apj, 581, 1019

\bibitem[{{Genzel} {et~al.}(2008){Genzel}, {Burkert}, {Bouch{\'e}}, {Cresci},
  {F{\"o}rster Schreiber}, {Shapley}, {Shapiro}, {Tacconi}, {Buschkamp},
  {Cimatti}, {Daddi}, {Davies}, {Eisenhauer}, {Erb}, {Genel}, {Gerhard},
  {Hicks}, {Lutz}, {Naab}, {Ott}, {Rabien}, {Renzini}, {Steidel}, {Sternberg},
  \& {Lilly}}]{Genzel08}
{Genzel}, R., {et~al.} 2008, \apj, 687, 59

\bibitem[{{Genzel} {et~al.}(2011){Genzel}, {Newman}, {Jones}, {F{\"o}rster
  Schreiber}, {Shapiro}, {Genel}, {Lilly}, {Renzini}, {Tacconi}, {Bouch{\'e}},
  {Burkert}, {Cresci}, {Buschkamp}, {Carollo}, {Ceverino}, {Davies}, {Dekel},
  {Eisenhauer}, {Hicks}, {Kurk}, {Lutz}, {Mancini}, {Naab}, {Peng},
  {Sternberg}, {Vergani}, \& {Zamorani}}]{Genzel11}
---. 2011, \apj, 733, 101

\bibitem[{{Grazian} {et~al.}(2007){Grazian}, {Salimbeni}, {Pentericci},
  {Fontana}, {Nonino}, {Vanzella}, {Cristiani}, {de Santis}, {Gallozzi},
  {Giallongo}, \& {Santini}}]{Grazian07}
{Grazian}, A., {et~al.} 2007, \aap, 465, 393

\bibitem[{{Hayashi} {et~al.}(2009){Hayashi}, {Motohara}, {Shimasaku},
  {Onodera}, {Uchimoto}, {Kashikawa}, {Yoshida}, {Okamura}, {Ly}, \&
  {Malkan}}]{Hayashi09}
{Hayashi}, M., {et~al.} 2009, \apj, 691, 140

\bibitem[{{Henry} {et~al.}(2010){Henry}, {Salvato}, {Finoguenov}, {Bouche},
  {Brunner}, {Burwitz}, {Buschkamp}, {Egami}, {F{\"o}rster-Schreiber},
  {Fotopoulou}, {Genzel}, {Hasinger}, {Mainieri}, {Rovilos}, \&
  {Szokoly}}]{HenryP10}
{Henry}, J.~P., {et~al.} 2010, \apj, 725, 615

\bibitem[{{Hopkins} \& {Beacom}(2006)}]{HopkinsAM06}
{Hopkins}, A.~M., \& {Beacom}, J.~F. 2006, \apj, 651, 142

\bibitem[{{Ichikawa} {et~al.}(2006){Ichikawa}, {Suzuki}, {Tokoku}, {Uchimoto},
  {Konishi}, {Yoshikawa}, {Yamada}, {Tanaka}, {Omata}, \&
  {Nishimura}}]{Ichikawa06}
{Ichikawa}, T., {et~al.} 2006, in Society of Photo-Optical Instrumentation
  Engineers (SPIE) Conference Series, Vol. 6269

\bibitem[{{Ilbert} {et~al.}(2009){Ilbert}, {Salvato}, {Le Floc'h}, {Aussel},
  {Capak}, {McCracken}, {Mobasher}, {Kartaltepe}, {Scoville}, {Sanders},
  {Arnouts}, {Bundy}, {Cassata}, {Kneib}, {Koekemoer}, {Le Fevre}, {Lilly},
  {Surace}, {Taniguchi}, {Tasca}, {Thompson}, {Tresse}, {Zamojski}, {Zamorani},
  \& {Zucca}}]{Ilbert09}
{Ilbert}, O., {et~al.} 2009, ArXiv e-prints

\bibitem[{{Jones} {et~al.}(2010){Jones}, {Ellis}, {Jullo}, \&
  {Richard}}]{Jones10b}
{Jones}, T., {Ellis}, R., {Jullo}, E., \& {Richard}, J. 2010, \apjl, 725, L176

\bibitem[{{Jones} {et~al.}(2012){Jones}, {Ellis}, {Richard}, \&
  {Jullo}}]{Jones12}
{Jones}, T., {Ellis}, R.~S., {Richard}, J., \& {Jullo}, E. 2012, ArXiv e-prints

\bibitem[{{Jullo} {et~al.}(2007){Jullo}, {Kneib}, {Limousin},
  {El{\'{\i}}asd{\'o}ttir}, {Marshall}, \& {Verdugo}}]{Jullo07}
{Jullo}, E., {Kneib}, J.-P., {Limousin}, M., {El{\'{\i}}asd{\'o}ttir}, {\'A}.,
  {Marshall}, P.~J., \& {Verdugo}, T. 2007, New Journal of Physics, 9, 447

\bibitem[{{Kelson}(2003)}]{Kelson03}
{Kelson}, D.~D. 2003, \pasp, 115, 688

\bibitem[{{Kennicutt}(1998)}]{Kennicutt98b}
{Kennicutt}, Jr., R.~C. 1998, \araa, 36, 189

\bibitem[{{Kewley} \& {Dopita}(2002)}]{Kewley02}
{Kewley}, L.~J., \& {Dopita}, M.~A. 2002, \apjs, 142, 35

\bibitem[{{Kewley} \& {Ellison}(2008)}]{Kewley08}
{Kewley}, L.~J., \& {Ellison}, S.~L. 2008, \apj, 681, 1183

\bibitem[{{Kewley} {et~al.}(2004){Kewley}, {Geller}, \& {Jansen}}]{Kewley04}
{Kewley}, L.~J., {Geller}, M.~J., \& {Jansen}, R.~A. 2004, \aj, 127, 2002

\bibitem[{{Kewley} {et~al.}(2006){Kewley}, {Groves}, {Kauffmann}, \&
  {Heckman}}]{Kewley06}
{Kewley}, L.~J., {Groves}, B., {Kauffmann}, G., \& {Heckman}, T. 2006, \mnras,
  372, 961

\bibitem[{{Kneib} {et~al.}(1993){Kneib}, {Mellier}, {Fort}, \&
  {Mathez}}]{Kneib93}
{Kneib}, J.~P., {Mellier}, Y., {Fort}, B., \& {Mathez}, G. 1993, \aap, 273, 367

\bibitem[{{Kobulnicky} \& {Kewley}(2004)}]{Kobulnicky04}
{Kobulnicky}, H.~A., \& {Kewley}, L.~J. 2004, \apj, 617, 240

\bibitem[{{K{\"o}ppen} \& {Edmunds}(1999)}]{Koppen99}
{K{\"o}ppen}, J., \& {Edmunds}, M.~G. 1999, \mnras, 306, 317

\bibitem[{{Kriek} {et~al.}(2007){Kriek}, {van Dokkum}, {Franx}, {Illingworth},
  {Coppi}, {F{\"o}rster Schreiber}, {Gawiser}, {Labb{\'e}}, {Lira},
  {Marchesini}, {Quadri}, {Rudnick}, {Taylor}, {Urry}, \& {van der
  Werf}}]{Kriek07}
{Kriek}, M., {et~al.} 2007, \apj, 669, 776

\bibitem[{{Lacey} \& {Fall}(1985)}]{Lacey85}
{Lacey}, C.~G., \& {Fall}, S.~M. 1985, \apj, 290, 154

\bibitem[{{Lamareille} {et~al.}(2009){Lamareille}, {Brinchmann}, {Contini},
  {Walcher}, {Charlot}, {P{\'e}rez-Montero}, {Zamorani}, {Pozzetti},
  {Bolzonella}, {Garilli}, {Paltani}, {Bongiorno}, {Le F{\`e}vre}, {Bottini},
  {Le Brun}, {Maccagni}, {Scaramella}, {Scodeggio}, {Tresse}, {Vettolani},
  {Zanichelli}, {Adami}, {Arnouts}, {Bardelli}, {Cappi}, {Ciliegi}, {Foucaud},
  {Franzetti}, {Gavignaud}, {Guzzo}, {Ilbert}, {Iovino}, {McCracken}, {Marano},
  {Marinoni}, {Mazure}, {Meneux}, {Merighi}, {Pell{\`o}}, {Pollo}, {Radovich},
  {Vergani}, {Zucca}, {Romano}, {Grado}, \& {Limatola}}]{Lamareille09}
{Lamareille}, F., {et~al.} 2009, \aap, 495, 53

\bibitem[{{Lara-Lopez} {et~al.}(2012){Lara-Lopez}, {Lopez-Sanchez}, \&
  {Hopkins}}]{Lara12}
{Lara-Lopez}, M.~A., {Lopez-Sanchez}, A.~R., \& {Hopkins}, A.~M. 2012, ArXiv
  e-prints

\bibitem[{{Larkin} {et~al.}(2006){Larkin}, {Barczys}, {Krabbe}, {Adkins},
  {Aliado}, {Amico}, {Brims}, {Campbell}, {Canfield}, {Gasaway}, {Honey},
  {Iserlohe}, {Johnson}, {Kress}, {Lafreniere}, {Magnone}, {Magnone},
  {McElwain}, {Moon}, {Quirrenbach}, {Skulason}, {Song}, {Spencer}, {Weiss}, \&
  {Wright}}]{Larkin06}
{Larkin}, J., {et~al.} 2006, New Astronomy Reviews, 50, 362

\bibitem[{{Law} {et~al.}(2009){Law}, {Steidel}, {Erb}, {Larkin}, {Pettini},
  {Shapley}, \& {Wright}}]{Law09}
{Law}, D.~R., {Steidel}, C.~C., {Erb}, D.~K., {Larkin}, J.~E., {Pettini}, M.,
  {Shapley}, A.~E., \& {Wright}, S.~A. 2009, \apj, 697, 2057

\bibitem[{{Lemoine-Busserolle} {et~al.}(2003){Lemoine-Busserolle}, {Contini},
  {Pell{\'o}}, {Le Borgne}, {Kneib}, \& {Lidman}}]{LB03}
{Lemoine-Busserolle}, M., {Contini}, T., {Pell{\'o}}, R., {Le Borgne}, J.-F.,
  {Kneib}, J.-P., \& {Lidman}, C. 2003, \aap, 397, 839

\bibitem[{{Lequeux} {et~al.}(1979){Lequeux}, {Peimbert}, {Rayo}, {Serrano}, \&
  {Torres-Peimbert}}]{Lequeux79}
{Lequeux}, J., {Peimbert}, M., {Rayo}, J.~F., {Serrano}, A., \&
  {Torres-Peimbert}, S. 1979, \aap, 80, 155

\bibitem[{{Limousin} {et~al.}(2007){Limousin}, {Richard}, {Jullo}, {Kneib},
  {Fort}, {Soucail}, {El{\'{\i}}asd{\'o}ttir}, {Natarajan}, {Ellis}, {Smail},
  {Czoske}, {Smith}, {Hudelot}, {Bardeau}, {Ebeling}, {Egami}, \&
  {Knudsen}}]{Limousin07}
{Limousin}, M., {et~al.} 2007, \apj, 668, 643

\bibitem[{{Liu} {et~al.}(2008){Liu}, {Shapley}, {Coil}, {Brinchmann}, \&
  {Ma}}]{Liu08}
{Liu}, X., {Shapley}, A.~E., {Coil}, A.~L., {Brinchmann}, J., \& {Ma}, C.-P.
  2008, \apj, 678, 758

\bibitem[{{Maiolino} {et~al.}(2008){Maiolino}, {Nagao}, {Grazian}, {Cocchia},
  {Marconi}, {Mannucci}, {Cimatti}, {Pipino}, {Ballero}, {Calura}, {Chiappini},
  {Fontana}, {Granato}, {Matteucci}, {Pastorini}, {Pentericci}, {Risaliti},
  {Salvati}, \& {Silva}}]{Maiolino08}
{Maiolino}, R., {et~al.} 2008, \aap, 488, 463

\bibitem[{{Mannucci} {et~al.}(2010){Mannucci}, {Cresci}, {Maiolino}, {Marconi},
  \& {Gnerucci}}]{Mannucci10}
{Mannucci}, F., {Cresci}, G., {Maiolino}, R., {Marconi}, A., \& {Gnerucci}, A.
  2010, \mnras, 408, 2115

\bibitem[{{McLean} {et~al.}(1998){McLean}, {Becklin}, {Bendiksen}, {Brims},
  {Canfield}, {Figer}, {Graham}, {Hare}, {Lacayanga}, {Larkin}, {Larson},
  {Levenson}, {Magnone}, {Teplitz}, \& {Wong}}]{McLean98}
{McLean}, I.~S., {et~al.} 1998, in Presented at the Society of Photo-Optical
  Instrumentation Engineers (SPIE) Conference, Vol. 3354, Society of
  Photo-Optical Instrumentation Engineers (SPIE) Conference Series, ed.
  {A.~M.~Fowler}, 566--578

\bibitem[{{Nagamine} {et~al.}(2001){Nagamine}, {Fukugita}, {Cen}, \&
  {Ostriker}}]{Nagamine01}
{Nagamine}, K., {Fukugita}, M., {Cen}, R., \& {Ostriker}, J.~P. 2001, \apj,
  558, 497

\bibitem[{{Noeske} {et~al.}(2007{\natexlab{a}}){Noeske}, {Faber}, {Weiner},
  {Koo}, {Primack}, {Dekel}, {Papovich}, {Conselice}, {Le Floc'h}, {Rieke},
  {Coil}, {Lotz}, {Somerville}, \& {Bundy}}]{Noeske07b}
{Noeske}, K.~G., {et~al.} 2007{\natexlab{a}}, \apjl, 660, L47

\bibitem[{{Noeske} {et~al.}(2007{\natexlab{b}}){Noeske}, {Weiner}, {Faber},
  {Papovich}, {Koo}, {Somerville}, {Bundy}, {Conselice}, {Newman},
  {Schiminovich}, {Le Floc'h}, {Coil}, {Rieke}, {Lotz}, {Primack}, {Barmby},
  {Cooper}, {Davis}, {Ellis}, {Fazio}, {Guhathakurta}, {Huang}, {Kassin},
  {Martin}, {Phillips}, {Rich}, {Small}, {Willmer}, \& {Wilson}}]{Noeske07a}
---. 2007{\natexlab{b}}, \apjl, 660, L43

\bibitem[{{Oke} {et~al.}(1995){Oke}, {Cohen}, {Carr}, {Cromer}, {Dingizian},
  {Harris}, {Labrecque}, {Lucinio}, {Schaal}, {Epps}, \& {Miller}}]{Oke95}
{Oke}, J.~B., {et~al.} 1995, \pasp, 107, 375

\bibitem[{{Oppenheimer} \& {Dav{\'e}}(2008)}]{Oppenheimer08}
{Oppenheimer}, B.~D., \& {Dav{\'e}}, R. 2008, \mnras, 387, 577

\bibitem[{{Osterbrock}(1989)}]{Osterbrock89}
{Osterbrock}, D.~E. 1989, {Astrophysics of gaseous nebulae and active galactic
  nuclei} (Research supported by the University of California, John Simon
  Guggenheim Memorial Foundation, University of Minnesota, et al.~Mill Valley,
  CA, University Science Books, 1989, 422 p.)

\bibitem[{{Pagel} \& {Edmunds}(1981)}]{Pagel81}
{Pagel}, B.~E.~J., \& {Edmunds}, M.~G. 1981, \araa, 19, 77

\bibitem[{{Pagel} \& {Patchett}(1975)}]{Pagel75}
{Pagel}, B.~E.~J., \& {Patchett}, B.~E. 1975, \mnras, 172, 13

\bibitem[{{Panter} {et~al.}(2008){Panter}, {Jimenez}, {Heavens}, \&
  {Charlot}}]{Panter08}
{Panter}, B., {Jimenez}, R., {Heavens}, A.~F., \& {Charlot}, S. 2008, \mnras,
  391, 1117

\bibitem[{{Pettini} \& {Pagel}(2004)}]{Pettini04}
{Pettini}, M., \& {Pagel}, B.~E.~J. 2004, \mnras, 348, L59

\bibitem[{{Pettini} {et~al.}(2001){Pettini}, {Shapley}, {Steidel}, {Cuby},
  {Dickinson}, {Moorwood}, {Adelberger}, \& {Giavalisco}}]{Pettini01}
{Pettini}, M., {Shapley}, A.~E., {Steidel}, C.~C., {Cuby}, J.-G., {Dickinson},
  M., {Moorwood}, A.~F.~M., {Adelberger}, K.~L., \& {Giavalisco}, M. 2001,
  \apj, 554, 981

\bibitem[{{Quider} {et~al.}(2009){Quider}, {Pettini}, {Shapley}, \&
  {Steidel}}]{Quider09}
{Quider}, A.~M., {Pettini}, M., {Shapley}, A.~E., \& {Steidel}, C.~C. 2009,
  \mnras, 1081

\bibitem[{{Rafelski} {et~al.}(2012){Rafelski}, {Wolfe}, {Prochaska},
  {Neeleman}, \& {Mendez}}]{Rafelski12}
{Rafelski}, M., {Wolfe}, A.~M., {Prochaska}, J.~X., {Neeleman}, M., \&
  {Mendez}, A.~J. 2012, \apj, 755, 89

\bibitem[{{Reddy} {et~al.}(2006){Reddy}, {Steidel}, {Erb}, {Shapley}, \&
  {Pettini}}]{Reddy06}
{Reddy}, N.~A., {Steidel}, C.~C., {Erb}, D.~K., {Shapley}, A.~E., \& {Pettini},
  M. 2006, \apj, 653, 1004

\bibitem[{{Reddy} {et~al.}(2008){Reddy}, {Steidel}, {Pettini}, {Adelberger},
  {Shapley}, {Erb}, \& {Dickinson}}]{Reddy08}
{Reddy}, N.~A., {Steidel}, C.~C., {Pettini}, M., {Adelberger}, K.~L.,
  {Shapley}, A.~E., {Erb}, D.~K., \& {Dickinson}, M. 2008, \apjs, 175, 48

\bibitem[{{Richard} {et~al.}(2011){Richard}, {Jones}, {Ellis}, {Stark},
  {Livermore}, \& {Swinbank}}]{Richard11a}
{Richard}, J., {Jones}, T., {Ellis}, R., {Stark}, D.~P., {Livermore}, R., \&
  {Swinbank}, M. 2011, \mnras, 126

\bibitem[{{Richard} {et~al.}(2007){Richard}, {Kneib}, {Jullo}, {Covone},
  {Limousin}, {Ellis}, {Stark}, {Bundy}, {Czoske}, {Ebeling}, \&
  {Soucail}}]{Richard07}
{Richard}, J., {et~al.} 2007, \apj, 662, 781

\bibitem[{{Rubin} {et~al.}(1984){Rubin}, {Ford}, \& {Whitmore}}]{Rubin84}
{Rubin}, V.~C., {Ford}, Jr., W.~K., \& {Whitmore}, B.~C. 1984, \apjl, 281, L21

\bibitem[{{Savaglio} {et~al.}(2005){Savaglio}, {Glazebrook}, {Le Borgne},
  {Juneau}, {Abraham}, {Chen}, {Crampton}, {McCarthy}, {Carlberg}, {Marzke},
  {Roth}, {J{\o}rgensen}, \& {Murowinski}}]{Savaglio05}
{Savaglio}, S., {et~al.} 2005, \apj, 635, 260

\bibitem[{{Searle} \& {Sargent}(1972)}]{Searle72}
{Searle}, L., \& {Sargent}, W.~L.~W. 1972, \apj, 173, 25

\bibitem[{{Shapley} {et~al.}(2005){Shapley}, {Coil}, {Ma}, \&
  {Bundy}}]{Shapley05}
{Shapley}, A.~E., {Coil}, A.~L., {Ma}, C.-P., \& {Bundy}, K. 2005, \apj, 635,
  1006

\bibitem[{{Skillman} {et~al.}(1989){Skillman}, {Kennicutt}, \&
  {Hodge}}]{Skillman89}
{Skillman}, E.~D., {Kennicutt}, R.~C., \& {Hodge}, P.~W. 1989, \apj, 347, 875

\bibitem[{{Sobral} {et~al.}(2009){Sobral}, {Best}, {Geach}, {Smail}, {Kurk},
  {Cirasuolo}, {Casali}, {Ivison}, {Coppin}, \& {Dalton}}]{Sobral09b}
{Sobral}, D., {et~al.} 2009, \mnras, 398, 75

\bibitem[{{Songaila} \& {Cowie}(2002)}]{Songaila02}
{Songaila}, A., \& {Cowie}, L.~L. 2002, \aj, 123, 2183

\bibitem[{{Stark} {et~al.}(2008){Stark}, {Swinbank}, {Ellis}, {Dye}, {Smail},
  \& {Richard}}]{Stark08}
{Stark}, D.~P., {Swinbank}, A.~M., {Ellis}, R.~S., {Dye}, S., {Smail}, I.~R.,
  \& {Richard}, J. 2008, \nat, 455, 775

\bibitem[{{Steidel} {et~al.}(2003){Steidel}, {Adelberger}, {Shapley},
  {Pettini}, {Dickinson}, \& {Giavalisco}}]{Steidel03}
{Steidel}, C.~C., {Adelberger}, K.~L., {Shapley}, A.~E., {Pettini}, M.,
  {Dickinson}, M., \& {Giavalisco}, M. 2003, \apj, 592, 728

\bibitem[{{Steidel} {et~al.}(1996){Steidel}, {Giavalisco}, {Pettini},
  {Dickinson}, \& {Adelberger}}]{Steidel96}
{Steidel}, C.~C., {Giavalisco}, M., {Pettini}, M., {Dickinson}, M., \&
  {Adelberger}, K.~L. 1996, \apjl, 462, L17

\bibitem[{{Steidel} {et~al.}(2004){Steidel}, {Shapley}, {Pettini},
  {Adelberger}, {Erb}, {Reddy}, \& {Hunt}}]{Steidel04}
{Steidel}, C.~C., {Shapley}, A.~E., {Pettini}, M., {Adelberger}, K.~L., {Erb},
  D.~K., {Reddy}, N.~A., \& {Hunt}, M.~P. 2004, \apj, 604, 534

\bibitem[{{Swinbank} {et~al.}(2009){Swinbank}, {Webb}, {Richard}, {Bower},
  {Ellis}, {Illingworth}, {Jones}, {Kriek}, {Smail}, {Stark}, \& {van
  Dokkum}}]{Swinbank09a}
{Swinbank}, A.~M., {et~al.} 2009, \mnras, 400, 1121

\bibitem[{{Tremonti} {et~al.}(2004){Tremonti}, {Heckman}, {Kauffmann},
  {Brinchmann}, {Charlot}, {White}, {Seibert}, {Peng}, {Schlegel}, {Uomoto},
  {Fukugita}, \& {Brinkmann}}]{Tremonti04}
{Tremonti}, C.~A., {et~al.} 2004, \apj, 613, 898

\bibitem[{{Veilleux} \& {Osterbrock}(1987)}]{Veilleux87}
{Veilleux}, S., \& {Osterbrock}, D.~E. 1987, \apjs, 63, 295

\bibitem[{{Wuyts} {et~al.}(2012){Wuyts}, {Rigby}, {Sharon}, \&
  {Gladders}}]{Wuyts12b}
{Wuyts}, E., {Rigby}, J.~R., {Sharon}, K., \& {Gladders}, M.~D. 2012, \apj,
  755, 73

\bibitem[{{Wuyts} {et~al.}(2011){Wuyts}, {F{\"o}rster Schreiber}, {van der
  Wel}, {Magnelli}, {Guo}, {Genzel}, {Lutz}, {Aussel}, {Barro}, {Berta},
  {Cava}, {Graci{\'a}-Carpio}, {Hathi}, {Huang}, {Kocevski}, {Koekemoer},
  {Lee}, {Le Floc'h}, {McGrath}, {Nordon}, {Popesso}, {Pozzi}, {Riguccini},
  {Rodighiero}, {Saintonge}, \& {Tacconi}}]{Wuyts_S11}
{Wuyts}, S., {et~al.} 2011, \apj, 742, 96

\bibitem[{{Yabe} {et~al.}(2012){Yabe}, {Ohta}, {Iwamuro}, {Yuma}, {Akiyama},
  {Tamura}, {Kimura}, {Takato}, {Moritani}, {Sumiyoshi}, {Maihara},
  {Silverman}, {Dalton}, {Lewis}, {Bonfield}, {Lee}, {Curtis Lake}, {Macaulay},
  \& {Clarke}}]{Yabe12}
{Yabe}, K., {et~al.} 2012, \pasj, 64, 60

\bibitem[{{Yates} {et~al.}(2012){Yates}, {Kauffmann}, \& {Guo}}]{Yates12}
{Yates}, R.~M., {Kauffmann}, G., \& {Guo}, Q. 2012, \mnras, 422, 215

\bibitem[{{Yuan} \& {Kewley}(2009)}]{Yuan09}
{Yuan}, T.-T., \& {Kewley}, L.~J. 2009, \apjl, 699, L161

\bibitem[{{Yuan} {et~al.}(2011){Yuan}, {Kewley}, {Swinbank}, {Richard}, \&
  {Livermore}}]{Yuan11}
{Yuan}, T.-T., {Kewley}, L.~J., {Swinbank}, A.~M., {Richard}, J., \&
  {Livermore}, R.~C. 2011, \apjl, 732, L14+

\bibitem[{{Zahid} {et~al.}(2012){Zahid}, {Dima}, {Kewley}, {Erb}, \&
  {Dave}}]{Zahid12b}
{Zahid}, H.~J., {Dima}, G.~I., {Kewley}, L.~J., {Erb}, D.~K., \& {Dave}, R.
  2012, ArXiv e-prints

\bibitem[{{Zahid} {et~al.}(2011){Zahid}, {Kewley}, \& {Bresolin}}]{Zahid11}
{Zahid}, H.~J., {Kewley}, L.~J., \& {Bresolin}, F. 2011, \apj, 730, 137

\bibitem[{{Zaritsky} {et~al.}(1994){Zaritsky}, {Kennicutt}, \&
  {Huchra}}]{Zaritsky94}
{Zaritsky}, D., {Kennicutt}, Jr., R.~C., \& {Huchra}, J.~P. 1994, \apj, 420, 87

\end{thebibliography}

\end{document}